\documentclass{article}
\input{tcilatex}

\begin{document}

\begin{center}
{\LARGE GRASSMANN VARIABLES AND\ \bigskip }

{\LARGE THE JAYNES-CUMMINGS MODEL}
\end{center}

\bigskip

\begin{center}
B. J. Dalton

ARC Centre for Quantum-Atom Optics

and

Centre for Atom Optics and Ultrafast Spectroscopy

Swinburne University of Technology, Melbourne, Victoria 3122,
Australia\bigskip

B. M. Garraway

Department of Physics and Astronomy

University of Sussex, Falmer, Brighton BN19QH, UK\bigskip

J. Jeffers

Department of Physics

University of Strathclyde, Glasgow, G40NG, UK\bigskip

S. M. Barnett

Department of Physics

University of Strathclyde, Glasgow, G40NG, UK\pagebreak 
\end{center}

\subsection{Abstract}

The Jaynes-Cummings model of a two-level atom in a single mode cavity is of
fundamental importance both in quantum optics and in quantum physics
generally, involving the interaction of two simple quantum systems - one
fermionic system (the TLA), the other bosonic (the cavity mode). Depending
on the initial conditions a variety of interesting effects occur, ranging
from ongoing oscillations of the atomic population difference at the Rabi
frequency when the atom is excited and the cavity is in an n-photon Fock
state, to collapses and revivals of these oscillations starting with the
atom unexcited and the cavity mode in a coherent state. The observation of
revivals for Rydberg atoms in a high-Q microwave cavity is key experimental
evidence for quantization of the EM field. Theoretical treatments of the
Jaynes-Cummings model based on expanding the state vector in terms on
products of atomic and n-photon states and deriving coupled equations for
the amplitudes are a well-known and simple method for determining the
effects. 

In quantum optics however, the behaviour of the bosonic quantum EM field is
often treated using phase space methods, where the bosonic mode annihilation
and creation operators are represented by c-number phase space variables,
with the density operator represented by a distribution function of these
variables. Fokker-Planck equations for the distribution function are
obtained, and either used directly to determine quantities of experimental
interest or used to develop c-number Langevin equations for stochastic
versions of the phase space variables from which experimental quantities are
obtained as stochastic averages. Phase space methods have also been
developed to include atomic systems, with the atomic spin operators being
represented by c--number phase space variables, and distribution functions
involving these variables and those for any bosonic modes being shown to
satisfy Fokker-Planck equations from which c-number Langevin equations are
often developed. However, atomic spin operators satisfy the standard angular
momentum commutation rules rather than the commutation rules for bosonic
annihilation and creation operators, and are in fact second order
combinations of fermion annihilation and creation operators. Though phase
space methods in which the fermion operators are represented directly by
c-number phase space variables have not been successful, the
anti-commutation rules for these operators suggests the possibility of using
Grassmann variables - which have similar anti-commutation properties.
However, in spite of the seminal work by Cahill and Glauber and a few
applications, the use of phase space methods in quantum optics to treat
fermionic systems by representing fermionic annihilation and creation
operators directly by Grassmann phase space variables is rather rare. This
paper shows that phase space methods using a positive P type distribution
function involving both c-number variables (for the cavity mode) and
Grassmann variables (for the TLA) can be used to treat the Jaynes-Cummings
model. Although it is a Grassmann function, the distribution function is
equivalent to six c-number functions of the two bosonic variables.
Experimental quantities are given as bosonic phase space integrals involving
the six functions. A Fokker-Planck equation involving both left and right
Grassmann differentiation can be obtained for the distribution function, and
is equivalent to six coupled equations for the six c-number functions. 

The approach used involves choosing the canonical form of the (non-unique)
positive P distribution function, in which the correspondence rules for the
bosonic operators are non-standard and hence the Fokker-Planck equation is
also unusual. Initial conditions, such as those above for initially
uncorrelated states, are discussed and used to determine the initial
distribution function. Transformations to new bosonic variables rotating at
the cavity frequency enables the six coupled equations for the new c-number
functions - that are also equivalent to the canonical Grassmann distribution
function - to be solved analytically, based on an ansatz from an earlier
paper by Stenholm. It is then shown that the distribution function is
exactly the same as that determined from the well-known solution based on
coupled amplitude equations.

In quantum-atom optics theories for many atom bosonic and fermionic systems
are needed. With large atom numbers, treatments must often take into account
many quantum modes - especially for fermions. Generalisations of phase space
distribution functions of phase space variables for a few modes to phase
space distribution functionals of field functions (which represent the field
operators, c-number fields for bosons, Grassmann fields for fermions) are
now being developed for large systems. For the fermionic case, the treatment
of the simple two mode problem represented by the Jaynes-Cummings model is a
useful test case for the future development of phase space Grassmann
distribution functional methods for fermionic applications in quantum-atom
optics.

\begin{center}
\pagebreak 
\end{center}

\section{Introduction}

\label{Section - Introduction}

The simplest theoretical model in quantum optics involves a two state atomic
system coupled to a single mode quantum electromagmetic (EM)\ field, in
which the system transition frequency is in near resonance with the mode
frequency. This system was first treated using the rotating wave
approximation and is referred to as the Jaynes-Cummings model \cite%
{JaynesCummings63a}. The model can be used not only in cavity quantum
electrodynamics (QED) to describe a two level atom (TLA)\ in a high-Q cavity
- which could be two Rydberg atom states in a microwave cavity, or two lower
lying atomic states in an optical cavity, but also in laser spectroscopy for
treating a two level atom in free space coupled to a single mode laser field
in a coherent state. The Jaynes-Cummings model is of fundamental importance
both in quantum optics and in quantum physics generally, involving the
interaction of two simple quantum systems - one being fermionic (the TLA),
the other bosonic (the cavity mode). In laser spectroscopy the prediction 
\cite{Mollow69a}, interpretation \cite{Cohen76a} and observation \cite%
{Schuda74a}, \cite{Wu75a}, \cite{Hartig76a} of the three peak resonance
fluorescence spectrum for a TLA strongly driven by a narrow bandwidth laser
field stimulated much interest in the new research field of quantum optics.%
\emph{\ }In the cavity QED\ case, a variety of interesting effects occur
depending on the initial conditions, ranging from ongoing oscillations of
the atomic population difference at the Rabi frequency when the atom is
excited and the cavity is in an n-photon Fock state, to collapses and
revivals of these oscillations starting with the atom unexcited and the
cavity mode in a coherent state\textbf{\ }\cite{Eberly80a}, \cite{Barnett86a}%
, \cite{Gerry05a}. The collapse time corresponds to the time scale for phase
factors with photon numbers at the extremes of the photon number
distribution to get out of phase by $\pi $. The revival time corresponds to
when the phase factors for neighboring photon numbers get out of phase by $%
2\pi $, and hence rephase. For the coherent state case the revival time is $%
\sqrt{\overline{n}}$ times longer than the collapse time, where $\overline{n}
$ is the mean photon number of the coherent field. The observation of
revivals \cite{Rempe87a}, \cite{Haroche96a} for Rydberg atoms in a high-Q
microwave cavity is key experimental evidence for quantization of the EM
field. Over 1200 papers have been written about the Jaynes-Cummings model
indicating its importance in quantum optics, and a comprehensive review is
provided by Shore and Knight \cite{Shore93a}.

Theoretical treatments of the Jaynes-Cummings model based on expanding the
state vector in terms of products of atomic and n-photon states and deriving
coupled equations for the coefficients are a well-known and simple method
for determining the effects \cite{Barnett97a}, \cite{Gerry05a}. Another
simple approach is based on using the actual atom-cavity mode energy
eigenstates - the dressed states \cite{Cohen76a} - to describe the
behaviour. In Stenholm's approach \cite{Stenholm81a} the state vector was
expanded in products of Bargmann states for the field and spinor states for
the TLA. However, in quantum optics the behaviour of the bosonic quantum EM
field is often treated using phase space methods \cite{Gardiner91a}, \cite%
{Walls94a}, \cite{Barnett97a}, \cite{Scully97a} with the density operator
represented by a \emph{phase space distribution function }and the bosonic
mode annihilation and creation operators represented by c-number \emph{phase
space variables}. \emph{Fokker-Planck equations} for the distribution
function are obtained, and either used directly to determine quantities of
experimental interest or used to develop c-number \emph{Langevin equations}
for stochastic versions of the phase space variables, such as when
distribution functions of the positive $P$ type \cite{Drummond80a} are used.
Phase space methods have also been developed to include atomic systems \cite%
{Gardiner91a}, \cite{Walls94a}, with the atomic spin operators being
represented by c--number phase space variables, and distribution functions
involving these variables and those for any bosonic modes being shown to
satisfy Fokker-Planck equations from which c-number Langevin equations are
often developed. However, atomic spin operators satisfy the standard angular
momentum commutation rules rather than the commutation rules for bosonic
annihilation and creation operators, and are in fact second order
combinations of fermion annihilation and creation operators. It would
perhaps be more natural to use a phase space method in which the phase space
variables represent the fermion operators directly. The problem however is
that it is not possible for fermion annihilation and creation operators to
be represented by c-number variables. The anti-commutation rules for these
operators suggests the use of variables which have similar anti-commutation
properties, and such variables are \emph{Grassmann numbers}. These numbers
were invented in the 19th century by the mathematician Hermann Grassmann,
but their properties would be unfamiliar to many physicists. A general
account of Grassmann variables is to be found in the book by Berezin \cite%
{Berezin66a}.

Apart from the seminal work by Cahill and Glauber \cite{Cahill99a} and a few
applications \cite{Plimak01a}, \cite{Anastopoulos00a}, \cite{Shresta05a}, 
\cite{Plimak09a} the use of phase space methods in quantum optics to treat
fermionic systems by representing fermionic annihilation and creation
operators with Grassmann variables \cite{Berezin66a}, \cite{Cahill99a} is
rather rare. In contrast, this approach is widely used in particle physics 
\cite{Zinn-Justin02a}, \cite{Rivers87a}, \cite{Blaizot86a}. This paper shows
how phase space methods using a distribution function involving both
c-number variables (for the cavity mode) and Grassmann variables (for the
TLA) can be used to treat the Jaynes-Cummings model. Quantities of
experimental interest such as atomic state populations and coherences or
mean numbers of photons are related to \emph{quantum correlation functions},
which are expectation values of normally ordered products of boson, fermion
annihhilation and creation operators. Accordingly, the distribution function
chosen is of the \emph{positive} $P$ type for the bosonic and is similar to
the \emph{complex P} type for the fermionic variables (for short, a positive 
$P$ distribution). Unlike the more usual Glauber-Sudarshan $P$, Husimi $Q$
or Wigner $W$ distributions, the positive and complex $P$ distributions
involve a \emph{double phase space}, with two phase space variables for each
bosonic or fermonic mode. The distribution function is \emph{not unique} but
the existence of the distribution function in a canonical form can be
established. The distribution function is a \emph{Grassmann function}, and
is specified by six c-number functions of the two bosonic variables.
Experimental quantities are given as bosonic phase space integrals involving
the six functions. The non-uniqueness of the positive P distribution is
reflected in different Fokker-Planck equations based on differing
correspondence rules. Fokker-Planck equations involving both left and right
Grassmann differentiation can be obtained for both the canonical and
standard distribution functions, and these are equivalent to six coupled
equations for the six c-number bosonic functions. Initial conditions, such
as for initially uncorrelated states, are discussed and used to determine
the initial canonical distribution function. The positive $P$ type
distribution functions would enable the phase space variables to be replaced
by stochastic variables and the Fokker-Planck equation replaced by Langevin
stochastic equations of the Ito type. However, this development is not
needed for the present paper as the Fokker-Planck equation for the canonical
distribution function is solved directly via an adaption of Stenholm's
method \cite{Stenholm81a}. The solution is entirely in agreement with the
standard quantum optics results. However, the analogous solution of the
Fokker-Planck equation for the standard distribution function leads to a
distribution function that diverges on the phase space boundary and for
large times, indicating that the original Fokker-Planck equation was not
valid. Applications of the positive $P$ distribution using the canonical
distribution function may be preferable to those where the derivation of the
Fokker-Planck equation is based on the standard correspondence rules. For
the canonical distribution function more general Fokker-Planck equations
involving derivatives higher than second order may occur \cite{Schack91a},
so that no replacement by Langevin stochastic equations is possible. The
Fokker-Planck equation may also not have a positive definite diffusion
matrix.

The use of phase space approaches to the Jaynes-Cummings model is itself
rather unusual, however the paper by Eiselt and Risken \cite{Eiselt91a} does
use such a method.\textbf{\ }A generalised Wigner distribution for the field
mode is constructed from four field mode operators given by the matrix
elements of the full density operator with the two atomic states, then
combined with the field mode annihilation and creation operators. A
generalised density operator for the field with six components is obtained
and used to define a six component characteristic function and hence a six
component Wigner function. Damping effects are also included. For zero
damping, collapse and revival effects are found for the population
difference of the atomic states. We note that their approach also requires
consideration of six field distribution functions, as is required in the
present work. However, their approach cannot be related in any simple way to
ours.

Although it is of interest to show that the Jaynes-Cummings model can be
treated via Grassmann phase space methods the work presented here has more
general relevance. In quantum-atom optics Bose-Einstein condensates and
degenerate Fermi gases are being studied and theories for many atom bosonic
and fermionic systems are needed. With large atom numbers, treatments must
often take into account many quantum modes - especially for fermions in view
of the Pauli exclusion principle. Phase space distribution function methods
based on separate variables for each mode then start to become unwieldy.
Generalisations of phase space distribution function methods involving phase
space variables for a few modes to \emph{phase space} \emph{distribution
functional} methods based on field functions which represent the entire
field operators are now being developed for large systems. The field
functions are c-number fields for the bosonic case, Grassmann fields for the
case of fermions. This development is more advanced for bosonic systems,
with applications already made to problems such as optical light in
dispersive non-linear media \cite{Drummond87a}, \cite{Kennedy88a}, spatial
squeezing in quantum EM\ fields \cite{Gatti97a} and quantum noise in
Bose-Einstein condensates \cite{Steel98a}. However, distribution functional
methods involving atomic media have also been formulated \cite{Graham70a},
based on a different approach in which the atomic system is treated in terms
of c-number fields representing the atomic spin field operators rather than
the particle field creation and annihilation operators. More recently, an
alternative phase space approach for fermions, the Gaussian operator basis
method \cite{Corney06a}, originally developed for bosonic systems \cite%
{Corney03a} has been published. Here the fermionic density operators are
represented as a positive distribution over a generalised phase space. The
phase space variables again are c-numbers, now associated with pairs of
fermion annihilation and/or creation operators, and Grassmann numbers are
only used to establish properties of the Gaussian operators. However, less
progress has been made for the different approach - analogous to that
applied to bosonic systems - in which the fermion particle field operators
are represented by Grassmann fields, and the treatment of the simple two
mode problem represented by the Jaynes-Cummings model is a useful first step
in the development of this type of phase space Grassmann distribution
functional methods for applications in quantum-atom optics. For such fermion
multimode systems, functional Fokker-Planck equations would be obtained for
the distribution functional, and these would be converted into equivalent
Langevin stochastic field equations of the Ito type. It is expected that the
latter step would be important for practical numerical applications, since
solving for the full distribution functional would likely be impractical,
and unnecessary if the required physical predictions can be made via less
computer intensive stochastic methods.

In section \ref{Section - Basic Physics of Jaynes Cummings Model} the basic
physics of the Jaynes-Cummings model is reviewed, with the Hamiltonian, the
physical and spin states, atomic population and transition operators, photon
number operators and spin operators being treated in terms of fermion and
boson creation, annihilation operators. The phase space approach is
developed in section \ref{Section - Phase Space Theory}, starting with
defining the quantum correlation functions then introducing the
characteristic function and its phase space integral relation to the
distribution function. The characteristic and distribution functions are
shown to be certain Grassmann functions with c-number coefficients that are
functions of the bosonic phase space variables, and the expression for the
canonical form of the distribution function is presented. Results for the
quantum correlation functions, atomic populations and coherences are
obtained in terms of phase space integrals. In section \ref{Section -
Fokker-Planck Equation} correspondence rules for the effects of boson and
fermion operators on the density operator are obtained, and used to derive
Fokker-Planck equations for the Jaynes-Cummings model. The standard and
canonical correspondence rules are presented and the Fokker-Planck equation
is obtained for the canonical distribution function. Sets of coupled
equations for the c-number distribution function coefficients are obtained
from the Fokker-Planck equations. The original c-number distribution
function coefficients are replaced by coefficients in a frame rotating at
the cavity freqency and expressed in terms of rotating phase space
variables. The initial form of the distribution function is found for the
case of uncorrelated initial states, such as those leading to effects such
as Rabi oscillations and collapse/revival phenomena. In section \ref{Section
- Fokker-Planck Equation Solution} the coupled c-number equations are solved
analytically for the canonical distribution case\textbf{\ }and shown to
predict the characteristic the Jaynes-Cummings mode effects. The conclusions
for the paper are set out in section \ref{Section - Conclusion}. Basic
results for Grassmann algebra and calculus and for boson and fermion
Bargmann states are presented in the Appendices.\medskip

\section{Basic Physics of Jaynes-Cummings Model}

\label{Section - Basic Physics of Jaynes Cummings Model}

\subsection{Atomic System and Cavity Mode}

The basic atomic system treated in the Jaynes-Cummings model is a \emph{\
single atom} system with just two internal states, denoted $\left\vert
1\right\rangle $, $\left\vert 2\right\rangle $, with energies $E_{1}$, $%
E_{2} $. Centre of mass degrees of freedom are ignored. There are therefore
just four atomic operators to consider, denoted $\widehat{\sigma }%
_{ij}=\left\vert i\right\rangle \left\langle j\right\vert $. Two of these
are \emph{population} operators, $\widehat{P}_{1}=\left\vert 1\right\rangle
\left\langle 1\right\vert $ and $\widehat{P}_{2}=\left\vert 2\right\rangle
\left\langle 2\right\vert $, and two are atomic \emph{transition} operators $%
\widehat{\sigma }_{+}=\left\vert 2\right\rangle \left\langle 1\right\vert $
and $\widehat{\sigma }_{-}=\left\vert 1\right\rangle \left\langle
2\right\vert $. In the simple Jaynes-Cummings model the two-level atom
interacts with a cavity mode. Let $\widehat{a},\widehat{a}^{\dag }$ be the
boson annihilation, creation operators for the field mode. The states for
the cavity mode are the usual $n$ photon states, denoted $\left\vert
n\right\rangle $, with $n=0,1,2,..$. If the atomic transition frequency is $%
\omega _{0}=(E_{2}-E_{1})/\hbar $, the frequency for the field mode is $%
\omega $ and the atom-field mode coupling constant (one-photon Rabi
frequency) is $\Omega $, then the \emph{Hamiltonian} for the \emph{one atom
Jaynes-Cummings model} is given by%
\begin{equation}
\widehat{H}_{JC}=E_{A}(\widehat{P}_{2}+\widehat{P}_{1})+\frac{1}{2}\hbar
\omega _{0}(\widehat{P}_{2}-\widehat{P}_{1})+\hbar \omega (\widehat{a}^{\dag
}\widehat{a})+\frac{1}{2}\hbar \Omega (\widehat{\sigma }_{+}\widehat{a}+%
\widehat{a}^{\dag }\widehat{\sigma }_{-})  \label{Eq.OneAtomJCHamiltonian}
\end{equation}%
where the average atomic energy is $E_{A}=(E_{2}+E_{1})/2$. The first term
is usually ignored as it only introduces a phase factor $\exp (-E_{A}t/\hbar
)$ into the evolution of all states in the one atom Jaynes-Cummimgs model.

The dynamics of the one atom Jaynes-Cummings model is treated in many
quantum optics textbooks and papers (see \cite{Stenholm73a}, \cite%
{Barnett97a}, \cite{Gerry05a}). The state vector is expanded in terms of $n$
photon states for the field and the two internal atomic states as 
\begin{eqnarray}
\left\lfloor \Psi \right\rangle &=&\exp (-E_{A}t/\hbar )
\label{Eq.JCStateVector} \\
&&\times \dsum\limits_{n}\left( A_{\overline{n-1}2}(t)\exp (-i(\overline{n-1}%
\omega +\frac{1}{2}\omega _{0})t)\;\left\vert 2\right\rangle \left\vert
n-1\right\rangle +A_{n1}(t)\exp (-i(n\omega -\frac{1}{2}\omega
_{0})t)\;\left\vert 1\right\rangle \left\vert n\right\rangle \right) 
\nonumber
\end{eqnarray}%
involving interaction picture amplitudes $A_{n1},A_{\overline{n-1}2}$. The
coupled equations for the amplitudes 
\begin{eqnarray}
i\frac{\partial }{\partial t}A_{\overline{n-1}2} &=&\frac{1}{2}\Omega \sqrt{n%
}\exp (+i\Delta t)\;A_{n1}  \nonumber \\
i\frac{\partial }{\partial t}A_{n1} &=&\frac{1}{2}\Omega \sqrt{n}\exp
(-i\Delta t)\;A_{\overline{n-1}2}  \label{Eq.JCAmpEqns}
\end{eqnarray}%
involve the detuning 
\begin{equation}
\Delta =\omega _{0}-\omega  \label{Eq.Detuning0}
\end{equation}%
and may be solved using Laplace transforms. The solution is%
\begin{eqnarray}
A_{n1}(t) &=&\exp (-\frac{1}{2}i\Delta t)\left( \cos (\frac{1}{2}\omega
_{n}t)\left\{ A_{n1}(0)\right\} +i\sin (\frac{1}{2}\omega _{n}t)\left\{ 
\frac{\Delta \;A_{n1}(0)-\Omega \sqrt{n}\;A_{\overline{n-1}2}(0)}{\omega _{n}%
}\right\} \right)  \nonumber \\
A_{\overline{n-1}2}(t) &=&\exp (+\frac{1}{2}i\Delta t)\left( \cos (\frac{1}{2%
}\omega _{n}t)\left\{ A_{\overline{n-1}2}(0)\right\} -i\sin (\frac{1}{2}%
\omega _{n}t)\left\{ \frac{\Omega \sqrt{n}\;A_{n1}(0)+\Delta \;A_{\overline{%
n-1}2}(0)}{\omega _{n}}\right\} \right)  \nonumber \\
&&  \label{Eq.JCTimeDepAmps}
\end{eqnarray}%
where 
\begin{equation}
\omega _{n}=\sqrt{\Delta ^{2}+n\Omega ^{2}}  \label{Eq.RabiFreq}
\end{equation}%
is the Rabi frequency. Population and coherence oscillations at various
frequencies $\omega _{n}$ are therefore predicted, and which frequencies are
observed depends on the inital conditions. For example, if the atom is not
excited and the field is in an $m$ photon state, then $A_{n1}(0)=\delta
_{nm} $, $A_{\overline{n-1}2}(0)=0$ and hence oscillations at a single
frequency $\omega _{m}$ will occur. On the other hand, if the field was in a
coherent state with amplitude $\eta $, then oscillation frequencies
clustered around the mean frequency $\omega _{\overline{n}}$ associated with
mean photon number $\overline{n}=|\eta |^{2}$ occur, the range of
frequencies being associated with the standard deviation $\Delta n=\sqrt{%
\overline{n}}$ in photon numbers. This results in the collapse, revival
effects discussed previously. In an alternative approach, Stenholm \cite%
{Stenholm81a} expanded the state vector in products of Bargman states for
the field and spinor states for the TLA and also demonstrated collapse,
revival effects. As we see, his solution can be adapted to determining the
Grassmann distribution function in the present treatment. \medskip

\subsection{Fermion and Bosonic Modes}

Instead of treating the Jaynes-Cummings model via the usual elementary
quantum optics approach - for example via matrix mechanics with basis states 
$\left\vert 1\right\rangle \left\vert n\right\rangle $, $\left\vert
2\right\rangle \left\vert n-1\right\rangle $, we can replace the original
one two-level atom plus cavity mode system with a somewhat \emph{enlarged
system} consisting of two fermionic modes $1$, $2$ interacting with one
bosonic mode. This enlarged system includes states that are in one-one
correspondence with the states for the original one two-level atom plus
single cavity mode system, and the general dynamics for the larger
Fermi-Bose system incorporates all possible behaviour for the original
Jaynes-Cummings model. Note that relating the two state atomic system to two
fermion modes has nothing to do with whether the atom itself is a fermion or
a boson.

This process involves the introduction of \emph{fermion annihilation,
creation} operators $\widehat{c}_i,\widehat{c}_i^{\dag }$ $(i=1,2)$ for the
two fermion modes, which satisfy standard fermion anti-commutation rules.
The commutation and anti-commutation rules for the fermion and boson
operators are%
\begin{eqnarray}
\{\widehat{c}_i,\widehat{c}_{j}^{\dag }\} &=&\delta _{ij}  \nonumber \\
\{\widehat{c}_i,\widehat{c}_{j}\} &=&\{\widehat{c}_i^{\dag },\widehat{c}
_{j}^{\dag }\}=0  \nonumber \\
\lbrack \widehat{a},\widehat{a}^{\dag }] &=&1  \label{Eq.CommAntiCommRules}
\end{eqnarray}
with the boson and fermion operators commuting.

We then introduce the four possible fermion \emph{Fock states} 
\begin{equation}
\left\vert m_{1};m_{2}\right\rangle =(\widehat{c}_{1}^{\dag })^{m_{1}}(%
\widehat{c}_{2}^{\dag })^{m_{2}}\left\vert 0\right\rangle
\label{Eq.FermionFockStates}
\end{equation}%
where $m_{1},m_{2}$ are equal to $0,1$ only and $\left\vert 0\right\rangle $
is the vacuum state. For the vacuum state $\widehat{c}_{i}\left\vert
0\right\rangle =0$, $\left\langle 0\right\vert \widehat{c}_{i}^{\dag }=0$.
Thus $\left\vert 0;0\right\rangle $ will be the vacuum state, with no
fermions in either mode, $\left\vert 1;0\right\rangle $ and $\left\vert
0;1\right\rangle $ will be the one fermion states with one fermion in modes $%
1,2$ respectively, and $\left\vert 1;1\right\rangle $ will be the two
fermion state with one fermion in each mode. Naturally, the Pauli exclusion
principle excludes states with more than one fermion in any mode.

The full \emph{basis states} for the Jaynes-Cummings model are Fock states
of the form%
\begin{equation}
\left\vert m_{1};m_{2};n\right\rangle =(\widehat{c}_{1}^{\dag })^{m_{1}}(%
\widehat{c}_{2}^{\dag })^{m_{2}}\frac{(\widehat{a}^{\dag })^{n}}{\sqrt{n!}}%
\left\vert 0\right\rangle  \label{Eq.JCBasisStates}
\end{equation}%
where $m_{i}=0,1$ give the number of atoms in state $i=1,2$ and $n=0,1,2,...$
gives the number of photons in the field mode. Obviously since $(\widehat{c}%
_{i}^{\dag })^{2}=0$ there can not be more than one atom in state $i$. The
state with no atoms or photons present is the vacuum state $\left\vert
0\right\rangle $.\medskip

\subsection{Quantum States}

\label{SubSection - Quantum States}

The most \emph{general mixed state} for which there may be either zero, one
or two atoms has a density operator%
\begin{eqnarray}
\widehat{\rho } &=&\dsum\limits_{n,m}(\rho _{00n;00m}\left\vert
0;0;n\right\rangle \left\langle 0;0,m\right\vert  \nonumber \\
&&+\dsum\limits_{n,m}(\rho _{10n;10m}\left\vert 1;0;n\right\rangle
\left\langle 1;0,m\right\vert +\rho _{10n;01m}\left\vert 1;0;n\right\rangle
\left\langle 0;1,m\right\vert  \nonumber \\
&&+\rho _{01n;10m}\left\vert 0;1;n\right\rangle \left\langle
1;0,m\right\vert +\rho _{01n;01m}\left\vert 0;1;n\right\rangle \left\langle
0;1,m\right\vert )  \nonumber \\
&&+\dsum\limits_{n,m}(\rho _{11n;11m}\left\vert 1;1;n\right\rangle
\left\langle 1;1,m\right\vert  \label{Eq.GeneralMixedState}
\end{eqnarray}%
where the $\rho _{ijn;klm}$ $(i,j,k,l=0,1)$ are density matrix elements.
This corresponds to a statistical mixture of states in which there are no
atoms, one atom or two atoms. Pure states which are quantum superpositions
of states with differering numbers of atoms are forbidden by the
super-selection rule based on conservation of atom number. Statistical
mixtures of such forbidden states would in general lead to the presence of
coherences between states with differing numbers of atoms. Thus there are no
coherences between states with different numbers of atoms in (\ref%
{Eq.GeneralMixedState}).

The condition that the trace of the density operator is one requires%
\begin{equation}
Tr\widehat{\rho }=\dsum\limits_{n}(\rho _{00n;00n}+\rho _{10n;10n}+\rho
_{01n;01n}+\rho _{11n;11n})=1  \label{Eq.TraceCondition}
\end{equation}

Note that in addition to atomic states corresponding to the one atom
Jaynes-Cummings model, we now have two extra states in our enlarged system.
and we need to consider whether these have any physical significance. The
vacuum state $\left\vert 0;0\right\rangle $ could correspond to an
experiment with only the cavity field present but with no atoms inside the
cavity - not a situation of great interest. The two fermion state $%
\left\vert 1;1\right\rangle $ could correspond to a two atom system inside
the cavity, with one atom in state $\left\vert 1\right\rangle $ and the
other in state $\left\vert 2\right\rangle $. All four fermion states $%
\left\vert m_{1};m_{2}\right\rangle $ also have a mathematical
interpretation in terms of \emph{spin states} in our enlarged system.\medskip

\subsection{Normal Ordering and Vacuum State Projector}

\label{SubSection - Vacuum Projector}

The vacuum state $\left\vert 0\right\rangle $ which contains no bosons or
fermions in any mode is associated with the \emph{vacuum state projector} $%
\left\vert 0\right\rangle \left\langle 0\right\vert $. It turns out that the
vacuum projector can be written in terms of \emph{normally ordered} forms of
products of exponential operators based on mode number operators. The
normally ordered form for a \emph{product} of annihilation, creation
operators involves placing all creation operators to the left of all
annihilation operators, whilst retaining all creation and all annihilation
operators in their original order, and multiplying by either $+1$ or $-1$
depending on whether the required permutation of \emph{fermion} operators
from the original product is even or odd \cite{Kaku93a}.

General \emph{functions} of the annihilation, creation operators (such as
the Hamiltonian) can always be written as linear combinations of products
with c-number coefficients, and the normally ordered form of such a function
is just the linear combination of the normally ordered form of the various
products. A useful theorem that applies to the normally ordered product of
two functions involving the number operators for two different fermion modes
is%
\begin{equation}
\mathcal{N}(F(\widehat{c}_{i}^{\dag }\widehat{c}_{i})\,G(\widehat{c}%
_{j}^{\dag }\widehat{c}_{j}))=\mathcal{N}(F(\widehat{c}_{i}^{\dag }\widehat{c%
}_{i}))\,\mathcal{N}(G(\widehat{c}_{j}^{\dag }\widehat{c}_{j}))\qquad (i\neq
j)  \label{Eq.NormThmFermions}
\end{equation}%
where $F,G$ are arbitrary functions. This result only depends on the
annihilation and creation operators anticommuting for $i\neq j$.

The vacuum state projector is given in terms of normally ordered operators as%
\begin{eqnarray}
\left\vert 0\right\rangle \left\langle 0\right\vert &=&\mathcal{N}%
(\dprod\limits_{i}\exp (-\widehat{a}_{i}^{\dag }\widehat{a}_{i}))=\mathcal{N}%
(\exp (-\dsum\limits_{i}\widehat{a}_{i}^{\dag }\widehat{a}_{i}))
\label{Eq.VacProjBosons} \\
&=&\mathcal{N}(\dprod\limits_{i}\exp (-\widehat{c}_{i}^{\dag }\widehat{c}%
_{i}))=\mathcal{N}(\exp (-\dsum\limits_{i}\widehat{c}_{i}^{\dag }\widehat{c}%
_{i}))  \label{Eq.VacProjFermions}
\end{eqnarray}%
for bosons and fermions respectively. The (\ref{Eq.VacProjBosons}) result
for bosons is given in \cite{Barnett97a} (see p.45). The result for fermions
follows from the vacuum state requirements that $\widehat{c}_{i}\left\vert
0\right\rangle =0$, $\left\langle 0\right\vert \widehat{c}_{i}^{\dag }=0$
are also satisfied by the expression in (\ref{Eq.VacProjFermions}), together
with the requirement that the vacuum state matrix element for the projector
is also unity.\medskip

\subsection{Population and Transition Operators}

\label{SubSection - Normal Ordering, Popn and Trans Oprs}

We now identify the two one fermion Fock states with the two internal atomic
states in the one atom Jaynes-Cummings model as follows:%
\begin{eqnarray}
\widehat{c}_{1}^{\dag }\,\left\vert 0\right\rangle &=&\left\vert
1;0\right\rangle \Longleftrightarrow \left\vert 1\right\rangle  \nonumber \\
\widehat{c}_{2}^{\dag }\,\left\vert 0\right\rangle &=&\left\vert
0;1\right\rangle \Longleftrightarrow \left\vert 2\right\rangle
\label{Eq.FockEquivJCStates}
\end{eqnarray}%
and identify the four different products involving an annihilation, a
creation operator and the vacuum state projector with the four \emph{one
atom atomic population }or \emph{projector operator }$\widehat{P}_{1,2}$ and 
\emph{transtion operators }$\widehat{\sigma }_{\pm }$\emph{\ }%
\begin{eqnarray}
\widehat{c}_{1}^{\dag }\,\left\vert 0\right\rangle \left\langle 0\right\vert
\,\widehat{c}_{1} &\Longleftrightarrow &\widehat{P}_{1}\qquad \widehat{c}%
_{2}^{\dag }\,\left\vert 0\right\rangle \left\langle 0\right\vert \,\widehat{%
c}_{2}\Longleftrightarrow \widehat{P}_{2}  \nonumber \\
\widehat{c}_{1}^{\dag }\,\left\vert 0\right\rangle \left\langle 0\right\vert
\,\widehat{c}_{2} &\Longleftrightarrow &\widehat{\sigma }_{-}\qquad \widehat{%
c}_{2}^{\dag }\,\left\vert 0\right\rangle \left\langle 0\right\vert \,%
\widehat{c}_{1}\Longleftrightarrow \widehat{\sigma }_{+}
\label{Eq.FermionOprsEquivAtomicOprs}
\end{eqnarray}

For the expanded Jaynes-Cummings model the \emph{two atom state population }%
or\emph{\ projector operator} $\widehat{P}_{12}$ with one atom in each of
the atomic states we have

\begin{equation}
\widehat{c}_{1}^{\dag }\widehat{c}_{2}^{\dag }\left\vert 0\right\rangle
=\left\vert 1;1\right\rangle  \label{Eq.FockEquivTwoAtomState}
\end{equation}%
\begin{equation}
\widehat{c}_{1}^{\dag }\widehat{c}_{2}^{\dag }\,\left\vert 0\right\rangle
\left\langle 0\right\vert \,\widehat{c}_{2}\widehat{c}_{1}%
\Longleftrightarrow \widehat{P}_{12}  \label{Eq.FermionOprsEquivTwoAtomOpr}
\end{equation}%
and for the \emph{zero atom population }or\emph{\ projector operator} $%
\widehat{P}_{0}$ with no atoms in either of the internal atomic states we
have

\begin{equation}
\left\vert 0\right\rangle =\left\vert 0;0\right\rangle
\label{Eq.FockEquivNoAtomState}
\end{equation}%
\begin{equation}
\,\left\vert 0\right\rangle \left\langle 0\right\vert \,\Longleftrightarrow 
\widehat{P}_{0}  \label{Eq.FermionOprsEquivZeroAtomOpr}
\end{equation}%
This of course is the same as the vacuum state projector.

Using Eq.(\ref{Eq.VacProjFermions}) and the result (\ref{Eq.NormThmFermions}%
) we can obtain expressions for all the atom state operators just in terms
of fermion annihilation, creation operators. We find for the one atom
population and transition operators 
\begin{eqnarray}
\widehat{P}_{1} &=&\widehat{c}_{1}^{\dag }\widehat{c}_{1}-\widehat{c}%
_{1}^{\dag }\widehat{c}_{2}^{\dag }\widehat{c}_{2}\widehat{c}_{1}\qquad 
\widehat{P}_{2}=\widehat{c}_{2}^{\dag }\widehat{c}_{2}-\widehat{c}_{1}^{\dag
}\widehat{c}_{2}^{\dag }\widehat{c}_{2}\widehat{c}_{1}
\label{Eq.OneAtomPopulationOprs} \\
\widehat{\sigma }_{-} &=&\widehat{c}_{1}^{\dag }\widehat{c}_{2}\qquad 
\widehat{\sigma }_{+}=\widehat{c}_{2}^{\dag }\widehat{c}_{1}
\label{Eq.OneAtomTransitionOprs}
\end{eqnarray}%
and for the two atom and zero atom population operators%
\begin{eqnarray}
\widehat{P}_{12} &=&\widehat{c}_{1}^{\dag }\widehat{c}_{2}^{\dag }\widehat{c}%
_{2}\widehat{c}_{1}  \label{Eq.TwoAtomPopulationOpr} \\
\widehat{P}_{0} &=&1-\widehat{c}_{1}^{\dag }\widehat{c}_{1}-\widehat{c}%
_{2}^{\dag }\widehat{c}_{2}+\widehat{c}_{1}^{\dag }\widehat{c}_{2}^{\dag }%
\widehat{c}_{2}\widehat{c}_{1}  \label{Eq.ZeroAtomPopulationOpr}
\end{eqnarray}%
The population operators thus satisfy the condition 
\begin{equation}
\widehat{P}_{0}+\widehat{P}_{1}+\widehat{P}_{2}+\widehat{P}_{12}=\widehat{1}
\label{Eq.PopulationConservn}
\end{equation}%
which (as will be seen below), just represents the total probability for any
general mixed state (\ref{Eq.GeneralMixedState}) that the system is found in
either a zero, one or two atom state being equal to unity\medskip

\subsection{Hamiltonian and Number Operators}

\label{SubSection - Hamiltonian and Number Operators}

The Hamiltonian needs to be modified for the \emph{enlarged system} to allow
for states with two or zero atoms. The one atom Jaynes-Cummings model
Hamiltonian $\widehat{H}_{JC}$ is replaced by an \emph{enlarged Hamiltonian }%
$\widehat{H}$ by adding a two atom energy term $(E_{2}+E_{1})\widehat{P}%
_{12} $ to the Hamiltonian in (\ref{Eq.OneAtomJCHamiltonian}). This enlarged
Hamiltonian has zero matrix elements between two atom states $\left\vert
1;1\right\rangle $, one atom states $\left\vert 0;1\right\rangle $, $%
\left\vert 1;0\right\rangle $ and zero atom states $\left\vert
0;0\right\rangle $. The two and zero atom states are eigenstates of the
Hamiltonian with atomic energies $E_{2}+E_{1}$ and $0$, as expected. The
atom-field coupling term also cannot cause a transtion between states with
differing \ total atom number, consistent with not violating the
super-selection rule.

In terms of the bosonic and fermionic annihilation, creation operators the 
\emph{enlarged Hamiltonian} for the Jaynes-Cummings model may be written as%
\begin{equation}
\widehat{H}=E_{A}(\widehat{c}_{2}^{\dag }\widehat{c}_{2}+\widehat{c}%
_{1}^{\dag }\widehat{c}_{1})+\frac{1}{2}\hbar \omega _{0}(\widehat{c}%
_{2}^{\dag }\widehat{c}_{2}-\widehat{c}_{1}^{\dag }\widehat{c}_{1})+\hbar
\omega (\widehat{a}^{\dag }\widehat{a})+\frac{1}{2}\hbar \Omega (\widehat{c}%
_{2}^{\dag }\widehat{c}_{1}\widehat{a}+\widehat{a}^{\dag }\widehat{c}%
_{1}^{\dag }\widehat{c}_{2})  \label{Eq.Hamiltonian}
\end{equation}%
where the results (\ref{Eq.OneAtomPopulationOprs}), (\ref%
{Eq.OneAtomTransitionOprs}) and (\ref{Eq.TwoAtomPopulationOpr}) for the one
and two atom operators have been substituted into the enlarged Hamiltonian.
Terms involving $\widehat{c}_{1}^{\dag }\widehat{c}_{2}^{\dag }\widehat{c}%
_{2}\widehat{c}_{1}$ cancel out.

The \emph{number of photons} present is determined from the operator 
\begin{equation}
\widehat{n}=\widehat{a}^{\dag }\widehat{a}  \label{Eq.PhotonNumberOpr}
\end{equation}%
whose eigenvalues are $n=0,1,2,.$We can also introduce \emph{number operators%
} for the two atomic states via%
\begin{equation}
\widehat{n}_{i}=\widehat{c}_{i}^{\dag }\widehat{c}_{i}\qquad (i=1,2)
\label{Eq.AtomNumberOpr}
\end{equation}%
where from the femion anticommutation rules the eigenvalues for the atomic
number operators are $0,1$ only.

The \emph{total number of atoms} present is given by the number operator%
\begin{eqnarray}
\widehat{N} &=&(\widehat{c}_{2}^{\dag }\widehat{c}_{2}+\widehat{c}_{1}^{\dag
}\widehat{c}_{1})  \label{Eq.NumberOpr} \\
&=&0\times \widehat{P}_{0}+1\times (\widehat{P}_{1}+\widehat{P}_{2})+2\times 
\widehat{P}_{12}  \label{Eq.NumberOprResult}
\end{eqnarray}%
the second expression for the number operator being related to the
projectors $\widehat{P}_{0}$, $\widehat{P}_{1}$, $\widehat{P}_{2}$ and $%
\widehat{P}_{12}$ being for zero, one, one and two atom states respectively.
The number operator can have eigenvalues $0,1,2$ and clearly%
\begin{equation}
\widehat{N\,}\left\vert m_{1};m_{2};n\right\rangle
=(m_{1}+m_{2})\,\left\vert m_{1};m_{2};n\right\rangle
\label{Eq.TotalAtomNumber}
\end{equation}

The number operator commutes with the Hamiltonian, so if we initially have a
physical state with a specified number of atoms, then state evolution does
not change the atom number. Thus one atom states do not change into two atom
states and if the initial density operator corresponds to the form in (\ref%
{Eq.GeneralMixedState}) with $\rho _{00n;00m}=\rho _{11n;11m}=0$ for one
atom states it will remain in this form. There is of course no conservation
law for the photon number. \emph{\medskip }.

\subsection{Probabilities and Coherences}

\label{SubSection - Prob and Coherences}

We can obtain expressions for the physical quantities in terms of the
annihilation, creation operators. For the \emph{one atom probability} of
finding one atom in state $\left\vert 1\right\rangle $ and none in state $%
\left\vert 2\right\rangle $ we have 
\begin{equation}
P_{1}=Tr(\widehat{P}_{1}\,\widehat{\rho })=\left\langle \widehat{c}%
_{1}^{\dag }\widehat{c}_{1}\right\rangle -\left\langle \widehat{c}_{1}^{\dag
}\widehat{c}_{2}^{\dag }\widehat{c}_{2}\widehat{c}_{1}\right\rangle =Tr(%
\widehat{c}_{1}\widehat{\rho }\widehat{c}_{1}^{\dag })-Tr(\widehat{c}_{2}%
\widehat{c}_{1}\widehat{\rho }\widehat{c}_{1}^{\dag }\widehat{c}_{2}^{\dag })
\label{Eq.ProbOneAtomOneZeroAtomTwo}
\end{equation}%
where $\left\langle \widehat{\Xi }\right\rangle =Tr(\widehat{\rho }\,%
\widehat{\Xi })$. Similarly the \emph{one atom probability} of finding one
atom in state $\left\vert 2\right\rangle $ and none in state $\left\vert
1\right\rangle $ is 
\begin{equation}
P_{2}=Tr(\widehat{P}_{2}\,\widehat{\rho })=\left\langle \widehat{c}%
_{2}^{\dag }\widehat{c}_{2}\right\rangle -\left\langle \widehat{c}_{1}^{\dag
}\widehat{c}_{2}^{\dag }\widehat{c}_{2}\widehat{c}_{1}\right\rangle =Tr(%
\widehat{c}_{2}\widehat{\rho }\widehat{c}_{2}^{\dag })-Tr(\widehat{c}_{2}%
\widehat{c}_{1}\widehat{\rho }\widehat{c}_{1}^{\dag }\widehat{c}_{2}^{\dag })
\label{Eq.ProbZeroAtomOneOneAtomTwo}
\end{equation}%
Note that these probabilities are not the same as $\left\langle \widehat{c}%
_{i}^{\dag }\widehat{c}_{i}\right\rangle $ as might be expected.

The \emph{one atom coherences} between state $\left\vert 1\right\rangle $
and state $\left\vert 2\right\rangle $ are given by 
\begin{eqnarray}
\rho _{12} &=&Tr(\widehat{\sigma }_{-}\,\widehat{\rho })=\left\langle 
\widehat{c}_{1}^{\dag }\widehat{c}_{2}\right\rangle =Tr(\widehat{c}_{2}\,%
\widehat{\rho }\,\widehat{c}_{1}^{\dag })  \nonumber \\
\rho _{21} &=&Tr(\widehat{\sigma }_{+}\,\widehat{\rho })=\left\langle 
\widehat{c}_{2}^{\dag }\widehat{c}_{1}\right\rangle =Tr(\widehat{c}_{1}\,%
\widehat{\rho }\,\widehat{c}_{2}^{\dag })  \label{Eq.OneAtomCoherences}
\end{eqnarray}

The \emph{mean number of photons} is 
\begin{equation}
\overline{n}=\left\langle \widehat{a}^{\dag }\widehat{a}\right\rangle
\label{Eq.MeanPhotonNumber}
\end{equation}

Also, the \emph{two atom probability }$P_{12}$\emph{\ }for finding one atom
in state $\left\vert 1\right\rangle $ and one in state $\left\vert
2\right\rangle $, and the \emph{zero atom probability} $P_{0}$ for finding
no atom either in state $\left\vert 1\right\rangle $ or in state $\left\vert
2\right\rangle $ are given by%
\begin{eqnarray}
P_{12} &=&Tr(\widehat{P}_{12}\,\widehat{\rho })=\left\langle \widehat{c}%
_{1}^{\dag }\widehat{c}_{2}^{\dag }\widehat{c}_{2}\widehat{c}%
_{1}\right\rangle =Tr(\widehat{c}_{2}\widehat{c}_{1}\widehat{\rho }\widehat{c%
}_{1}^{\dag }\widehat{c}_{2}^{\dag })  \label{Eq.ProbTwoAtoms} \\
P_{0} &=&Tr(\widehat{P}_{0}\,\widehat{\rho })=1-P_{1}-P_{2}-P_{12}
\label{Eq.ProbZeroAtoms}
\end{eqnarray}%
the last result just expressing the fact that all the probabilities must add
up to one. Expressions involving the \emph{density matrix elements} for 
\emph{general mixed states} can be easily obtained for all these results.

Clearly, for \emph{one atom states} where $\rho _{00n;00m}=\rho _{11n;11m}=0$
we have using (\ref{Eq.TraceCondition}) 
\begin{eqnarray}
P_{0} &=&P_{12}=0\qquad P_{1}+P_{2}=1  \label{Eq.OneAtomStateProbs} \\
\overline{n} &=&\dsum\limits_{n}n(\rho _{10n;10n}+\rho _{01n;01n})
\label{Eq.MeanPhotonOneAtomState}
\end{eqnarray}

For general mixed states as in (\ref{Eq.GeneralMixedState}) we see that
expectation values of odd numbers of fermionic creation and destruction
operators are zero. Thus%
\begin{equation}
\left\langle \widehat{c}_{i}\right\rangle =\left\langle \widehat{c}%
_{i}^{\dag }\right\rangle =\left\langle \widehat{c}_{1}^{\dag }\widehat{c}%
_{2}^{\dag }\widehat{c}_{i}\right\rangle =\left\langle \widehat{c}_{i}^{\dag
}\widehat{c}_{2}\widehat{c}_{1}\right\rangle =0
\label{Eq.ExpnValuesPhysicalStates}
\end{equation}%
For one atom states the only non-zero expectation values are those that
involve one fermion annihilation operator and one creation operator of the
form $\left\langle \widehat{c}_{i}^{\dag }\widehat{c}_{j}\right\rangle $.
Thus in addition to the last results the expectation value involving four
fermion operators is zero 
\begin{equation}
\left\langle \widehat{c}_{1}^{\dag }\widehat{c}_{2}^{\dag }\widehat{c}_{2}%
\widehat{c}_{1}\right\rangle =P_{12}=0  \label{Eq.ExpnValuesPhysicalStates2}
\end{equation}%
since $\rho _{11n;11n}=0$ for one atom states. This corresponds to the
probability of finding two atoms present being zero in the one atom
Jaynes-Cummings model. For general mixed states the quantity $\left\langle 
\widehat{c}_{1}^{\dag }\widehat{c}_{2}^{\dag }\widehat{c}_{2}\widehat{c}%
_{1}\right\rangle $ is conserved, since $\widehat{c}_{1}^{\dag }\widehat{c}%
_{2}^{\dag }\widehat{c}_{2}\widehat{c}_{1}$ commutes with the Hamiltonian

Although we will be mainly focused on states corresponding to one atom
states in the Jaynes-Cummings model, it will be convenient to introduce 
\emph{non-physical states} involving coherent superpositions of different
total numbers of atoms and even states where the expansion coefficients are
Grassmann numbers. For non-physical states involving superpositions of
different atom numbers there may be off diagonal elements of the density
matrix between states of different atom numbers, and the results in Eq.(\ref%
{Eq.ExpnValuesPhysicalStates}) may not apply.\medskip

\section{Phase Space Theory}

\label{Section - Phase Space Theory}

There are several ways to set out \emph{phase space theory}. Here we start
with the characteristic function - which is equivalent to the set of all
normally ordered quantum correlation functions - and then relate this to the
distribution function - whose phase space integral with products of phase
space variables determines the correlation functions. The \emph{normally
ordered quantum correlation functions} are defined as 
\begin{eqnarray}
&&G(m_{1},m_{2},n;\,p,l_{2},l_{1})  \nonumber \\
&=&\left\langle (\widehat{c}_{1}^{\dag })^{m_{1}}(\widehat{c}_{2}^{\dag
})^{m_{2}}(\widehat{a}^{\dag })^{n}(\widehat{a})^{p}(\widehat{c}%
_{2})^{l_{2}}(\widehat{c}_{1})^{l_{1}}\right\rangle  \nonumber \\
&=&\mbox{Tr}((\widehat{c}_{2})^{l_{2}}(\widehat{c}_{1})^{l_{1}}\,(\widehat{a}%
)^{p}\,\hat{\rho}\,(\widehat{a}^{\dag })^{n}\,(\widehat{c}_{1}^{\dag
})^{m_{1}}(\widehat{c}_{2}^{\dag })^{m_{2}})  \label{Eq.QCorrFn}
\end{eqnarray}%
where $m_{i},l_{i}=0,1$ only. Normally ordered correlation functions in
which the density operator appears in the middle of the trace with
annihilation operators on the left and creation operators on the right
appear in fundamental treatments of boson and fermion detection processes
(see for example \cite{Glauber65a}).

\subsection{Characteristic Function}

We define the \emph{characteristic function} $\chi (\xi ,\xi
^{+},h_{i},h_{i}^{+})$ via%
\begin{eqnarray}
\chi (\xi ,\xi ^{+},h,h^{+}) &=&\mbox{Tr}(\,\hat{\Omega}_{b}^{+}(\xi ^{+})\,%
\hat{\Omega}_{f}^{+}(h^{+})\,\hat{\rho}\,\hat{\Omega}_{f}^{-}(h)\,\hat{\Omega%
}_{b}^{-}(\xi ))  \label{Eq.Charfunction1} \\
\hat{\Omega}_{b}^{+}(\xi ^{+}) &=&\exp i\hat{a}\xi ^{+}\qquad \hat{\Omega}%
_{b}^{-}(\xi )=\exp i\xi \hat{a}^{\dag }  \label{Eq.BosonOmegaOprs} \\
\hat{\Omega}_{f}^{+}(h^{+}) &=&\exp i\sum\limits_{i=1,2}\hat{c}%
_{i}h_{i}^{+}\qquad \hat{\Omega}_{f}^{-}(h)=\exp i\sum\limits_{i=1,2}h_{i}%
\hat{c}_{i}^{\dag }  \label{Eq.FermionOmegaOprs}
\end{eqnarray}%
For the bosonic mode we associate a pair of \emph{c-numbers} $\xi ,\xi ^{+}$
. For each fermionic mode $i=1,2$ we associate a pair of \emph{\
Grassmann-numbers} $h_{i},h_{i}^{+}$ and $h=\{h_{1},h_{2}\}$, $%
h^{+}=\{h_{1}^{+},h_{2}^{+}\}$. The characteristic function will be a
c-number \emph{analytic} function of $\xi ,\xi ^{+}$ and a \emph{Grassmann
function} of $h_{1},h_{1}^{+},h_{2},h_{2}^{+}$. The factors for the
different modes in the last two expressions for $\hat{\Omega}_{f}^{+}(h^{+})$
and $\hat{\Omega}_{f}^{-}(h)$ may be put in any order since they commute,
but by convention the order will be $2,1$ for $\hat{\Omega}_{f}^{+}(h^{+})$
and $,1,2$ for $\hat{\Omega}_{f}^{-}(h)$. This leads to a quasi distribution
function of the \emph{positive P} type, which is generally better suited for
obtaining Ito stochastic equations.

For the \emph{physical states} where the state vector involves a single
number of atoms $(N=0,1,2)$, expectation values for cases where the number
of fermion annihilation operators differs from the number of creation
operators are zero. Using $\hat{\Omega}_{f}^{+}(h^{+})=(1+\hat{c}%
_{2}h_{2}^{+})(1+\hat{c}_{1}h_{1}^{+})$ and $\hat{\Omega}_{f}^{-}(h)=(1+h_{1}%
\hat{c}_{1}^{\dag })(1+h_{2}\hat{c}_{2}^{\dag })$ we see that the
characteristic function is of the form%
\begin{equation}
\chi (\xi ,\xi ^{+},h,h^{+})=\chi _{0}(\xi ,\xi
^{+})+\sum\limits_{i,j=1,2}\chi _{2}^{i;j}(\xi ,\xi
^{+})\,h_{j}^{+}h_{i}+\chi _{4}^{12;21}(\xi ,\xi
^{+})\,h_{2}^{+}h_{1}^{+}h_{1}h_{2}  \label{Eq.CharFunction2}
\end{equation}%
and the relationship with the coefficients for $\xi ,\xi ^{+}=0,0$ is 
\begin{eqnarray}
\chi _{0}(0,0) &=&1  \nonumber \\
\chi _{2}^{i,j}(0,0) &=&i^{2}\left\langle \widehat{c}_{i}^{\dag }\widehat{c}%
_{j}\right\rangle  \nonumber \\
\chi _{4}^{12,21}(0,0) &=&i^{4}\left\langle \widehat{c}_{1}^{\dag }\widehat{c%
}_{2}^{\dag }\widehat{c}_{2}\widehat{c}_{1}\right\rangle
\label{Eq.CharFnCoeffts2}
\end{eqnarray}%
Thus for one atom physical states at most six ($%
2.2!/(2!2!)=(C_{0}^{2})^{2}+(C_{1}^{2})^{2}+(C_{2}^{2})^{2}$ \cite%
{Gradshteyn65a}) c-number coefficients are required to define the
characteristic function $\chi (\xi ,\xi ^{+},h,h^{+})$ as a Grassmann
function.

The various normally ordered quantum correlation functions can be expressed
as c-number and \emph{Grassmann derivatives} of the characteristic function%
\begin{eqnarray}
&&G(m_{1},m_{2},n;\,p,l_{2},l_{1})  \nonumber \\
&=&\left( \frac{(\overrightarrow{\partial })^{l_{2}}}{\partial
(ih_{2}^{+})^{l_{2}}}\frac{(\overrightarrow{\partial })^{l_{1}}}{\partial
(ih_{1}^{+})^{l_{1}}}\left( \frac{\partial ^{n}}{\partial (i\xi )^{n}}\frac{%
\partial ^{p}}{\partial (i\xi ^{+})^{p}}\chi (\xi ,\xi ^{+},h,h^{+})\right) 
\frac{(\overleftarrow{\partial })^{m_{1}}}{\partial (ih_{1})^{m_{1}}}\frac{(%
\overleftarrow{\partial })^{m_{2}}}{\partial (ih_{2})^{m_{2}}}\right) _{\xi
,\xi ^{+},h,h^{+}=0}  \label{Eq.QCorrFnKeyResult1}
\end{eqnarray}%
where $m_{i},l_{i}=0,1$ only, and in the case where $l_{i}$ or $m_{i}$ is
zero, then no differentiation takes place.\medskip

\subsection{Distribution Function}

The characteristic function $\chi (\xi ,\xi ^{+},h,h^{+})$ is related to the 
\emph{distribution function} $P(\alpha ,\alpha ^{+},\alpha ^{\ast },\alpha
^{+\ast },g,g^{+})$ via \emph{phase space integrals}, which are c-number
integrals for the field mode and \emph{Grassmann integrals} for the two
atomic modes. The formula is%
\begin{eqnarray}
&&\chi (\xi ,\xi ^{+},h,h^{+})  \nonumber \\
&=&\int \prod\limits_{i=1,2}dg_{i}^{+}dg_{i}\,\int d^{2}\alpha
^{+}d^{2}\alpha \,\,  \nonumber \\
&&\times \exp i\sum\limits_{i=1}^{2}\{g_{i}h_{i}^{+}\}\,\exp i\{\alpha \,\xi
^{+}\}\cdot P(\alpha ,\alpha ^{+},\alpha ^{\ast },\alpha ^{+\ast
},g,g^{+})\cdot \exp i\{\xi \alpha ^{+}\}\,\exp
i\sum\limits_{i=1}^{2}\{h_{i}g_{i}^{+}\}  \nonumber \\
&&  \label{Eq.BoseFermiDistnFn}
\end{eqnarray}%
where for the field mode we associate another pair of c-numbers $\alpha
,\alpha ^{+}$, and for the two atomic modes $i$ we associate another pair of
g-numbers $g_{i},g_{i}^{+}$ and $g=\{g_{1},g_{2}\}$, $g^{+}=%
\{g_{1}^{+},g_{2}^{+}\}$. As previously the characteristic function $\chi
(\xi ,\xi ^{+},h,h^{+})$ is a function of the variables $\xi ,\xi
^{+},h_{i},h_{i}^{+}$ and $g_{i},g_{i}^{+}$ but \emph{not} of their complex
conjugates, whereas the distribution function $P(\alpha ,\alpha ^{+},\alpha
^{\ast },\alpha ^{+\ast },g,g^{+})\,$also depends on the complex conjugates $%
\alpha ^{\ast },\alpha ^{+\ast }$. Unlike the characteristic function, the
distribution function is \emph{non analytic}. The distribution function will
be a c-number function of $\alpha ,\alpha ^{+},\alpha ^{\ast },\alpha
^{+\ast }$ and a Grassmann function of $g_{1},g_{1}^{+},g_{2},g_{2}^{+}$,
the highest order monomial being $g_{1}g_{2}g_{2}^{+}g_{1}^{+}$. The
c-number integrations $d^{2}\alpha ^{+}$ and $d^{2}\alpha $ are over the two 
\emph{\ complex planes} $\alpha ,\alpha ^{+}$ and thus $d^{2}\alpha \equiv
d\alpha _{x}d\alpha _{y}$ and $d^{2}\alpha ^{+}\equiv d\alpha
_{x}^{+}d\alpha _{y}^{+}$. The Grassmann integrals $dg_{i}^{+}$ and $dg_{i}$
are over the \emph{single variables} $g_{i}^{+}$ and $g_{i}$ only, and not
over $g_{i}^{+\ast }$ and $g_{i}^{\ast }$ as well. The distribution function
is of the \emph{positive P} type for the bosonic variables $\alpha ,\alpha
^{+}$ and similar to the \emph{complex P} type for the Grassmann variables $%
g_{i},g_{i}^{+}$. Although pairs of differentials $dg_{i}^{+}dg_{i}$ commute
with other pairs, the convention used here is to write $\prod%
\limits_{i}dg_{i}^{+}dg_{i}\,=dg_{2}^{+}dg_{2}\,dg_{1}^{+}dg_{1}$.

The distribution function is of the form

\begin{eqnarray}
&&P(\alpha ,\alpha ^{+},\alpha ^{\ast },\alpha ^{+\ast },g,g^{+})  \nonumber
\\
&=&P_{0}(\alpha ,\alpha ^{+},\alpha ^{\ast },\alpha ^{+\ast
})+\sum\limits_{i;j}P_{2}^{i,j}(\alpha ,\alpha ^{+},\alpha ^{\ast },\alpha
^{+\ast })\,g_{i}g_{j}^{+}  \nonumber \\
&&+P_{2n}^{12,21}(\alpha ,\alpha ^{+},\alpha ^{\ast },\alpha ^{+\ast
})\,g_{1}g_{2}g_{2}^{+}g_{1}^{+}  \label{Eq.FermiBoseDistnFn2}
\end{eqnarray}%
Other possible terms do not lead to the characteristic function is given by (%
\ref{Eq.CharFunction2}) when the Grassmann integration in (\ref%
{Eq.BoseFermiDistnFn}) is carried out using (\ref{Eq.CompleteGrassInteg})
after expanding the Grassmann exponential factors. An ordering convention in
which the $g_{i}$ are arranged in ascending order and the $g_{j}^{+}$ in
descending order has been used. The coefficients are functions of the
bosonic variables $\alpha ,\alpha ^{+}$. There are six $%
(2.2)!/(2!2!)=(C_{0}^{2})^{2}+(C_{1}^{2})^{2}+(C_{2}^{2})^{2}$ \cite%
{Gradshteyn65a}) c-number coefficients to define the distribution function $%
P(\alpha ,\alpha ^{+},g,g^{+})$ as a Grassmann function, the same number of
course as the characteristic.that it determines. The distribution function
is thus an \emph{even Grassmann function} of the order $2^{2}=4$ in the
variables $g_{1},g_{1}^{+}$, $g_{2},g_{2}^{+}$, a feature that is needed
later.

The hermiticity of the density operator leads to relationships between the
coefficients 
\begin{eqnarray}
P_{0}(\widetilde{\mathbf{\alpha }})^{\ast } &=&P_{0}(\widetilde{\mathbf{\
\alpha }})  \nonumber \\
P_{2}^{1,1}(\widetilde{\mathbf{\alpha }})^{\ast } &=&P_{2}^{1,1}(\widetilde{%
\mathbf{\alpha }})\qquad P_{2}^{1,2}(\widetilde{\mathbf{\alpha }})^{\ast
}=P_{2}^{2,1}(\widetilde{\mathbf{\alpha }})  \nonumber \\
P_{2}^{2,1}(\widetilde{\mathbf{\alpha }})^{\ast } &=&P_{2}^{1;2}(\widetilde{%
\mathbf{\alpha }})\qquad P_{2}^{2,2}(\widetilde{\mathbf{\alpha }})^{\ast
}=P_{2}^{2,2}(\widetilde{\mathbf{\alpha }})  \nonumber \\
P_{4}^{12,21}(\widetilde{\mathbf{\alpha }})^{\ast } &=&P_{4}^{12,21}(%
\widetilde{\mathbf{\alpha }})  \label{Eq.JCDuistnFnConjResults}
\end{eqnarray}%
where for short we write $\widetilde{\mathbf{\alpha }}\equiv \{\alpha
,\alpha ^{+},\alpha ^{\ast },\alpha ^{+\ast }\}$. This shows that four of
the bosonic coefficients are real and the other two $P_{2}^{1,2}(\widetilde{%
\mathbf{\alpha }}),P_{2}^{2,1}(\widetilde{\mathbf{\alpha }})$ are complex
conjugates.

For the Jaynes-Cummings model case the quantum correlation functions are
obtained from the characteristic function from Eq. (\ref%
{Eq.QCorrFnKeyResult1}). Applying this result by carrying out the
differentiations on the formula (\ref{Eq.BoseFermiDistnFn})\ relating the
characteristic and distribution functions gives the \emph{quantum
correlation functions} in terms of \emph{phase space integrals} 
\begin{eqnarray}
&&G(m_{1},m_{2},n;\,p,l_{2},l_{1})  \nonumber \\
&=&\int d^{2}\alpha ^{+}d^{2}\alpha \,\int
\prod\limits_{i=1,2}dg_{i}^{+}dg_{i}\,  \nonumber \\
&&\times (g_{2})^{l_{2}}(g_{1})^{l_{1}}\,(\alpha )^{p}\,P(\alpha ,\alpha
^{+},\alpha ^{\ast },\alpha ^{+\ast },g,g^{+})\,(\alpha
^{+})^{n}\,(g_{1}^{+})^{m_{1}}(g_{2}^{+})^{m_{2}}
\label{Eq.BoseFermiCorrFnPhaseSpace}
\end{eqnarray}%
which involves both c-number and Grassmann number phase space integrals with
the Bose-Fermi distribution function. Note that the numbers of fermion
annihilation and creation operators are the same. Alternative forms for the
quantum correlation function with the distribution function as the left
factor in the integrand and all the $\alpha _{i}$ and $g_{j}$ placed to the
right of the $\alpha _{i}^{+}$ and $g_{j}^{+}$ can be easily found.

\subsection{Existence of Distribution Function - Canonical Form}

The distribution function $P(\alpha ,\alpha ^{+},\alpha ^{\ast },\alpha
^{+\ast },g,g^{+})$ is not required to be \emph{unique}, only that it
generates the characteristic function $\chi (\xi ,\xi ^{+},h,h^{+})$ - which
is unique - via the phase space integral (\ref{Eq.BoseFermiDistnFn}).
However, it is important to be able to show that a distribution function
always \emph{exists}.

The existence of the positive P type distribution function can be shown by
introducing the so-called \emph{canonical representation} of the density
operator, which involves the fermion and boson \emph{Bargmann coherent
states }(see Appendix \ref{Appendix Bargmann Coherent States}) 
\begin{eqnarray}
\hat{\rho} &=&\int \int dg^{+}dg\,\int \int d^{2}\alpha ^{+}d^{2}\alpha
\,\,P_{canon}(\alpha ,\alpha ^{+},\alpha ^{\ast },\alpha ^{+\ast },g,g^{+})\,%
\widehat{\Lambda }(g,g^{+},\alpha ,\alpha ^{+})  \nonumber \\
&=&\int \int dg^{+}dg\,\int \int d^{2}\alpha ^{+}d^{2}\alpha \,\widehat{%
\Lambda }(g,g^{+},\alpha ,\alpha ^{+})\,P_{canon}(\alpha ,\alpha ^{+},\alpha
^{\ast },\alpha ^{+\ast },g,g^{+})\,  \nonumber \\
&&  \label{Eq.CanonicalRepnBoseFermiDensityOpr}
\end{eqnarray}%
where 
\begin{equation}
\widehat{\Lambda }(g,g^{+},\alpha ,\alpha ^{+})=\frac{\left\vert
g\right\rangle _{B}\left\langle g^{+\ast }\right\vert _{B}}{Tr(\left\vert
g\right\rangle _{B}\left\langle g^{+\ast }\right\vert _{B})}\frac{\left\vert
\alpha \right\rangle _{B}\left\langle \alpha ^{+\ast }\right\vert _{B}}{%
Tr(\left\vert \alpha \right\rangle _{B}\left\langle \alpha ^{+\ast
}\right\vert _{B})}  \label{Eq.BoseFermiBargmanStatesProjector}
\end{equation}%
is a normalised \emph{projector} and

\begin{eqnarray}
&&P_{canon}(\alpha ,\alpha ^{+},\alpha ^{\ast },\alpha ^{+\ast },g,g^{+}) 
\nonumber \\
&=&\left( \frac{1}{4\pi ^{2}}\right) ^{n}\int \int dg^{+\ast }dg^{\ast }%
\mathbf{\,\exp (}\sum_{i}(g_{i}g_{i}^{\ast }+g_{i}^{+\ast
}g_{i}^{+}+g_{i}g_{i}^{+}\mathbf{))\exp (-}\frac{1}{2}\sum_{i}(\alpha
_{i}\alpha _{i}^{\ast }+\alpha _{i}^{+\ast }\alpha _{i}^{+}\mathbf{))} 
\nonumber \\
&&\times \left\langle g\right\vert _{B}\left\langle \frac{\alpha +\alpha
^{+\ast }}{2}\right\vert _{B}\widehat{\rho }\left\vert \frac{\alpha +\alpha
^{+\ast }}{2}\right\rangle _{B}\left\vert g^{+\ast }\right\rangle _{B}
\label{Eq.CanonicalRepnFnBoseFermiDensityOpr}
\end{eqnarray}%
is a canonical representation function for the density operator $\widehat{%
\rho }$. The two forms for the density operator are the same because the
normalised projector $\widehat{\Lambda }(g,g^{+},\alpha ,\alpha ^{+})$ is an
even Grassmann operator and thus commutes with $P_{canon}(\alpha ,\alpha
^{+},\alpha ^{\ast },\alpha ^{+\ast },g,g^{+})$. By substituting for the
canonical form of the density operator (\ref%
{Eq.CanonicalRepnBoseFermiDensityOpr}) into the expression (\ref%
{Eq.Charfunction1}) for the characteristic function we can easily show that
the characteristic function and distribution function are related as in (\ref%
{Eq.BoseFermiDistnFn}) with the distribution function given by the canonical
form (\ref{Eq.CanonicalRepnFnBoseFermiDensityOpr}). For bosons the proof of
the existence of the canonical form is given by Drummond and Gardiner (\cite%
{Gardiner91a}, \cite{Drummond80a}). For fermions the proof is outlined in
papers by Cahill and Glauber \cite{Cahill99a} and Plimak et al \cite%
{Plimak01a}.

Note also that \emph{irrespective }of whether or not the detailed form for $%
P(\alpha ,\alpha ^{+},\alpha ^{\ast },\alpha ^{+\ast },g,g^{+})$ is given by
the canonical form (\ref{Eq.CanonicalRepnFnBoseFermiDensityOpr}), the
expression (\ref{Eq.CanonicalRepnBoseFermiDensityOpr}) is often used to 
\emph{define} the positive $P$ type distribution function \cite{Gardiner91a}%
, rather than the characteristic function expression (\ref%
{Eq.BoseFermiDistnFn}). If the density operator can be written as 
\begin{equation}
\widehat{\rho }=\int \int dg^{+}dg\,\int \int d^{2}\alpha ^{+}d^{2}\alpha \,%
\widehat{\Lambda }(g,g^{+},\alpha ,\alpha ^{+})\,P(\alpha ,\alpha
^{+},\alpha ^{\ast },\alpha ^{+\ast },g,g^{+})  \label{Eq.DensityOprPosP}
\end{equation}%
then this defines \emph{a }(non-unique) positive $P$ distribution function.
This form of the density operator leads to the previous relationship (\ref%
{Eq.BoseFermiDistnFn}) between the distribution and characteristic
functions.\medskip

\subsection{Probabilities and Coherences as Phase Space Integrals}

\label{SubSection - Probabilities and Coherences as Phase Space Integrals}

The Grassmann phase space integrals can be evaluated using the non-zero
result $\dint
\dprod\limits_{i}dg_{i}^{+}dg_{i}\,g_{1}g_{2}g_{2}^{+}g_{1}^{+}=1$, giving
the probabilities and coherences as bosonic phase space integrals involving
the six coefficients that determine the distribution function (\ref%
{Eq.FermiBoseDistnFn2}).

For the \emph{two atom probability} which is given by the fourth order
quantum correlation function, we obtain the result 
\begin{equation}
P_{12}=\dint d^{2}\alpha ^{+}d^{2}\alpha \,P_{0}(\alpha ,\alpha ^{+},\alpha
^{\ast },\alpha ^{+\ast })  \label{Eq.BoseFermiCorrelnPhaseSpace4}
\end{equation}%
This correlation function may be zero without $P_{0}(\alpha ,\alpha
^{+},\alpha ^{\ast },\alpha ^{+\ast })$ being zero.

The \emph{one atom probabilities} for states $1,2$ are given by 
\begin{eqnarray}
P_{1} &=&\dint d^{2}\alpha ^{+}d^{2}\alpha \,\left( P_{2}^{2;2}(\alpha
,\alpha ^{+},\alpha ^{\ast },\alpha ^{+\ast })-P_{0}(\alpha ,\alpha
^{+},\alpha ^{\ast },\alpha ^{+\ast })\right)
\label{Eq.MeanAtomNumberState1Phase2} \\
P_{2} &=&\dint d^{2}\alpha ^{+}d^{2}\alpha \,\left( P_{2}^{1;1}(\alpha
,\alpha ^{+},\alpha ^{\ast },\alpha ^{+\ast })-P_{0}(\alpha ,\alpha
^{+},\alpha ^{\ast },\alpha ^{+\ast })\right)
\label{Eq.MeanAtomNumberState2Phase2}
\end{eqnarray}%
These quantities are of course real, consistent with (\ref%
{Eq.JCDuistnFnConjResults}).

For the \emph{atomic coherences} we have the results%
\begin{eqnarray}
\rho _{12} &=&-\dint d^{2}\alpha ^{+}d^{2}\alpha \,P_{2}^{1;2}(\alpha
,\alpha ^{+},\alpha ^{\ast },\alpha ^{+\ast })  \label{Eq.Coherence12Phase2}
\\
\rho _{21} &=&-\dint d^{2}\alpha ^{+}d^{2}\alpha \,P_{2}^{2;1}(\alpha
,\alpha ^{+},\alpha ^{\ast },\alpha ^{+\ast })  \label{Eq.Coherence21Phase2}
\end{eqnarray}%
These quantities are complex conjugates, consistent with (\ref%
{Eq.JCDuistnFnConjResults}).

The \emph{mean number of photons} is 
\begin{equation}
\overline{n}=\dint d^{2}\alpha ^{+}d^{2}\alpha \,(\alpha
)\,P_{4}^{12;21}(\alpha ,\alpha ^{+})\,(\alpha ^{+})\,
\label{Eq.MeanPhotonNumberPhaseInt2}
\end{equation}%
This quantity is of course real, consistent with (\ref%
{Eq.JCDuistnFnConjResults}).

Note that the mean photon number result involves the fourth order expansion
coefficient $P_{4}^{12;21}(\alpha ,\alpha ^{+},\alpha ^{\ast },\alpha
^{+\ast })$, and the one atom probabilities for states $1,2$ involve the
opposite second order expansion coefficients - $P_{2}^{2;2}(\alpha ,\alpha
^{+},\alpha ^{\ast },\alpha ^{+\ast })$ and $P_{2}^{1;1}(\alpha ,\alpha
^{+},\alpha ^{\ast },\alpha ^{+\ast })$ respectively, minus the quantity $%
P_{0}(\alpha ,\alpha ^{+},\alpha ^{\ast },\alpha ^{+\ast })$ that determines
the two atom probability. The coherences involve the second expansion
coefficients $P_{2}^{1;2}(\alpha ,\alpha ^{+},\alpha ^{\ast },\alpha ^{+\ast
})$ and $P_{2}^{2;1}(\alpha ,\alpha ^{+},\alpha ^{\ast },\alpha ^{+\ast })$
respectively, multiplied by $-1$. Thus in their final form, only c-number
integrations are needed to determine the quantum correlation functions.

The \emph{normalisation integral} for the combined bose-fermi distribution
function is given by%
\begin{equation}
Tr\widehat{\rho }=\int \int d^{2}\alpha ^{+}d^{2}\alpha
\,\prod\limits_{i=1}^{2}dg_{i}^{+}dg_{i}\,P(\alpha ,\alpha ^{+},\alpha
^{\ast },\alpha ^{+\ast },g,g^{+})=1  \label{Eq.TraceRho}
\end{equation}%
If we substitute the specific ordered form (\ref{Eq.FermiBoseDistnFn2}) for
the distribution function, we see that the normalisation integral gives%
\begin{equation}
\int d^{2}\alpha ^{+}d^{2}\alpha \,P_{4}^{12;21}(\alpha ,\alpha ^{+},\alpha
^{\ast },\alpha ^{+\ast })=1  \label{Eq.FourthOrderCoeft2}
\end{equation}%
\medskip

\section{Fokker-Planck Equation}

\label{Section - Fokker-Planck Equation}

The derivation of the Fokker-Planck equation is based on using the \emph{%
correspondence rules}. As has been noted previously the distribution
function is not unique and is a non-analytic function of the bosonic phase
space variables. The correspondence rules are also non-unique, but the
application of the standard correspondence rules (\ref{Eq.Corr1A}) - (\ref%
{Eq.Corr8A}) is required to lead to a distribution function which correctly
determines the quantum correlation functions. It turns out however that the
Fokker-Planck equation based on the standard correspondence rules leads to a
distribution function that is not satisfactory for the Jaynes-Cummings
model. However, the canonical form of the distribution function always
exists and can be applied to the initial conditions. Furthermore, it turns
out to lead to a Fokker-Planck equation that can be solved analytically so
the Fokker-Planck equation for the \emph{canonical distribution function}
will now be obtained. The relevant correspondence rules will be those given
in Eqs. (\ref{Eq.CorrespRulesCanonicalForm2}) and the distribution function
will be written in termsof the new variables (\ref{Eq.PhaseVariableChange}).

\subsection{Correspondence Rules}

We replace the Liouville-von Neumann equation for the density operator by a 
\emph{Fokker-Planck equation} for the distribution function. To do this we
make use of so-called \emph{correspondence rules}, which are presented here
for the general case of a combined system of bosons and fermions. The
correspondence rules state what happens to the distribution function $%
P(\alpha ,\alpha ^{+},\alpha ^{\ast },\alpha ^{+\ast },g,g^{+})$ when the
density operator $\hat{\rho}$ is replaced by the product of the density
operator with an annihilation or creation operator.

The\textbf{\ }\emph{standard} correspondence rules are: 
\begin{eqnarray}
\hat{\rho} &\Rightarrow &\hat{a}_{i}\,\hat{\rho}\qquad P(\alpha ,\alpha
^{+},\alpha ^{\ast },\alpha ^{+\ast },g,g^{+})\Rightarrow \alpha _{i}\,P
\label{Eq.Corr1A} \\
\hat{\rho} &\Rightarrow &\hat{\rho}\,\hat{a}_{i}\qquad P(\alpha ,\alpha
^{+},\alpha ^{\ast },\alpha ^{+\ast },g,g^{+})\Rightarrow \left( \alpha _{i}-%
\frac{\partial }{\partial \alpha _{i}^{+}}\right) P  \label{Eq.Corr2A} \\
\hat{\rho} &\Rightarrow &\hat{a}_{i}^{\dag }\,\hat{\rho}\qquad P(\alpha
,\alpha ^{+},\alpha ^{\ast },\alpha ^{+\ast },g,g^{+})\Rightarrow \left(
\alpha _{i}^{+}-\frac{\partial }{\partial \alpha _{i}}\right) P
\label{Eq.Corr3A} \\
\hat{\rho} &\Rightarrow &\hat{\rho}\,\hat{a}_{i}^{\dag }\,\qquad P(\alpha
,\alpha ^{+},\alpha ^{\ast },\alpha ^{+\ast },g,g^{+})\Rightarrow \alpha
_{i}^{+}P  \label{Eq.Corr4a} \\
\hat{\rho} &\Rightarrow &\widehat{c}_{i}\,\hat{\rho}\qquad P(\alpha ,\alpha
^{+},\alpha ^{\ast },\alpha ^{+\ast },g,g^{+})\Rightarrow g_{i}\,P=P\,g_{i}
\label{Eq.Corr5A} \\
\hat{\rho} &\Rightarrow &\hat{\rho}\,\widehat{c}_{i}\qquad P(\alpha ,\alpha
^{+},\alpha ^{\ast },\alpha ^{+\ast },g,g^{+})\Rightarrow P\,\left( +\frac{%
\overleftarrow{\partial }}{\partial g_{i}^{+}}-g_{i}\right)
\label{Eq.Corr6A} \\
&=&\left( -\frac{\overrightarrow{\partial }}{\partial g_{i}^{+}}%
-g_{i}\right) \,P  \nonumber \\
\hat{\rho} &\Rightarrow &\widehat{c}_{i}^{\dag }\,\hat{\rho}\qquad P(\alpha
,\alpha ^{+},\alpha ^{\ast },\alpha ^{+\ast },g,g^{+})\Rightarrow \left( +%
\frac{\overrightarrow{\partial }}{\partial g_{i}}-g_{i}^{+}\right) \,P
\label{Eq.Corr7A} \\
&=&P\left( -\frac{\overleftarrow{\partial }}{\partial g_{i}}-g_{i}^{+}\right)
\nonumber \\
\hat{\rho} &\Rightarrow &\hat{\rho}\widehat{c}_{i}^{\dag }\qquad P(\alpha
,\alpha ^{+},\alpha ^{\ast },\alpha ^{+\ast },g,g^{+})\Rightarrow
\,P\,g_{i}^{+}=\,g_{i}^{+}\,P  \label{Eq.Corr8A} \\
\frac{\partial \hat{\rho}}{\partial t} &\rightarrow &\frac{\partial P(\alpha
,\alpha ^{+},\alpha ^{\ast },\alpha ^{+\ast },g,g^{+})}{\partial t}
\label{Eq.Corr9A}
\end{eqnarray}%
The proof of these results can be obtained either starting from the
expression (\ref{Eq.BoseFermiDistnFn}) for the \emph{characteristic function}
or from the expression (\ref{Eq.DensityOprPosP}) for the \emph{canonical form%
} of the density operator. The latter approach is simpler and is based on
the effect of the annihilation or creation operators on the \emph{Bargmann
state projectors} given in Eqs. (\ref{Eq.BosonBargProjectorResults}) and (%
\ref{Eq.FermionBargProjectorResults}). Both proofs involve an \emph{%
integration by parts} step, which requires the distribution function to go
to zero rapidly enough on the bosonic phase space boundary.

The proof of the first expressions for the \emph{fermion results} do not
depend on the distribution function $P(\alpha ,\alpha ^{+},\alpha ^{\ast
},\alpha ^{+\ast },g,g^{+})$ being an even Grassmann function. However, the
second form of the fermion results uses the feature that the distribution
function $P(\alpha ,\alpha ^{+},\alpha ^{\ast },\alpha ^{+\ast },g,g^{+})$
is an even Grassmann function. If it had been odd, then the second forms
would have had all their signs reversed. This point becomes important when 
\emph{applying the correspondence rules} to derive Fokker-Planck equation
terms associated with the effects of \emph{several} fermionic annihilation
or creation operators on the density operator. The correspondence rules are
carried out \emph{in succession} to give the required term in the
Fokker-Planck equation. However, multiplying or differentiating a Grassmann
distribution function changes it between being even and being odd, and this
has to be taken into account when applying the correspondence rules in
succession when dealing with fermionic operators. The safest procedure is to
use the correspondence rule form that does not depend on the evenness or
oddness of the distribution function, which is why the \emph{first forms}
for the fermion rules in Eqs. (\ref{Eq.Corr5A}) - (\ref{Eq.Corr8A}) should
be used.

The non-uniqueness of the distribution function is associated with the
bosonic phase space variables. The standard \emph{bosonic results} for the
correspondence rules in Eqs. (\ref{Eq.Corr1A}) - (\ref{Eq.Corr4a}) can be
generalised. Since the bosonic projectors $\widehat{\Lambda }_{b}(\alpha
,\alpha ^{+})$ are analytic it follows that 
\begin{eqnarray}
0 &=&\int \int dg^{+}dg\,\int \int d^{2}\alpha ^{+}d^{2}\alpha \,\left( 
\frac{\partial }{\partial \alpha _{i}^{\ast }}\widehat{\Lambda }%
(g,g^{+},\alpha ,\alpha ^{+})\right) \,P(\alpha ,\alpha ^{+},\alpha ^{\ast
},\alpha ^{+\ast },g,g^{+})  \nonumber \\
&=&\int \int dg^{+}dg\,\int \int d^{2}\alpha ^{+}d^{2}\alpha \,\left( \frac{%
\partial }{\partial \alpha _{i}^{+\ast }}\widehat{\Lambda }(g,g^{+},\alpha
,\alpha ^{+})\right) \,P(\alpha ,\alpha ^{+},\alpha ^{\ast },\alpha ^{+\ast
},g,g^{+})  \nonumber \\
&&
\end{eqnarray}%
and hence in the correspondence rule proof based on the canonical form of
the density operator we see that arbitrary linear combinations 
\begin{equation}
\lambda \frac{\partial }{\partial \alpha _{i}^{\ast }}+\lambda ^{+}\frac{%
\partial }{\partial \alpha _{i}^{+\ast }}
\label{Eq.AdditionalBoseCorresTerms}
\end{equation}%
of the derivatives with respect to $\alpha _{i}^{\ast },\alpha _{i}^{+\ast }$
may be added to \emph{each} of the standard bosonic correspondence rules in
Eqs. (\ref{Eq.Corr1A}) -(\ref{Eq.Corr4a}) at the applying integration by
parts step. A similar situation occurs for the proof based on the
characteristic function. Hence for example, the correspondence rule (\ref%
{Eq.Corr2A}) can be replaced by 
\begin{eqnarray}
\hat{\rho} &\Rightarrow &\hat{\rho}\,\hat{a}_{i}\qquad P(\alpha ,\alpha
^{+},\alpha ^{\ast },\alpha ^{+\ast },g,g^{+})\Rightarrow \left( \alpha _{i}-%
\frac{\partial }{\partial \alpha _{i}^{+}}-\frac{\partial }{\partial \alpha
_{i}^{+\ast }}\right) P=\left( \alpha _{i}-\frac{\partial }{\partial \alpha
_{ix}^{+}}\right) P\,  \nonumber \\
&&  \label{Eq.Corr2A1} \\
\hat{\rho} &\Rightarrow &\hat{\rho}\,\hat{a}_{i}\qquad P(\alpha ,\alpha
^{+},\alpha ^{\ast },\alpha ^{+\ast },g,g^{+})\Rightarrow \left( \alpha _{i}-%
\frac{\partial }{\partial i\alpha _{i}^{+}}+\frac{\partial }{\partial \alpha
_{i}^{+\ast }}\right) P=\left( \alpha _{i}-\frac{\partial }{\partial
(i\alpha _{iy}^{+})}\right) P  \nonumber \\
&&  \label{Eq.Corr2AB}
\end{eqnarray}%
choosing $\lambda =0,\lambda ^{+}=-1$ or $\lambda =0,\lambda ^{+}=+1$
respectively.

There are however, further possibilities - a feature not widely commented
upon in other work but which ultimately reflects the non-uniqueness of the
positive $P$ distribution for bosons. In particular, the \emph{flexibility}
in the correspondence rules is even greater that merely replacing $\alpha
_{i}$, $\alpha _{i}^{+}$, $\frac{{\LARGE \partial }}{{\LARGE \partial \alpha 
}_{i}}$ or $\frac{{\LARGE \partial }}{{\LARGE \partial \alpha }_{i}^{+}}$ by
these quantities plus a particular linear combination of $\frac{{\LARGE %
\partial }}{{\LARGE \partial \alpha }_{i}^{\ast }}$ and $\frac{{\LARGE %
\partial }}{{\LARGE \partial \alpha }_{i}^{+\ast }}$ as in (\ref%
{Eq.AdditionalBoseCorresTerms}) for every term when $\hat{\rho}\Rightarrow 
\hat{a}_{i}\,\hat{\rho}$, $\hat{\rho}\,\hat{a}_{i}$, $\hat{a}_{i}^{\dag }\,%
\hat{\rho}$ or $\hat{\rho}\,\hat{a}_{i}^{\dag }$ for the creation,
anihilation operators associated with a specific mode $i$. In fact, the
linear combination used can be \emph{different} for each $\alpha _{i}$, $%
\alpha _{i}^{+}$, $\frac{{\LARGE \partial }}{{\LARGE \partial \alpha }_{i}}$
or $\frac{{\LARGE \partial }}{{\LARGE \partial \alpha }_{i}^{+}}$ \emph{%
wherever} it occurs! So for example, if in one term where $\hat{\rho}%
\Rightarrow \hat{\rho}\,\hat{a}_{i}$ we replace $\left( \alpha _{i}-\frac{%
{\LARGE \partial }}{{\LARGE \partial \alpha }_{i}^{+}}\right) $ by $\left(
\alpha _{i}-\frac{{\LARGE \partial }}{{\LARGE \partial \alpha }_{i}^{+}}-%
\frac{{\LARGE \partial }}{{\LARGE \partial \alpha }_{i}^{+\ast }}\right) $
to give $\left( \alpha _{i}-\frac{{\LARGE \partial }}{{\LARGE \partial
\alpha }_{ix}^{+}}\right) $, in another term where $\hat{\rho}\Rightarrow 
\hat{\rho}\,\hat{a}_{i}$ we may replace $\left( \alpha _{i}-\frac{{\LARGE %
\partial }}{{\LARGE \partial \alpha }_{i}^{+}}\right) $ by $\left( \alpha
_{i}-\frac{{\LARGE \partial }}{{\LARGE \partial \alpha }_{i}^{+}}+\frac{%
{\LARGE \partial }}{{\LARGE \partial \alpha }_{i}^{+\ast }}\right) $ to give 
$\left( \alpha _{i}-\frac{{\LARGE \partial }}{{\LARGE \partial (i\alpha }%
_{iy}^{+}{\LARGE )}}\right) $. The reason why this is possible is that the
only requirement is that the equation for the distribution function gives
the correct equation for the characteristic function (or the density
operator). Additional terms of the form in (\ref%
{Eq.AdditionalBoseCorresTerms}) acting on the distribution function produce
zero when in the integration by parts step they act back on either the
exponential factor $\exp i\sum\limits_{j}\{\alpha _{j}\,\xi _{j}^{+}\}\,$and 
$\exp i\sum\limits_{j}\{\xi _{j}\alpha _{j}^{+}\}$ or the Bargmann state
projector $\widehat{\Lambda }_{b}(\alpha ,\alpha ^{+})$, both of which are
analytic functions of the $\alpha _{i}$, $\alpha _{i}^{+}$ and hence yield
zero when $\frac{{\LARGE \partial }}{{\LARGE \partial \alpha }_{i}^{\ast }}$
or $\frac{{\LARGE \partial }}{{\LARGE \partial \alpha }_{i}^{+\ast }}$ are
applied. This flexibility is important in being able to convert the
Fokker-Planck equation based on the standard correspondence rules into a
form with a \emph{positive definite} diffusion matrix (see \cite{Gardiner91a}%
, \cite{Drummond80a}).

For the \emph{canonical distribution function} the flexibility described in
the previous paragraph is \emph{not} available for the correspondence rules.
They do however represent particular choices of the forms (\ref%
{Eq.AdditionalBoseCorresTerms}). In terms of \emph{new variables} $\gamma
_{i}$, $\gamma _{i}^{\ast }$, $\delta _{i}$, $\delta _{i}^{\ast }$ which
replace $\alpha _{i},\alpha _{i}^{+},\alpha _{i}^{\ast },\alpha _{i}^{+\ast
} $ via 
\begin{eqnarray}
\gamma _{i} &=&\frac{1}{2}(\alpha _{i}+\alpha _{i}^{+\ast })\qquad \gamma
_{i}^{\ast }=\frac{1}{2}(\alpha _{i}^{\ast }+\alpha _{i}^{+})  \nonumber \\
\delta _{i} &=&\frac{1}{2}(\alpha _{i}-\alpha _{i}^{+\ast })\qquad \delta
_{i}^{\ast }=\frac{1}{2}(\alpha _{i}^{\ast }-\alpha _{i}^{+})  \nonumber \\
\alpha _{i} &=&\gamma _{i}+\delta _{i}\qquad \alpha _{i}^{\ast }=\gamma
_{i}^{\ast }+\delta _{i}^{\ast }  \nonumber \\
\alpha _{i}^{+} &=&\gamma _{i}^{\ast }-\delta _{i}^{\ast }\qquad \alpha
_{i}^{+\ast }=\gamma _{i}-\delta _{i}  \label{Eq.PhaseVariableChange}
\end{eqnarray}%
the canonical distribution function can be written in the form%
\begin{eqnarray}
&&P_{canon}(\gamma ,\gamma ^{\ast },\delta ,\delta ^{\ast },g,g^{+})\, 
\nonumber \\
&=&\left( \frac{1}{4\pi ^{2}}\right) ^{n}\mathbf{\exp (-}\sum_{i}\delta
_{i}\delta _{i}^{\ast })\;\mathbf{\exp (-}\sum_{i}\gamma _{i}\gamma
_{i}^{\ast })\;  \nonumber \\
&&\times \int \int dg^{+\ast }dg^{\ast }\mathbf{\,\exp (}%
\sum_{i}(g_{i}g_{i}^{\ast }+g_{i}^{+\ast }g_{i}^{+}+g_{i}g_{i}^{+}\mathbf{))}%
\left\langle g\right\vert _{B}\left\langle \gamma \right\vert _{B}\widehat{%
\rho }\left\vert \gamma \right\rangle _{B}\left\vert g^{+\ast }\right\rangle
_{B}  \nonumber \\
&&  \label{Eq.CanonicalDistnForm}
\end{eqnarray}%
The phase space integration is changed: 
\begin{equation}
\diint d^{2}\alpha ^{+}\,d^{2}\alpha \Rightarrow 4\diint d^{2}\delta
\,d^{2}\gamma  \label{Eq.PhaseSpaceIntnChange}
\end{equation}

The \emph{canonical correspondence rules} are 
\begin{eqnarray}
\hat{\rho} &\Rightarrow &\hat{a}_{i}\,\hat{\rho}\qquad P_{canon}(\gamma
,\gamma ^{\ast },\delta ,\delta ^{\ast },g,g^{+})\Rightarrow \left( \frac{%
\partial }{\partial \gamma _{i}^{\ast }}+\gamma _{i}\right) \,P_{canon} 
\nonumber \\
\hat{\rho} &\Rightarrow &\hat{\rho}\,\hat{a}_{i}\qquad P_{canon}(\gamma
,\gamma ^{\ast },\delta ,\delta ^{\ast },g,g^{+})\Rightarrow \left( \gamma
_{i}\right) \,P_{canon}  \nonumber \\
\hat{\rho} &\Rightarrow &\hat{a}_{i}^{\dag }\,\hat{\rho}\qquad
P_{canon}(\gamma ,\gamma ^{\ast },\delta ,\delta ^{\ast
},g,g^{+})\Rightarrow \left( \gamma _{i}^{\ast }\right) \,P_{canon} 
\nonumber \\
\hat{\rho} &\Rightarrow &\hat{\rho}\,\hat{a}_{i}^{\dag }\,\qquad
P_{canon}(\gamma ,\gamma ^{\ast },\delta ,\delta ^{\ast
},g,g^{+})\Rightarrow \left( \frac{\partial }{\partial \gamma _{i}}+\gamma
_{i}^{\ast }\right) \,P_{canon}  \label{Eq.CorrespRulesCanonicalForm2}
\end{eqnarray}%
using the results in Eqs. (\ref{Eq.BoseBargEigenv1}), (\ref%
{Eq.BoseBargEigenv2}), (\ref{Eq.BoseBargDeriv1}) and (\ref%
{Eq.BoseBargEigenv2}) directly in the expression (\ref{Eq.CanonicalDistnForm}%
) for the distribution function. These correspondence rules can also be
obtained from the standard bosonic correspondence rules by adding the
additional terms in Eq. (\ref{Eq.AdditionalBoseCorresTerms}). We see that 
\begin{eqnarray}
\alpha _{i}+2\frac{\partial }{\partial \alpha _{i}^{\ast }} &=&\left( \frac{%
\partial }{\partial \gamma _{i}^{\ast }}+\gamma _{i}\right) +\left( \frac{%
\partial }{\partial \delta _{i}^{\ast }}+\delta _{i}\right)  \nonumber \\
\left( \alpha _{i}-\frac{\partial }{\partial \alpha _{i}^{+}}\right) +2\frac{%
\partial }{\partial \alpha _{i}^{\ast }} &=&\gamma _{i}+\left( \frac{%
\partial }{\partial \delta _{i}^{\ast }}+\delta _{i}\right)  \nonumber \\
\left( \alpha _{i}^{+}-\frac{\partial }{\partial \alpha _{i}}\right) +2\frac{%
\partial }{\partial \alpha _{i}^{+\ast }} &=&\gamma _{i}^{\ast }-\left( 
\frac{\partial }{\partial \delta _{i}}+\delta _{i}^{\ast }\right)  \nonumber
\\
\alpha _{i}^{+}+2\frac{\partial }{\partial \alpha _{i}^{+\ast }} &=&\left( 
\frac{\partial }{\partial \gamma _{i}}+\gamma _{i}^{\ast }\right) -\left( 
\frac{\partial }{\partial \delta _{i}}+\delta _{i}^{\ast }\right)
\label{Eq.ModifiedBoseCorrRules}
\end{eqnarray}%
using 
\begin{eqnarray}
\frac{\partial }{\partial \gamma _{i}^{\ast }} &=&\frac{\partial }{\partial
\alpha _{i}^{\ast }}+\frac{\partial }{\partial \alpha _{i}^{+}}\qquad \frac{%
\partial }{\partial \gamma _{i}}=\frac{\partial }{\partial \alpha _{i}}+%
\frac{\partial }{\partial \alpha _{i}^{+\ast }}  \nonumber \\
\frac{\partial }{\partial \delta _{i}^{\ast }} &=&\frac{\partial }{\partial
\alpha _{i}^{\ast }}-\frac{\partial }{\partial \alpha _{i}^{+}}\qquad \frac{%
\partial }{\partial \delta _{i}}=\frac{\partial }{\partial \alpha _{i}}-%
\frac{\partial }{\partial \alpha _{i}^{+\ast }}
\end{eqnarray}%
It follows that applying the modified correspondence rules in Eq. (\ref%
{Eq.ModifiedBoseCorrRules}) to the canonical form (\ref%
{Eq.CanonicalDistnForm}) for the distribution function gives the same result
as in (\ref{Eq.CorrespRulesCanonicalForm2}), since the effect of the
operators involving $\delta _{i}$, $\delta _{i}^{\ast }$ is zero. Written in
terms of the original variables $\alpha _{i},\alpha _{i}^{+},\alpha
_{i}^{\ast },\alpha _{i}^{+\ast }$ the canonical correspondence rules (\ref%
{Eq.CorrespRulesCanonicalForm2}) were originally obtained by Schack and
Schenzle \cite{Schack91a} (see the Appendix).\medskip

\subsection{Canonical Distribution Function - Fokker-Planck Equation}

The \emph{Fokker-Planck equation} for the canonical distribution function $%
P_{canon}(\gamma ,\gamma ^{\ast },\delta ,\delta ^{\ast },g,g^{+})$ is%
\begin{eqnarray}
&&\frac{\partial P_{canon}(\gamma ,\gamma ^{\ast },\delta ,\delta ^{\ast
},g,g^{+})}{\partial t}  \nonumber \\
&=&-i\frac{E_{A}}{\hbar }\left( \frac{\overrightarrow{\partial }}{\partial
g_{2}}(g_{2})P_{canon}(\gamma ,\gamma ^{\ast },\delta ,\delta ^{\ast
},g,g^{+})+\frac{\overrightarrow{\partial }}{\partial g_{1}}%
(g_{1})P_{canon}(\gamma ,\gamma ^{\ast },\delta ,\delta ^{\ast
},g,g^{+})\right)  \nonumber \\
&&+i\frac{E_{A}}{\hbar }\left( P_{canon}(\gamma ,\gamma ^{\ast },\delta
,\delta ^{\ast },g,g^{+})(g_{2}^{+})\frac{\overleftarrow{\partial }}{%
\partial g_{2}^{+}}+P_{canon}(\gamma ,\gamma ^{\ast },\delta ,\delta ^{\ast
},g,g^{+})(g_{1}^{+})\frac{\overleftarrow{\partial }}{\partial g_{1}^{+}}%
\right)  \nonumber \\
&&-\frac{1}{2}i\omega _{0}\left( \frac{\overrightarrow{\partial }}{\partial
g_{2}}(g_{2})P_{canon}(\gamma ,\gamma ^{\ast },\delta ,\delta ^{\ast
},g,g^{+}))-\frac{\overrightarrow{\partial }}{\partial g_{1}}%
(g_{1})P_{canon}(\gamma ,\gamma ^{\ast },\delta ,\delta ^{\ast
},g,g^{+})\right)  \nonumber \\
&&+\frac{1}{2}i\omega _{0}\left( P_{canon}(\gamma ,\gamma ^{\ast },\delta
,\delta ^{\ast },g,g^{+})(g_{2}^{+})\frac{\overleftarrow{\partial }}{%
\partial g_{2}^{+}}-P_{canon}(\gamma ,\gamma ^{\ast },\delta ,\delta ^{\ast
},g,g^{+})(g_{1}^{+})\frac{\overleftarrow{\partial }}{\partial g_{1}^{+}}%
\right)  \nonumber \\
&&-i\omega \,\left( \frac{\partial }{\partial \gamma ^{\ast }}(\gamma ^{\ast
}\,P_{canon}(\gamma ,\gamma ^{\ast },\delta ,\delta ^{\ast },g,g^{+}))\,-%
\frac{\partial }{\partial \gamma }(\gamma \,P_{canon}(\gamma ,\gamma ^{\ast
},\delta ,\delta ^{\ast },g,g^{+}))\right)  \nonumber \\
&&-\frac{1}{2}i\Omega \left( \frac{\overrightarrow{\partial }}{\partial g_{1}%
}(g_{2})(\gamma ^{\ast }P_{canon}(\gamma ,\gamma ^{\ast },\delta ,\delta
^{\ast },g,g^{+}))\right)  \nonumber \\
&&+\,\frac{1}{2}i\Omega \left( (\gamma \,P_{canon}(\gamma ,\gamma ^{\ast
},\delta ,\delta ^{\ast },g,g^{+}))(g_{2}^{+}\,)\frac{\overleftarrow{%
\partial }}{\partial g_{1}^{+}}\right)  \nonumber \\
&&-\,\frac{1}{2}i\Omega \,\frac{\overrightarrow{\partial }}{\partial g_{2}}%
(g_{1})(\left( \gamma +\frac{\partial }{\partial \gamma ^{\ast }}\right)
P_{canon}(\gamma ,\gamma ^{\ast },\delta ,\delta ^{\ast },g,g^{+})) 
\nonumber \\
&&+\,\frac{1}{2}i\Omega \,\,\left( (\left( \gamma ^{\ast }+\frac{\partial }{%
\partial \gamma }\right) \,P_{canon}(\gamma ,\gamma ^{\ast },\delta ,\delta
^{\ast },g,g^{+}))\,(g_{1}^{+})\frac{\overleftarrow{\partial }}{\partial
g_{2}^{+}}\right)  \nonumber \\
&&+\,\frac{1}{2}i\Omega \,(g_{2}^{+}g_{1})(\left( \frac{\partial }{\partial
\gamma ^{\ast }}\right) P_{canon}(\gamma ,\gamma ^{\ast },\delta ,\delta
^{\ast },g,g^{+}))  \nonumber \\
&&-\,\frac{1}{2}i\Omega \,\,\left( (\left( \frac{\partial }{\partial \gamma }%
\right) \,P_{canon}(\gamma ,\gamma ^{\ast },\delta ,\delta ^{\ast
},g,g^{+}))\,(g_{1}^{+}g_{2})\right)  \nonumber \\
&&  \label{Eq.CanonicalFokkerPlanckJCModel}
\end{eqnarray}%
where the bosonic correspondence rules (\ref{Eq.CorrespRulesCanonicalForm2})
and the first versions of fermionic correspondence rules (\ref{Eq.Corr5A}) -
(\ref{Eq.Corr8A}) have been used. The Fokker-Planck equation based on the
standard correspondence rules is set out in Appendix \ref{Appendix - Results
for Standard Distn Fn}.\medskip

\subsection{Coupled Distribution Function Coefficients}

Writing $\gamma ,\delta $ for $\gamma ,\gamma ^{\ast },\delta ,\delta ^{\ast
}$ the expression (\ref{Eq.FermiBoseDistnFn2}) for the distribution function
can now be substituted into the Fokker-Planck equation (\ref%
{Eq.CanonicalFokkerPlanckJCModel}) for the \emph{canonical} $P$+
distribution to obtain coupled equations for the six c-number coefficients $%
P_{0}(\gamma ,\delta )$, $P_{2}^{i;j}(\gamma ,\delta )\,$\ and $%
P_{4}^{12;21}(\gamma ,\delta )\,$that specify the distribution function. For
convenience we use the same terminology for these coefficients as in the
general case and leave the canonical label understood. In the derivation the
Grassmann differentiations of various products of Grassmann variables are
first carried out using the results in Appendix \ref{Appendix Grassmann} and
we then equate terms involving the the zeroth, second and fourth order
monomials in the Grassmann variables $g_{1}$, $g_{2}$, $g_{2}^{+}$ and $%
g_{1}^{+}$ to arrive at six separate coupled equations for the\ c-number
coefficients $P_{0}(\gamma ,\delta )$, $P_{2}^{i;;j}(\gamma ,\delta )\,$\
and $P_{4}^{12;21}(\gamma ,\delta )\,$that specify the distribution function.

For the \emph{zeroth }order terms we have 
\begin{equation}
\frac{\partial }{\partial t}P_{0}(\gamma ,\delta )=-i\omega \,\left( \frac{%
\partial }{\partial \gamma ^{\ast }}\gamma ^{\ast }-\frac{\partial }{%
\partial \gamma }\gamma \right) P_{0}(\gamma ,\delta )
\label{Eq.CanonicalFPEZeroOrder}
\end{equation}

For the \emph{second }order terms we have four equations. 
\begin{eqnarray}
&&\frac{\partial }{\partial t}P_{2}^{1;1}(\gamma ,\delta )  \nonumber \\
&=&-i\omega \,\left( \frac{\partial }{\partial \gamma ^{\ast }}\gamma ^{\ast
}-\frac{\partial }{\partial \gamma }\gamma \right) P_{2}^{1;1}(\gamma
,\delta )  \nonumber \\
&&+\,\frac{1}{2}i\Omega \,\left( \gamma +\frac{\partial }{\partial \gamma
^{\ast }}\right) P_{2}^{2;1}(\gamma ,\delta )-\,\frac{1}{2}i\Omega
\,\,\left( \gamma ^{\ast }+\frac{\partial }{\partial \gamma }\right)
P_{2}^{1;2}(\gamma ,\delta )  \label{Eq.CanonicalFPESecondOrder11}
\end{eqnarray}%
\begin{eqnarray}
&&\frac{\partial }{\partial t}P_{2}^{1;2}(\gamma ,\delta )  \nonumber \\
&=&-i\omega \,\left( \frac{\partial }{\partial \gamma ^{\ast }}\gamma ^{\ast
}-\frac{\partial }{\partial \gamma }\gamma \right) P_{2}^{1;2}(\gamma
,\delta )  \nonumber \\
&&-i\omega _{0}P_{2}^{1;2}(\gamma ,\delta )  \nonumber \\
&&-\,\frac{1}{2}i\Omega \,\gamma \,\left( P_{2}^{1;1}(\gamma ,\delta
)-P_{2}^{2;2}(\gamma ,\delta )\right) +\,\frac{1}{2}i\Omega \,\left( \frac{%
\partial }{\partial \gamma ^{\ast }}\right) \left( P_{2}^{2;2}(\gamma
,\delta )-P_{0}(\gamma ,\delta )\right)  \nonumber \\
&&  \label{Eq.CanonicalFPESecondOrder12}
\end{eqnarray}%
\begin{eqnarray}
&&\frac{\partial }{\partial t}P_{2}^{2;1}(\gamma ,\delta )  \nonumber \\
&=&-i\omega \,\left( \frac{\partial }{\partial \gamma ^{\ast }}\gamma ^{\ast
}-\frac{\partial }{\partial \gamma }\gamma \right) P_{2}^{2;1}(\gamma
,\delta )  \nonumber \\
&&+i\omega _{0}P_{2}^{2;1}(\gamma ,\delta )  \nonumber \\
&&+\frac{1}{2}i\Omega \,\gamma ^{\ast }\left( P_{2}^{1;1}(\gamma ,\delta
)-P_{2}^{2;2}(\gamma ,\delta )\right) \,-\,\frac{1}{2}i\Omega \,\,\left( 
\frac{\partial }{\partial \gamma }\right) \left( P_{2}^{2;2}(\gamma ,\delta
)-P_{0}(\gamma ,\delta )\right)  \nonumber \\
&&  \label{Eq.CanonicalFPESecondOrder21}
\end{eqnarray}%
\begin{eqnarray}
&&\frac{\partial }{\partial t}P_{2}^{2;2}(\gamma ,\delta )  \nonumber \\
&=&-i\omega \,\left( \frac{\partial }{\partial \gamma ^{\ast }}\gamma ^{\ast
}-\frac{\partial }{\partial \gamma }\gamma \right) P_{2}^{2;2}(\gamma
,\delta )  \nonumber \\
&&+\frac{1}{2}i\Omega \left( \gamma ^{\ast }P_{2}^{1;2}(\gamma ,\delta
)\right) \,-\,\frac{1}{2}i\Omega \left( \gamma \,P_{2}^{2;1}(\gamma ,\delta
)\right)  \label{Eq.CanonicaFPESecondOrder22}
\end{eqnarray}

For the \emph{fourth }order term%
\begin{eqnarray}
&&\frac{\partial }{\partial t}P_{4}^{12;21}(\gamma ,\delta )  \nonumber \\
&=&-i\omega \,\left( \frac{\partial }{\partial \gamma ^{\ast }}\gamma ^{\ast
}-\frac{\partial }{\partial \gamma }\gamma \right) P_{4}^{12;21}(\gamma
,\delta )  \nonumber \\
&&+\frac{1}{2}i\Omega \,\left( \frac{\partial }{\partial \gamma ^{\ast }}%
\right) P_{2}^{2;1}(\gamma ,\delta )-\frac{1}{2}i\Omega \left( \frac{%
\partial }{\partial \gamma }\right) P_{2}^{1;2}(\gamma ,\delta )
\label{Eq.CanonicalFPEFourthOrder}
\end{eqnarray}%
The corresponding coupled equations for the general distribution function
are set out in Appendix \ref{Appendix - Results for Standard Distn Fn}%
.\medskip

\subsection{Initial Conditions for Uncorrelated Case}

For the case of \emph{uncorrelated} initial states the density operator is a
product $\widehat{\rho }=\widehat{\rho }_{f}\,\widehat{\rho }_{b}$ and the
overall \emph{initial distribution function} is just the product of the
atomic term $P_{f}(g,g^{+})$ with the cavity term $P_{b}(\alpha ,\alpha
^{+},\alpha ^{\ast },\alpha ^{+\ast })$ so that at the initial time the
distribution function is 
\begin{equation}
P(\alpha ,\alpha ^{+},\alpha ^{\ast },\alpha ^{+\ast
},g,g^{+})_{0}=P_{f}(g,g^{+})P_{b}(\alpha ,\alpha ^{+},\alpha ^{\ast
},\alpha ^{+\ast })  \label{Eq.InitialDistnFn}
\end{equation}%
and therefore the coefficients are given by 
\begin{eqnarray}
P_{0} &=&\left\langle \widehat{c}_{1}^{\dag }\widehat{c}_{2}^{\dag }\widehat{%
c}_{2}\widehat{c}_{1}\right\rangle P_{b}(\alpha ,\alpha ^{+},\alpha ^{\ast
},\alpha ^{+\ast })=0  \nonumber \\
P_{2}^{1;1} &=&\left\langle \widehat{c}_{2}^{\dag }\widehat{c}%
_{2}\right\rangle P_{b}(\alpha ,\alpha ^{+},\alpha ^{\ast },\alpha ^{+\ast
})\qquad P_{2}^{1;2}=-\left\langle \widehat{c}_{1}^{\dag }\widehat{c}%
_{2}\right\rangle P_{b}(\alpha ,\alpha ^{+},\alpha ^{\ast },\alpha ^{+\ast })
\nonumber \\
P_{2}^{2;1} &=&-\left\langle \widehat{c}_{2}^{\dag }\widehat{c}%
_{1}\right\rangle P_{b}(\alpha ,\alpha ^{+},\alpha ^{\ast },\alpha ^{+\ast
})\qquad P_{2}^{2;2}=\left\langle \widehat{c}_{1}^{\dag }\widehat{c}%
_{1}\right\rangle P_{b}(\alpha ,\alpha ^{+},\alpha ^{\ast },\alpha ^{+\ast })
\nonumber \\
P_{4}^{12;21} &=&P_{b}(\alpha ,\alpha ^{+},\alpha ^{\ast },\alpha ^{+\ast })
\label{Eq.InitialDistnFn Coeffts}
\end{eqnarray}%
where the initial distribution function for the cavity mode is 
\begin{equation}
P_{b}(\alpha ,\alpha ^{+},\alpha ^{\ast },\alpha ^{+\ast })=\left( \frac{1}{%
4\pi ^{2}}\right) \mathbf{\exp (-}\frac{1}{4}|\alpha -\alpha ^{+\ast
}|^{2})\left\langle \frac{\alpha +\alpha ^{+\ast }}{2},\frac{\alpha ^{\ast
}+\alpha ^{+}}{2}\right\vert \widehat{\rho }_{b}\left\vert \frac{\alpha
+\alpha ^{+\ast }}{2},\frac{\alpha ^{\ast }+\alpha ^{+}}{2}\right\rangle
\label{Eq.InitialBoseDistFn}
\end{equation}%
\medskip

\subsection{Rotating Phase Variables and Coefficients}

To proceed further it is useful to introduce \emph{rotating phase variables}
defined via the transformation%
\begin{eqnarray}
\alpha &=&\beta \exp (-i\omega t)\qquad \alpha ^{\ast }=\beta ^{\ast }\exp
(+i\omega t)  \nonumber \\
\alpha ^{+} &=&\beta ^{+}\exp (+i\omega t)\qquad \alpha ^{+\ast }=\beta
^{+\ast }\exp (-i\omega t)  \label{Eq.RotatingPhaseVariables}
\end{eqnarray}%
This then gives%
\begin{eqnarray}
\gamma &=&\frac{1}{2}(\beta +\beta ^{+\ast })\exp (-i\omega t)=\widetilde{%
\gamma }\exp (-i\omega t)  \nonumber \\
\gamma ^{\ast } &=&\frac{1}{2}(\beta ^{\ast }+\beta ^{+})\exp (+i\omega t)=%
\widetilde{\gamma }^{\ast }\exp (+i\omega t)  \nonumber \\
\delta &=&\frac{1}{2}(\beta -\beta ^{+\ast })\exp (-i\omega t)=\widetilde{%
\delta }\exp (-i\omega t)  \nonumber \\
\delta ^{\ast } &=&\frac{1}{2}(\beta ^{\ast }-\beta ^{+})\exp (+i\omega t)=%
\widetilde{\delta }^{\ast }\exp (+i\omega t)  \label{Eq.NewGammaDelta}
\end{eqnarray}%
so that we will replace $\gamma ,\gamma ^{\ast },\delta ,\delta ^{\ast }$ by 
$\widetilde{\gamma },\widetilde{\gamma }^{\ast },\widetilde{\delta },%
\widetilde{\delta }^{\ast }$, or $\widetilde{\gamma },\widetilde{\delta }$
for short. The canonical distribution function coefficients will be of the
form $\widetilde{P}(\widetilde{\gamma },\widetilde{\delta })$, where we now
change the labelling of the coefficients to $\widetilde{P}$ so as to reflect
this variable change.

This transformation enables the elimination of the cavity field term from
the equations for the coefficients. Explicit time dependences will be left
understood. For any $\widetilde{P}(\widetilde{\gamma },\widetilde{\delta })$%
\begin{equation}
\frac{\partial }{\partial t}P(\gamma ,\delta )=i\omega \left( \frac{\partial 
}{\partial \widetilde{\gamma }}\widetilde{\gamma }\,\widetilde{P}(\widetilde{%
\gamma },\widetilde{\delta })-\frac{\partial }{\partial \widetilde{\gamma }%
^{\ast }}\widetilde{\gamma }^{\ast }\,\widetilde{P}(\widetilde{\gamma },%
\widetilde{\delta })\right) +\frac{\partial }{\partial t}\widetilde{P}(%
\widetilde{\gamma },\widetilde{\delta })
\end{equation}%
where since each canonical distribution coefficient is of the form $\exp (-%
\widetilde{\delta }\,\widetilde{\delta }^{\ast })\,F(\widetilde{\gamma },%
\widetilde{\gamma }^{\ast })$ it follows that the derivative terms involving 
$\widetilde{\delta },\widetilde{\delta }^{\ast }$ can be ignored because 
\begin{equation}
\frac{\partial }{\partial \widetilde{\delta }}\widetilde{P}(\widetilde{%
\gamma },\widetilde{\delta })\frac{\partial \widetilde{\delta }}{\partial t}+%
\frac{\partial }{\partial \widetilde{\delta }^{\ast }}\widetilde{P}(%
\widetilde{\gamma },\widetilde{\delta })\frac{\partial \widetilde{\delta }%
^{\ast }}{\partial t}=0
\end{equation}%
Note also that the original $\gamma ,\gamma ^{\ast },\delta ,\delta ^{\ast }$
were time independent. On the right side of each coefficient equation we
have a term%
\begin{equation}
-i\omega \,\left( \frac{\partial }{\partial \gamma ^{\ast }}\gamma ^{\ast }-%
\frac{\partial }{\partial \gamma }\gamma \right) P(\gamma ,\delta )=-i\omega
\,\left( \frac{\partial }{\partial \widetilde{\gamma }^{\ast }}\widetilde{%
\gamma }^{\ast }\,\widetilde{P}(\widetilde{\gamma },\widetilde{\delta })-%
\frac{\partial }{\partial \widetilde{\gamma }}\widetilde{\gamma }\,%
\widetilde{P}(\widetilde{\gamma },\widetilde{\delta })\right)
\end{equation}%
This cancels out the equivalent term on the left side so we now have the
following results.

For the \emph{zeroth }order terms we have 
\begin{equation}
\frac{\partial }{\partial t}\widetilde{P}_{0}(\widetilde{\gamma },\widetilde{%
\delta })=0  \label{Eq.CanonicalFPEZeroOrderB}
\end{equation}

For the \emph{second }order terms we have four equations. 
\begin{eqnarray}
&&\frac{\partial }{\partial t}\widetilde{P}_{2}^{1;1}(\widetilde{\gamma },%
\widetilde{\delta })  \nonumber \\
&=&+\,\frac{1}{2}i\Omega \,\left( \widetilde{\gamma }+\frac{\partial }{%
\partial \widetilde{\gamma }^{\ast }}\right) \widetilde{P}_{2}^{2;1}(%
\widetilde{\gamma },\widetilde{\delta })\exp (-i\omega t)  \nonumber \\
&&-\,\frac{1}{2}i\Omega \,\,\left( \widetilde{\gamma }^{\ast }+\frac{%
\partial }{\partial \widetilde{\gamma }}\right) \widetilde{P}_{2}^{1;2}(%
\widetilde{\gamma },\widetilde{\delta })\exp (+i\omega t)  \nonumber \\
&&  \label{Eq.CanonicalFPESecondOrder11B}
\end{eqnarray}%
\begin{eqnarray}
&&\frac{\partial }{\partial t}\widetilde{P}_{2}^{1;2}(\widetilde{\gamma },%
\widetilde{\delta })  \nonumber \\
&=&-i\omega _{0}\widetilde{P}_{2}^{1;2}(\widetilde{\gamma },\widetilde{%
\delta })  \nonumber \\
&&-\,\frac{1}{2}i\Omega \,\widetilde{\gamma }\,\exp (-i\omega t)\left( 
\widetilde{P}_{2}^{1;1}(\widetilde{\gamma },\widetilde{\delta })-\widetilde{P%
}_{2}^{2;2}(\widetilde{\gamma },\widetilde{\delta })\right)  \nonumber \\
&&+\,\frac{1}{2}i\Omega \,\exp (-i\omega t)\left( \frac{\partial }{\partial 
\widetilde{\gamma }^{\ast }}\right) \left( \widetilde{P}_{2}^{2;2}(%
\widetilde{\gamma },\widetilde{\delta })-\widetilde{P}_{0}(\widetilde{\gamma 
},\widetilde{\delta })\right)  \nonumber \\
&&  \label{Eq.CanonicalFPESecondOrder12B}
\end{eqnarray}%
\begin{eqnarray}
&&\frac{\partial }{\partial t}\widetilde{P}_{2}^{2;1}(\widetilde{\gamma },%
\widetilde{\delta })  \nonumber \\
&=&+i\omega _{0}\widetilde{P}_{2}^{2;1}(\widetilde{\gamma },\widetilde{%
\delta })  \nonumber \\
&&+\frac{1}{2}i\Omega \,\widetilde{\gamma }^{\ast }\exp (+i\omega t)\left( 
\widetilde{P}_{2}^{1;1}(\widetilde{\gamma },\widetilde{\delta })-\widetilde{P%
}_{2}^{2;2}(\widetilde{\gamma },\widetilde{\delta })\right) \,  \nonumber \\
&&-\,\frac{1}{2}i\Omega \,\,\exp (+i\omega t)\left( \frac{\partial }{%
\partial \widetilde{\gamma }}\right) \left( \widetilde{P}_{2}^{2;2}(%
\widetilde{\gamma },\widetilde{\delta })-\widetilde{P}_{0}(\widetilde{\gamma 
},\widetilde{\delta })\right)  \nonumber \\
&&  \label{Eq.CanonicalFPESecondOrder21B}
\end{eqnarray}%
\begin{eqnarray}
&&\frac{\partial }{\partial t}\widetilde{P}_{2}^{2;2}(\widetilde{\gamma },%
\widetilde{\delta })  \nonumber \\
&=&+\frac{1}{2}i\Omega \left( \widetilde{\gamma }^{\ast }\widetilde{P}%
_{2}^{1;2}(\widetilde{\gamma },\widetilde{\delta })\exp (+i\omega t)\right)
\,  \nonumber \\
&&-\,\frac{1}{2}i\Omega \left( \widetilde{\gamma }\,\widetilde{P}_{2}^{2;1}(%
\widetilde{\gamma },\widetilde{\delta })\exp (-i\omega t)\right)  \nonumber
\\
&&  \label{Eq.CanonicalFPESecondOrder22B}
\end{eqnarray}

For the \emph{fourth }order term%
\begin{eqnarray}
&&\frac{\partial }{\partial t}\widetilde{P}_{4}^{12;21}(\widetilde{\gamma },%
\widetilde{\delta })  \nonumber \\
&=&+\frac{1}{2}i\Omega \,\left( \frac{\partial }{\partial \widetilde{\gamma }%
^{\ast }}\right) \widetilde{P}_{2}^{2;1}(\widetilde{\gamma },\widetilde{%
\delta })\exp (-i\omega t)-\frac{1}{2}i\Omega \left( \frac{\partial }{%
\partial \widetilde{\gamma }}\right) \widetilde{P}_{2}^{1;2}(\widetilde{%
\gamma },\widetilde{\delta })\exp (+i\omega t)  \nonumber \\
&&  \label{Eq.CanonicalFPEFourthOrderB}
\end{eqnarray}

Also, equations with $\emph{no}$ $\emph{explicit}$ $\emph{time}$ $\emph{%
dependence}$ can be obtained via the change of coefficients from $\widetilde{%
P}$ to $\widetilde{S}$ by incorporating $\exp (\pm i\omega t)$ in the $%
\widetilde{P}_{2}^{1;2}(\widetilde{\gamma },\widetilde{\delta })$, $%
\widetilde{P}_{2}^{2;1}(\widetilde{\gamma },\widetilde{\delta })$
coefficients, and we can also \emph{factor out} the explicit dependence on $%
\widetilde{\delta }\,$\ via the overall factor $\exp (-\widetilde{\delta }\,%
\widetilde{\delta }^{\ast })=\exp (-\delta \,\delta ^{\ast })$. We also note
that the canonical form (\ref{Eq.CanonicalDistnForm}) of the distribution
function (and hence all the coefficients) contains the factor $\exp (-\gamma
\gamma ^{\ast })=\exp (-\widetilde{\gamma }\widetilde{\gamma }^{\ast })$, so
it is useful to \emph{eliminate }this factor via the change from $\widetilde{%
P}(\widetilde{\gamma },\widetilde{\delta })$ to $\widetilde{S}(\widetilde{%
\gamma })$. The new coefficients are:%
\begin{eqnarray}
\widetilde{P}_{2}^{1;1}(\widetilde{\gamma },\widetilde{\delta }) &=&%
\widetilde{S}_{2}^{1;1}(\widetilde{\gamma })\,\exp (-\widetilde{\gamma }%
\widetilde{\gamma }^{\ast })\,\exp (-\widetilde{\delta }\,\widetilde{\delta }%
^{\ast })  \nonumber \\
\widetilde{P}_{2}^{2;2}(\widetilde{\gamma },\widetilde{\delta }) &=&%
\widetilde{S}_{2}^{2;2}(\widetilde{\gamma })\,\exp (-\widetilde{\gamma }%
\widetilde{\gamma }^{\ast })\,\exp (-\widetilde{\delta }\,\widetilde{\delta }%
^{\ast })  \nonumber \\
\widetilde{P}_{2}^{1;2}(\widetilde{\gamma },\widetilde{\delta }) &=&%
\widetilde{S}_{2}^{1;2}(\widetilde{\gamma })\,\exp (-\widetilde{\gamma }%
\widetilde{\gamma }^{\ast })\,\exp (-\widetilde{\delta }\,\widetilde{\delta }%
^{\ast })\exp (-i\omega t)  \nonumber \\
\widetilde{P}_{2}^{2;1}(\widetilde{\gamma },\widetilde{\delta }) &=&%
\widetilde{S}_{2}^{2;1}(\widetilde{\gamma })\,\exp (-\widetilde{\gamma }%
\widetilde{\gamma }^{\ast })\,\exp (-\widetilde{\delta }\,\widetilde{\delta }%
^{\ast })\exp (+i\omega t)  \nonumber \\
\widetilde{P}_{0}(\widetilde{\gamma },\widetilde{\delta }) &=&\widetilde{S}%
_{0}(\widetilde{\gamma })\,\exp (-\widetilde{\gamma }\widetilde{\gamma }%
^{\ast })\,\exp (-\widetilde{\delta }\,\widetilde{\delta }^{\ast }) 
\nonumber \\
\widetilde{P}_{4}^{12;21}(\widetilde{\gamma },\widetilde{\delta }) &=&%
\widetilde{S}_{4}^{12;21}(\widetilde{\gamma })\,\exp (-\widetilde{\gamma }%
\widetilde{\gamma }^{\ast })\,\exp (-\widetilde{\delta }\,\widetilde{\delta }%
^{\ast })  \label{Eq.CanonicalDistnCoeftFinal}
\end{eqnarray}%
The separate cavity and atomic transition frequencies are then incorporated
into the \emph{detuning} 
\begin{equation}
\Delta =\omega _{0}-\omega  \label{Eq.DetuningB}
\end{equation}%
Also, $\widetilde{\delta }\,$\ now plays no further role in the dynamics and
the $\widetilde{S}$ only depend on $\widetilde{\gamma },\widetilde{\gamma }%
^{\ast }$ (or $\widetilde{\gamma }$ for short).

With this substitution the equations for the coefficients are as follows:

For the \emph{zeroth }order term 
\begin{equation}
\frac{\partial }{\partial t}\widetilde{S}_{0}(\widetilde{\gamma })=0
\label{Eq.CanonicalDistnS0}
\end{equation}

The four second \emph{second }order equations are. 
\begin{eqnarray}
&&\frac{\partial }{\partial t}\left( \widetilde{S}_{2}^{1;1}(\widetilde{%
\gamma })-\widetilde{S}_{0}(\widetilde{\gamma })\right)  \nonumber \\
&=&+\,\frac{1}{2}i\Omega \,\left( \frac{\partial }{\partial \widetilde{%
\gamma }^{\ast }}\right) \widetilde{S}_{2}^{2;1}(\widetilde{\gamma })-\,%
\frac{1}{2}i\Omega \,\,\left( \frac{\partial }{\partial \widetilde{\gamma }}%
\right) \widetilde{S}_{2}^{1;2}(\widetilde{\gamma })
\label{Eq.CanonicalDistnS11}
\end{eqnarray}%
\begin{eqnarray}
&&\frac{\partial }{\partial t}\widetilde{S}_{2}^{1;2}(\widetilde{\gamma }) 
\nonumber \\
&=&-i\Delta \,\widetilde{S}_{2}^{1;2}(\widetilde{\gamma })  \nonumber \\
&&-\,\frac{1}{2}i\Omega \,\widetilde{\gamma }\,\left( \widetilde{S}%
_{2}^{1;1}(\widetilde{\gamma })-\widetilde{S}_{0}(\widetilde{\gamma }%
)\right) +\,\frac{1}{2}i\Omega \,\left( \frac{\partial }{\partial \widetilde{%
\gamma }^{\ast }}\right) \left( \widetilde{S}_{2}^{2;2}(\widetilde{\gamma })-%
\widetilde{S}_{0}(\widetilde{\gamma })\right)  \label{Eq.CanonicalDistnS12}
\end{eqnarray}%
\begin{eqnarray}
&&\frac{\partial }{\partial t}\widetilde{S}_{2}^{2;1}(\widetilde{\gamma }) 
\nonumber \\
&=&+i\Delta \,\widetilde{S}_{2}^{2;1}(\widetilde{\gamma })  \nonumber \\
&&+\frac{1}{2}i\Omega \,\widetilde{\gamma }^{\ast }\left( \widetilde{S}%
_{2}^{1;1}(\widetilde{\gamma })-\widetilde{S}_{0}(\widetilde{\gamma }%
)\right) \,-\,\frac{1}{2}i\Omega \,\,\left( \frac{\partial }{\partial 
\widetilde{\gamma }}\right) \left( \widetilde{S}_{2}^{2;2}(\widetilde{\gamma 
})-\widetilde{S}_{0}(\widetilde{\gamma })\right)
\label{Eq.CanonicalDistnS21}
\end{eqnarray}%
\begin{eqnarray}
&&\frac{\partial }{\partial t}\left( \widetilde{S}_{2}^{2;2}(\widetilde{%
\gamma })-\widetilde{S}_{0}(\widetilde{\gamma })\right)  \nonumber \\
&=&+\frac{1}{2}i\Omega \left( \widetilde{\gamma }^{\ast }\widetilde{S}%
_{2}^{1;2}(\widetilde{\gamma })\right) \,-\,\frac{1}{2}i\Omega \left( 
\widetilde{\gamma }\,\widetilde{S}_{2}^{2;1}(\widetilde{\gamma })\right)
\label{Eq.CanonicalDistnS22}
\end{eqnarray}%
where we have substracted the zero quantity $\frac{{\LARGE \partial }}{%
{\LARGE \partial t}}\widetilde{S}_{0}(\widetilde{\gamma })$ from each side
of the first and fourth equation. From subsection \ref{SubSection -
Probabilities and Coherences as Phase Space Integrals} we see that the
quantities $\widetilde{S}_{2}^{i;i}(\widetilde{\gamma })-\widetilde{S}_{0}(%
\widetilde{\gamma })$ determine the one atom probabilities. However, from (%
\ref{Eq.CanonicalDistnS0}) and the initial conditions (\ref%
{Eq.InitialDistnFn Coeffts}) we see that $\widetilde{S}_{0}(\widetilde{%
\gamma })$ will be zero for the one atom Jaynes-Cummings model, so we can 
\emph{ignore} $\widetilde{S}_{0}(\widetilde{\gamma })$ henceforth.

The \emph{fourth} order equation is 
\begin{eqnarray}
&&\frac{\partial }{\partial t}\widetilde{S}_{4}^{12;21}(\widetilde{\gamma })
\nonumber \\
&=&+\frac{1}{2}i\Omega \,\left( -\widetilde{\gamma }+\frac{\partial }{%
\partial \widetilde{\gamma }^{\ast }}\right) \widetilde{S}_{2}^{2;1}(%
\widetilde{\gamma })-\frac{1}{2}i\Omega \left( -\widetilde{\gamma }^{\ast }+%
\frac{\partial }{\partial \widetilde{\gamma }}\right) \widetilde{S}%
_{2}^{1;2}(\widetilde{\gamma })  \label{Eq.CanonicalDistnS1221}
\end{eqnarray}%
Analogous equations for the standard distribution function are set out in
Appendix \ref{Appendix - Results for Standard Distn Fn}. \medskip

\section{Solution to Fokker-Planck Equation}

\label{Section - Fokker-Planck Equation Solution}

The approach used by Stenholm \cite{Stenholm81a} can be adapted to provide
an analytical solution for the canonical distribution function for the one
atom Jaynes-Cummings model for any initial conditions. The equations (\ref%
{Eq.CanonicalDistnS11}) - (\ref{Eq.CanonicalDistnS22}) for the $\widetilde{S}%
_{2}^{i;j}(\widetilde{\gamma })$ can be solved via the substitution 
\begin{equation}
\widetilde{S}_{2}^{i;j}(\widetilde{\gamma })=\Psi _{i}^{\ast }(\widetilde{%
\gamma }^{\ast })\,\Psi _{j}(\widetilde{\gamma })  \label{Eq.FactorSolution}
\end{equation}%
where the $\Psi _{i}^{\ast }(\widetilde{\gamma }^{\ast })$ are functions of
the $\widetilde{\gamma }^{\ast }$ and the $\Psi _{i}(\widetilde{\gamma })$
are functions of the $\widetilde{\gamma }$, and where $\Psi _{i}(\widetilde{%
\gamma })$ satisfy the coupled equations%
\begin{eqnarray}
\frac{\partial }{\partial t}\Psi _{1}(\widetilde{\gamma }) &=&\frac{1}{2}%
i\Delta \,\Psi _{1}(\widetilde{\gamma })-\frac{1}{2}i\Omega \,\left( \frac{%
\partial }{\partial \widetilde{\gamma }}\right) \Psi _{2}(\widetilde{\gamma }%
)  \nonumber \\
\frac{\partial }{\partial t}\Psi _{2}(\widetilde{\gamma }) &=&-\frac{1}{2}%
i\Delta \,\Psi _{2}(\widetilde{\gamma })-\frac{1}{2}i\Omega \,\widetilde{%
\gamma }\,\Psi _{1}(\widetilde{\gamma })  \label{Eq.SolutionAnsatz}
\end{eqnarray}%
This ansatz is consistent with the original equations (\ref%
{Eq.CanonicalDistnS11}) - (\ref{Eq.CanonicalDistnS22}) for the $\widetilde{S}%
_{2}^{i;j}(\widetilde{\gamma })$. As indicated previously we have set $%
\widetilde{S}_{0}(\widetilde{\gamma })=0$ for the one atom case.

Differentiating the second equation and substituting from the first gives 
\begin{eqnarray}
\frac{\partial ^{2}}{\partial t^{2}}\Psi _{2}(\widetilde{\gamma })+\frac{1}{4%
}\Delta ^{2}\,\Psi _{2}(\widetilde{\gamma }) &=&-\frac{1}{4}\Omega ^{2}\,%
\widetilde{\gamma }\,\left( \frac{\partial }{\partial \widetilde{\gamma }}%
\right) \Psi _{2}(\widetilde{\gamma })  \nonumber \\
&=&-\frac{1}{4}\Omega ^{2}\,\left( \frac{\partial }{\partial s}\right) \Psi
_{2}(\widetilde{\gamma })  \label{Eq.Psi2}
\end{eqnarray}%
where the substitution 
\begin{equation}
s=\lg \,\widetilde{\gamma }\qquad \widetilde{\gamma }=\exp \,s
\label{Eq.SVariable}
\end{equation}%
has been made.

A solution of the equation (\ref{Eq.Psi2}) can be obtained using \emph{%
separation} of the variables 
\begin{equation}
\Psi _{2}(\widetilde{\gamma })=T(t)\,K(s)  \label{Eq.SepnVar}
\end{equation}%
whence we find that 
\begin{equation}
\frac{1}{T(t)}\frac{d^{2}}{dt^{2}}T(t)+\frac{1}{4}\Delta ^{2}=-\frac{1}{4}%
\Omega ^{2}\,\frac{1}{K(s)}\left( \frac{\partial }{\partial s}\right)
K(s)=-\lambda  \label{Eq.SeparatedEqns}
\end{equation}%
where since the left side is a function of $t$ and the right side is a
function of $s$ the quantity $\lambda $ must be a constant.

The solution of these two equations is straightforward. We have 
\begin{eqnarray}
K(s) &=&C\exp (\frac{4\lambda }{\Omega ^{2}}s)  \nonumber \\
&=&C\left( \widetilde{\gamma }\right) ^{{\LARGE (4\lambda }/\;{\LARGE \Omega 
}^{{\LARGE 2}}{\LARGE )}}  \label{Eq.SolnK}
\end{eqnarray}%
where $C$ is a constant, and 
\begin{equation}
T(t)=A\cos \left( \sqrt{\lambda +\Delta ^{2}/4}\;t\right) +B\sin \left( 
\sqrt{\lambda +\Delta ^{2}/4}\;t\right)  \label{Eq.SolnT}
\end{equation}%
with $A$ and $B$ also constant.

Since we require the overall distribution function to be a non-singular
single-valued function of the phase space variables we see from Eq.(\ref%
{Eq.SolnK}) that there is a restriction on $\lambda $ such that 
\begin{equation}
\frac{4\lambda }{\Omega ^{2}}=n\qquad (n=0,1,2,..)
\label{Eq.RestrictionLambda}
\end{equation}%
where $n$ is an \emph{integer}.

Combining the variables to eliminate $\lambda $ and absorbing $C$ into the
other constants we see that a solution for $\Psi _{2}(\widetilde{\gamma })$
for a particular integer $n$ is 
\begin{equation}
\Psi _{2}(\widetilde{\gamma })=\widetilde{\gamma }^{n}\left( A_{n}\cos \frac{%
1}{2}\omega _{n}t+B_{n}\sin \frac{1}{2}\omega _{n}t\right)
\label{Eqn.SolnPsi2n}
\end{equation}%
where 
\begin{equation}
\omega _{n}=\sqrt{n\,\Omega ^{2}+\Delta ^{2}}  \label{Eq.JCOscillnFreq}
\end{equation}%
is the frequency associated with population and coherence oscillations in
the one atom Jaynes-Cummings model. The corresponding solution for $\Psi
_{1}(\widetilde{\gamma })$ is then obtained from Eq.(\ref{Eq.SolutionAnsatz}%
) and thus%
\begin{eqnarray}
\Psi _{1}(\widetilde{\gamma }) &=&i\;\widetilde{\gamma }^{(n-1)}\left( \cos 
\frac{1}{2}\omega _{n}t\left\{ \frac{\omega _{n}B_{n}+i\Delta A_{n}}{\Omega }%
\right\} +\sin \frac{1}{2}\omega _{n}t\left\{ \frac{-\omega
_{n}A_{n}+i\Delta B_{n}}{\Omega }\right\} \right)  \nonumber \\
&&  \label{Eq.SolnPsi1n}
\end{eqnarray}

However, a solution with $n=0$ leads to a singular $\widetilde{\gamma }^{-1}$
behaviour, so it follows that $n$ is restricted to the positive integers.
Also, as the ansatz equations are linear the general solution is a sum of
terms with differing $n$ so that we finally have the solution in the form 
\begin{eqnarray}
\Psi _{1}(\widetilde{\gamma }) &=&i\;\dsum\limits_{n=1}^{\infty }\widetilde{%
\gamma }^{(n-1)}\left( \left\{ \frac{\omega _{n}B_{n}+i\Delta A_{n}}{\Omega }%
\right\} \cos \frac{1}{2}\omega _{n}t+\left\{ \frac{-\omega
_{n}A_{n}+i\Delta B_{n}}{\Omega }\right\} \sin \frac{1}{2}\omega _{n}t\right)
\nonumber \\
\Psi _{2}(\widetilde{\gamma }) &=&\dsum\limits_{n=1}^{\infty }\widetilde{%
\gamma }^{n}\left( A_{n}\cos \frac{1}{2}\omega _{n}t+B_{n}\sin \frac{1}{2}%
\omega _{n}t\right)  \label{Eq.FinalSolnPsiS}
\end{eqnarray}%
The constants $A_{n}$, $B_{n}$ are chosen to fit the initial conditions.

We can now express the \emph{original distribution function} coefficients in
terms of these function using Eqs. (\ref{Eq.FactorSolution}) and (\ref%
{Eq.CanonicalDistnCoeftFinal}). Expressions from which $\widetilde{S}%
_{4}^{12;21}(\widetilde{\gamma })$ could be obtained, but these are of
little interest. The results can be written in terms of the original phase
variables by substituting 
\begin{equation}
\widetilde{\gamma }=\frac{1}{2}(\alpha +\alpha ^{+\ast })\exp (+i\omega
t)\qquad \widetilde{\delta }=\frac{1}{2}(\alpha -\alpha ^{+\ast })\exp
(+i\omega t)  \label{Eq.PhaseVariablesInversion}
\end{equation}%
from (\ref{Eq.RotatingPhaseVariables}) and (\ref{Eq.NewGammaDelta}) into the
above results.

The analogous solution for the general distribution function is set out in
Appendix \ref{Appendix - Results for Standard Distn Fn}. It turns out that
the Fokker-Planck equation based on the standard correspondence rules leads
to equations for the coefficients that can also be solved by a similar
ansatz. However, the solutions lead to a distribution function that diverges
on the phase space boundary and in general diverge at large $t$, with
dependences on hyperbolic functions of $\frac{1}{2}\sqrt{n}\Omega t$ in the
case of zero detuning. This then throws the original derivation of the
standard Fokker-Planck equation into doubt because the integration by parts
step fails. Other cases where this occurs have been studied by Gilchrist et
al \cite{Gilchrist97a}.

Finally, as we will see in subsection \ref{SubSection - Comparison with
Standard Result}, the solutions based on the canonical distribution function
(\ref{Eq.CanonicalDistnCoeftFinal}) agree with the standard quantum optics
result, and in particular the quantities $A_{n}$, $B_{n}$ can be chosen so
that the initial conditions are the same as for the canonical distribution
function determined from the standard quantum optics solution.\medskip

\subsection{Comparison with Standard Quantum Optics Result}

\label{SubSection - Comparison with Standard Result}

As a comparison, we now calculate the canonical distribution function (\ref%
{Eq.CanonicalRepnFnBoseFermiDensityOpr}) as determined from the state vector
given in Eq. (\ref{Eq.JCStateVector}) and (\ref{Eq.JCTimeDepAmps}) obtained
from standard quantum optics methods.

The density operator is 
\begin{eqnarray}
\widehat{\rho } &=&\dsum\limits_{n,m}B_{n1}(t)B_{m1}^{\ast }(t)\;\exp (-i%
\overline{n-m}\omega t)\;\widehat{c}_{1}^{\dag }\left\vert 0\right\rangle
\left\langle 0\right\vert \widehat{c}_{1}\;\left\vert n\right\rangle
\left\langle m\right\vert  \nonumber \\
&&+\dsum\limits_{n,m}B_{\overline{n-1}2}(t)B_{\overline{m-1}2}^{\ast
}(t)\;\exp (-i\overline{n-m}\omega t)\;\widehat{c}_{2}^{\dag }\left\vert
0\right\rangle \left\langle 0\right\vert \widehat{c}_{2}\;\left\vert 
\overline{n-1}\right\rangle \left\langle \overline{m-1}\right\vert  \nonumber
\\
&&+\dsum\limits_{n,m}B_{n1}(t)B_{\overline{m-1}2}^{\ast }(t)\;\exp (-i%
\overline{n-m}\omega t)\;\widehat{c}_{1}^{\dag }\left\vert 0\right\rangle
\left\langle 0\right\vert \widehat{c}_{2}\;\left\vert n\right\rangle
\left\langle \overline{m-1}\right\vert  \nonumber \\
&&+\dsum\limits_{n,m}B_{\overline{n-1}2}(t)B_{m1}^{\ast }(t)\;\exp (-i%
\overline{n-m}\omega t)\;\widehat{c}_{2}^{\dag }\left\vert 0\right\rangle
\left\langle 0\right\vert \widehat{c}_{1}\;\left\vert \overline{n-1}%
\right\rangle \left\langle m\right\vert  \label{Eq.DensityOprJCModel}
\end{eqnarray}%
where the new amplitudes are given by%
\begin{eqnarray}
B_{n1}(t) &=&\left( \cos (\frac{1}{2}\omega _{n}t)\left\{ A_{n1}(0)\right\}
+i\sin (\frac{1}{2}\omega _{n}t)\left\{ \frac{\Delta \;A_{n1}(0)-\Omega 
\sqrt{n}\;A_{\overline{n-1}2}(0)}{\omega _{n}}\right\} \right)  \nonumber \\
B_{\overline{n-1}2}(t) &=&\left( \cos (\frac{1}{2}\omega _{n}t)\left\{ A_{%
\overline{n-1}2}(0)\right\} -i\sin (\frac{1}{2}\omega _{n}t)\left\{ \frac{%
\Omega \sqrt{n}\;A_{n1}(0)+\Delta \;A_{\overline{n-1}2}(0)}{\omega _{n}}%
\right\} \right)  \nonumber \\
&&  \label{Eq.NewJCAmplitudes}
\end{eqnarray}%
noting that%
\begin{eqnarray}
\exp (-\frac{1}{2}i\Delta t)\exp (-i(n\omega -\frac{1}{2}\omega _{0})t)
&=&\exp (-i(n-\frac{1}{2})\omega t)  \nonumber \\
\exp (+\frac{1}{2}i\Delta t)\exp (-i(\overline{n-1}\omega +\frac{1}{2}\omega
_{0})t) &=&\exp (-i(n-\frac{1}{2})\omega t)
\end{eqnarray}

Then with $\lambda =\frac{1}{2}(\alpha +\alpha ^{+\ast })$, $\delta =\frac{1%
}{2}(\alpha -\alpha ^{+\ast })$ and $\widetilde{\gamma }=\gamma \exp
(+i\omega t)$, $\widetilde{\delta }=\delta \exp (+i\omega t)$, as in Eqs. (%
\ref{Eq.PhaseVariableChange}) and (\ref{Eq.NewGammaDelta}) we have on
substituting into Eq. (\ref{Eq.CanonicalRepnFnBoseFermiDensityOpr}) and
using expressions for the bosonic Bargmann states from Appendix \ref%
{Appendix Bargmann Coherent States}. 
\begin{eqnarray}
&&_{B}\left\langle \frac{1}{2}(\alpha +\alpha ^{+\ast })\right\vert \,%
\widehat{\rho }\,\left\vert \frac{1}{2}(\alpha +\alpha ^{+\ast
})\right\rangle _{B}  \nonumber \\
&=&\dsum\limits_{n,m}B_{n1}(t)B_{m1}^{\ast }(t)\;\widehat{c}_{1}^{\dag
}\left\vert 0\right\rangle \left\langle 0\right\vert \widehat{c}_{1}\;\frac{(%
\widetilde{\gamma }^{\ast })^{n}}{\sqrt{n!}}\frac{(\widetilde{\gamma })^{m}}{%
\sqrt{m!}}  \nonumber \\
&&+\dsum\limits_{n,m}B_{\overline{n-1}2}(t)B_{\overline{m-1}2}^{\ast }(t)\;%
\widehat{c}_{2}^{\dag }\left\vert 0\right\rangle \left\langle 0\right\vert 
\widehat{c}_{2}\;\frac{(\widetilde{\gamma }^{\ast })^{n-1}}{\sqrt{(n-1)!}}%
\frac{(\widetilde{\gamma })^{m-1}}{\sqrt{(m-1)!}}  \nonumber \\
&&+\dsum\limits_{n,m}B_{n1}(t)B_{\overline{m-1}2}^{\ast }(t)\;\widehat{c}%
_{1}^{\dag }\left\vert 0\right\rangle \left\langle 0\right\vert \widehat{c}%
_{2}\;\frac{(\widetilde{\gamma }^{\ast })^{n}}{\sqrt{n!}}\frac{(\widetilde{%
\gamma })^{m=1}}{\sqrt{(m-1)!}}\exp (+i\omega t)  \nonumber \\
&&+\dsum\limits_{n,m}B_{\overline{n-1}2}(t)B_{m1}^{\ast }(t)\;\widehat{c}%
_{2}^{\dag }\left\vert 0\right\rangle \left\langle 0\right\vert \widehat{c}%
_{1}\;\frac{(\widetilde{\gamma }^{\ast })^{n-1}}{\sqrt{(n-1)!}}\frac{(%
\widetilde{\gamma })^{m}}{\sqrt{m!}}\exp (-i\omega t)
\end{eqnarray}%
and 
\begin{equation}
\exp (-\frac{1}{2}(\alpha \alpha ^{\ast }+\alpha ^{+\ast }\alpha ^{+})=\exp
(-\widetilde{\gamma }\,\widetilde{\gamma }^{\ast })\,\exp (-\widetilde{%
\delta }\,\widetilde{\delta }^{\ast })
\end{equation}

Also using expressions for the fermionic Bargmann states from Appendix \ref%
{Appendix Bargmann Coherent States} 
\begin{eqnarray}
&&_{B}\left\langle g\;\right\vert \,\left( \widehat{c}_{l}^{\dag }\left\vert
0\right\rangle \left\langle 0\right\vert \widehat{c}_{k}\right) \,\left\vert
\;g\right\rangle _{B}  \nonumber \\
&=&\dsum\limits_{i,j}g_{i}^{\ast }g_{j}^{+\ast }\;\delta _{il}\delta _{jk}
\end{eqnarray}%
so that%
\begin{eqnarray}
&&_{B}\left\langle g\;\right\vert \,_{B}\left\langle \frac{1}{2}(\alpha
+\alpha ^{+\ast })\right\vert \,\widehat{\rho }\,\left\vert \frac{1}{2}%
(\alpha +\alpha ^{+\ast })\right\rangle _{B}\,\left\vert \;g\right\rangle
_{B}  \nonumber \\
&=&\dsum\limits_{n,m}B_{n1}(t)B_{m1}^{\ast }(t)\;g_{1}^{\ast }g_{1}^{+\ast
}\;\frac{(\widetilde{\gamma }^{\ast })^{n}}{\sqrt{n!}}\frac{(\widetilde{%
\gamma })^{m}}{\sqrt{m!}}  \nonumber \\
&&+\dsum\limits_{n,m}B_{\overline{n-1}2}(t)B_{\overline{m-1}2}^{\ast
}(t)\;g_{2}^{\ast }g_{2}^{+\ast }\;\frac{(\widetilde{\gamma }^{\ast
})^{(n-1)}}{\sqrt{(n-1)!}}\frac{(\widetilde{\gamma })^{(m-1)}}{\sqrt{(m-1)!}}
\nonumber \\
&&+\dsum\limits_{n,m}B_{n1}(t)B_{\overline{m-1}2}^{\ast }(t)\;g_{1}^{\ast
}g_{2}^{+\ast }\;\frac{(\widetilde{\gamma }^{\ast })^{n}}{\sqrt{n!}}\frac{(%
\widetilde{\gamma })^{(m=1)}}{\sqrt{(m-1)!}}\exp (+i\omega t)  \nonumber \\
&&+\dsum\limits_{n,m}B_{\overline{n-1}2}(t)B_{m1}^{\ast }(t)\;g_{2}^{\ast
}g_{1}^{+\ast }\;\frac{(\widetilde{\gamma }^{\ast })^{(n-1)}}{\sqrt{(n-1)!}}%
\frac{(\widetilde{\gamma })^{m}}{\sqrt{m!}}\exp (-i\omega t)
\end{eqnarray}

The Grassmann phase space integrations can be carried out using results from
Appendix \ref{Appendix Grassmann} and give 
\begin{eqnarray}
&&\int \int dg^{+\ast }dg^{\ast }\mathbf{\,\exp (}\sum_{i}(g_{i}g_{i}^{\ast
}+g_{i}^{+\ast }g_{i}^{+}+g_{i}g_{i}^{+}\mathbf{))\;}g_{k}^{\ast
}g_{l}^{+\ast }  \nonumber \\
&=&g_{1}g_{1}^{+}\,(\delta _{k2}\delta _{l2})+g_{1}g_{2}^{+}\,(-\delta
_{k2}\delta _{l1})+g_{2}g_{1}^{+}\,(-\delta _{k1}\delta
_{l2})+g_{2}g_{2}^{+}\,(\delta _{k1}\delta _{l1})  \nonumber \\
&&+g_{1}g_{2}g_{2}^{+}g_{1}^{+}\,(\delta _{k2}\delta _{l2}+\delta
_{k1}\delta _{l1})
\end{eqnarray}%
so combining the results we find that the canonical distribution function is%
\begin{eqnarray}
&&P_{canon}(\alpha ,\alpha ^{+},\alpha ^{\ast },\alpha ^{+\ast },g,g^{+}) 
\nonumber \\
&=&\left( \frac{1}{4\pi ^{2}}\right) \exp (-\widetilde{\gamma }\,\widetilde{%
\gamma }^{\ast })\,\exp (-\widetilde{\delta }\,\widetilde{\delta }^{\ast }) 
\nonumber \\
&&\times \lbrack \dsum\limits_{n,m}B_{n1}(t)B_{m1}^{\ast }(t)\;\left\{
g_{2}g_{2}^{+}+g_{1}g_{2}g_{2}^{+}g_{1}^{+}\right\} \;\frac{(\widetilde{%
\gamma }^{\ast })^{n}}{\sqrt{n!}}\frac{(\widetilde{\gamma })^{m}}{\sqrt{m!}}
\nonumber \\
&&+\dsum\limits_{n,m}B_{\overline{n-1}2}(t)B_{\overline{m-1}2}^{\ast
}(t)\;\left\{ g_{1}g_{1}^{+}+g_{1}g_{2}g_{2}^{+}g_{1}^{+}\right\} \;\frac{(%
\widetilde{\gamma }^{\ast })^{(n-1)}}{\sqrt{(n-1)!}}\frac{(\widetilde{\gamma 
})^{(m-1)}}{\sqrt{(m-1)!}}  \nonumber \\
&&+\dsum\limits_{n,m}B_{n1}(t)B_{\overline{m-1}2}^{\ast }(t)\;\left\{
-g_{2}g_{1}^{+}\right\} \;\frac{(\widetilde{\gamma }^{\ast })^{n}}{\sqrt{n!}}%
\frac{(\widetilde{\gamma })^{(m=1)}}{\sqrt{(m-1)!}}\exp (+i\omega t) 
\nonumber \\
&&+\dsum\limits_{n,m}B_{\overline{n-1}2}(t)B_{m1}^{\ast }(t)\;\left\{
-g_{1}g_{2}^{+}\right\} \;\frac{(\widetilde{\gamma }^{\ast })^{(n-1)}}{\sqrt{%
(n-1)!}}\frac{(\widetilde{\gamma })^{m}}{\sqrt{m!}}\exp (-i\omega t)]
\label{Eq.CanonicalDistnStandardJCQuantumOpt}
\end{eqnarray}

From this result we can identify the coefficients%
\begin{eqnarray}
\widetilde{P}_{2}^{1;1}(\widetilde{\gamma },\widetilde{\delta }) &=&\left( 
\frac{1}{4\pi ^{2}}\right) \exp (-\widetilde{\gamma }\,\widetilde{\gamma }%
^{\ast })\,\exp (-\widetilde{\delta }\,\widetilde{\delta }^{\ast
})\dsum\limits_{n,m}B_{\overline{n-1}2}(t)B_{\overline{m-1}2}^{\ast }(t)\;%
\frac{(\widetilde{\gamma }^{\ast })^{(n-1)}}{\sqrt{(n-1)!}}\frac{(\widetilde{%
\gamma })^{(m-1)}}{\sqrt{(m-1)!}}  \nonumber \\
\widetilde{P}_{2}^{1;2}(\widetilde{\gamma },\widetilde{\delta }) &=&-\exp
(-i\omega t)\left( \frac{1}{4\pi ^{2}}\right) \exp (-\widetilde{\gamma }\,%
\widetilde{\gamma }^{\ast })\,\exp (-\widetilde{\delta }\,\widetilde{\delta }%
^{\ast })\dsum\limits_{n,m}B_{\overline{n-1}2}(t)B_{m1}^{\ast }(t)\;\frac{(%
\widetilde{\gamma }^{\ast })^{(n-1)}}{\sqrt{(n-1)!}}\frac{(\widetilde{\gamma 
})^{m}}{\sqrt{m!}}  \nonumber \\
\widetilde{P}_{2}^{2;1}(\widetilde{\gamma },\widetilde{\delta }) &=&-\exp
(+i\omega t)\left( \frac{1}{4\pi ^{2}}\right) \exp (-\widetilde{\gamma }\,%
\widetilde{\gamma }^{\ast })\,\exp (-\widetilde{\delta }\,\widetilde{\delta }%
^{\ast })\dsum\limits_{n,m}B_{n1}(t)B_{\overline{m-1}2}^{\ast }(t)\;\frac{(%
\widetilde{\gamma }^{\ast })^{n}}{\sqrt{n!}}\frac{(\widetilde{\gamma }%
)^{(m=1)}}{\sqrt{(m-1)!}}  \nonumber \\
\widetilde{P}_{2}^{2;2}(\widetilde{\gamma },\widetilde{\delta }) &=&\left( 
\frac{1}{4\pi ^{2}}\right) \exp (-\widetilde{\gamma }\,\widetilde{\gamma }%
^{\ast })\,\exp (-\widetilde{\delta }\,\widetilde{\delta }^{\ast
})\dsum\limits_{n,m}B_{n1}(t)B_{m1}^{\ast }(t)\;\frac{(\widetilde{\gamma }%
^{\ast })^{n}}{\sqrt{n!}}\frac{(\widetilde{\gamma })^{m}}{\sqrt{m!}}
\label{Eq.DistnFnCoeftsJCStandardQOpt}
\end{eqnarray}%
and 
\begin{eqnarray}
\widetilde{P}_{0}(\widetilde{\gamma },\widetilde{\delta }) &=&0  \nonumber \\
\widetilde{P}_{4}^{12;21}(\widetilde{\gamma },\widetilde{\delta }) &=&\left( 
\frac{1}{4\pi ^{2}}\right) \exp (-\widetilde{\gamma }\,\widetilde{\gamma }%
^{\ast })\,\exp (-\widetilde{\delta }\,\widetilde{\delta }^{\ast })
\label{Eq.DistnFnCoeftsStandardQO2} \\
&&\times \lbrack \dsum\limits_{n,m}B_{n1}(t)B_{m1}^{\ast }(t)\;\frac{(%
\widetilde{\gamma }^{\ast })^{n}}{\sqrt{n!}}\frac{(\widetilde{\gamma })^{m}}{%
\sqrt{m!}}+\dsum\limits_{n,m}B_{\overline{n-1}2}(t)B_{\overline{m-1}2}^{\ast
}(t)\;\frac{(\widetilde{\gamma }^{\ast })^{(n-1)}}{\sqrt{(n-1)!}}\frac{(%
\widetilde{\gamma })^{(m-1)}}{\sqrt{(m-1)!}}]  \nonumber
\end{eqnarray}

If we write%
\begin{eqnarray}
\Phi _{1}(\widetilde{\gamma }) &=&\left( \frac{1}{2\pi }\right)
\dsum\limits_{m}B_{\overline{m-1}2}^{\ast }(t)\;\frac{(\widetilde{\gamma }%
)^{(m-1)}}{\sqrt{(m-1)!}}  \nonumber \\
\Phi _{2}(\widetilde{\gamma }) &=&-\left( \frac{1}{2\pi }\right)
\dsum\limits_{m}B_{m1}^{\ast }(t)\;\frac{(\widetilde{\gamma })^{m}}{\sqrt{m!}%
}  \label{Eq.FunctionsStandardQO}
\end{eqnarray}%
then%
\begin{eqnarray}
\widetilde{P}_{2}^{1;1}(\widetilde{\gamma },\widetilde{\delta }) &=&\exp (-%
\widetilde{\gamma }\,\widetilde{\gamma }^{\ast })\,\exp (-\widetilde{\delta }%
\,\widetilde{\delta }^{\ast })\;\Phi _{1}^{\ast }(\widetilde{\gamma }^{\ast
})\,\Phi _{1}(\widetilde{\gamma })  \nonumber \\
\widetilde{P}_{2}^{1;2}(\widetilde{\gamma },\widetilde{\delta }) &=&\exp (-%
\widetilde{\gamma }\,\widetilde{\gamma }^{\ast })\,\exp (-\widetilde{\delta }%
\,\widetilde{\delta }^{\ast })\;\Phi _{1}^{\ast }(\widetilde{\gamma }^{\ast
})\,\Phi _{2}(\widetilde{\gamma })\;\exp (-i\omega t)  \nonumber \\
\widetilde{P}_{2}^{2;1}(\widetilde{\gamma },\widetilde{\delta }) &=&\exp (-%
\widetilde{\gamma }\,\widetilde{\gamma }^{\ast })\,\exp (-\widetilde{\delta }%
\,\widetilde{\delta }^{\ast })\;\Phi _{2}^{\ast }(\widetilde{\gamma }^{\ast
})\,\Phi _{1}(\widetilde{\gamma })\;\exp (+i\omega t)  \nonumber \\
\widetilde{P}_{2}^{2;2}(\widetilde{\gamma },\widetilde{\delta }) &=&\exp (-%
\widetilde{\gamma }\,\widetilde{\gamma }^{\ast })\,\exp (-\widetilde{\delta }%
\,\widetilde{\delta }^{\ast })\;\Phi _{2}^{\ast }(\widetilde{\gamma }^{\ast
})\,\Phi _{2}(\widetilde{\gamma })  \label{Eq.FactorFormSecOrderCoeftsQO}
\end{eqnarray}

To have agreement with the previous results in Eq. (\ref%
{Eq.CanonicalDistnCoeftFinal}) so that 
\begin{eqnarray}
\Phi _{1}(\widetilde{\gamma }) &=&\Psi _{1}(\widetilde{\gamma })  \nonumber
\\
\Phi _{2}(\widetilde{\gamma }) &=&\Psi _{2}(\widetilde{\gamma })
\label{Eq.GrassStandardQOCompare}
\end{eqnarray}%
where $\Psi _{1}(\widetilde{\gamma })$ and $\Psi _{2}(\widetilde{\gamma })$
are as in Eqs. (\ref{Eq.FinalSolnPsiS}), we require%
\begin{eqnarray}
i\;\left\{ \frac{\omega _{n}B_{n}+i\Delta A_{n}}{\Omega }\right\} &=&\left( 
\frac{1}{2\pi }\right) \left\{ A_{\overline{n-1}2}(0)\right\} ^{\ast }\frac{1%
}{\sqrt{(n-1)!}}  \nonumber \\
i\;\left\{ \frac{-\omega _{n}A_{n}+i\Delta B_{n}}{\Omega }\right\} &=&\left( 
\frac{1}{2\pi }\right) (-i)^{\ast }\left\{ \frac{\Omega \sqrt{n}%
\;A_{n1}(0)+\Delta \;A_{\overline{n-1}2}(0)}{\omega _{n}}\right\} ^{\ast }%
\frac{1}{\sqrt{(n-1)!}}  \nonumber \\
A_{n} &=&\left( \frac{1}{2\pi }\right) (-)\left\{ A_{n1}(0)\right\} ^{\ast }%
\frac{1}{\sqrt{(n)!}}  \nonumber \\
B_{n} &=&\left( \frac{1}{2\pi }\right) (-)(+i)^{\ast }\left\{ \frac{\Delta
\;A_{n1}(0)-\Omega \sqrt{n}\;A_{\overline{n-1}2}(0)}{\omega _{n}}\right\}
^{\ast }\frac{1}{\sqrt{(n)!}}  \nonumber \\
&&  \label{Eq.ConsistencyRequired}
\end{eqnarray}%
The last two equations give explicit expressions for $A_{n}$ and $B_{n}$.
Substituting these expressions into the left side of the first two equations
gives the right hand sides, showing that the four equations are consistent.
Hence we see that for \emph{any} initial conditions for the one atom
Jaynes-Cummings model, the solution given by the Grassmann phase space
approach is the same as that from the standard quantum optics
treatment.\medskip

\subsection{Application of Results}

As an illustration of how to apply the above results for the canonical
distribution function we consider the case where the atom is initially in
the \emph{lower} state and the field is in a \emph{coherent} state of
amplitude $\eta $. In ths case we have from subsection \ref{SubSection -
Initial Conditions}%
\begin{eqnarray}
P_{2}^{1;1} &=&0\qquad P_{2}^{1;2}=0  \nonumber \\
P_{2}^{2;1} &=&0\qquad P_{2}^{2;2}=P_{b}(\alpha ,\alpha ^{+})
\end{eqnarray}%
with 
\[
P_{b}(\alpha ,\alpha ^{+})=\frac{1}{4\pi ^{2}}\exp (-\frac{{\small |}\alpha
-\alpha ^{+\ast }{\small |}^{2}}{4})\exp \left( -\left\vert \frac{1}{2}%
{\small (}\alpha +\alpha ^{+\ast })-\eta \right\vert ^{2}\right) 
\]%
Hence 
\begin{eqnarray}
\widetilde{P}_{2}^{2;2}(\widetilde{\gamma },\widetilde{\delta }) &=&\frac{%
{\small 1}}{4\pi ^{2}}\exp (-\widetilde{\delta }\,\widetilde{\delta }^{\ast
})\exp \left( -\left\vert \widetilde{\gamma }-\eta \right\vert ^{2}\right)
=\exp (-\widetilde{\delta }\,\widetilde{\delta }^{\ast })\exp (-\widetilde{%
\gamma }\widetilde{\gamma }^{\ast })\,\widetilde{S}_{2}^{2;2}(\widetilde{%
\gamma })  \nonumber \\
\widetilde{S}_{2}^{2;2}(\widetilde{\gamma }) &=&\frac{{\small 1}}{4\pi ^{2}}%
\exp \left( -(\widetilde{\gamma }-\eta )(\widetilde{\gamma }^{\ast }-\eta
^{\ast })\right) \exp (+\widetilde{\gamma }\widetilde{\gamma }^{\ast }) 
\nonumber \\
&=&\left( \frac{1}{2\pi }\exp (-\frac{1}{2}\eta \eta ^{\ast })\exp (-%
\widetilde{\gamma }^{\ast }\eta )\right) \left( \frac{1}{2\pi }\exp (-\frac{1%
}{2}\eta \eta ^{\ast })\exp (-\widetilde{\gamma }\eta ^{\ast })\right) 
\nonumber \\
&&
\end{eqnarray}

At $t=0$ we have 
\begin{eqnarray}
\Psi _{1}(\widetilde{\gamma }) &=&i\;\dsum\limits_{n=1}^{\infty }\widetilde{%
\gamma }^{(n-1)}\left( \left\{ \frac{\omega _{n}B_{n}+i\Delta A_{n}}{\Omega }%
\right\} \right) =0  \nonumber \\
\Psi _{2}(\widetilde{\gamma }) &=&\dsum\limits_{n=1}^{\infty }\widetilde{%
\gamma }^{n}\left( A_{n}\right) =\frac{1}{2\pi }\exp (-\frac{1}{2}\eta \eta
^{\ast })\exp (-\widetilde{\gamma }\eta ^{\ast })
\end{eqnarray}%
so that if we choose 
\begin{eqnarray}
A_{n} &=&\frac{1}{2\pi }\exp (-\frac{1}{2}\eta \eta ^{\ast })\frac{(-\eta
^{\ast })^{n}}{n!}  \nonumber \\
B_{n} &=&\frac{-i\Delta }{\omega _{n}}A_{n}
\end{eqnarray}%
the solutions for $\Psi _{1}(\widetilde{\gamma })$ and $\Psi _{2}(\widetilde{%
\gamma })$ determines the time dependent distribution function.\medskip

\section{Conclusion}

\label{Section - Conclusion}

We have shown that a phase space approach using Grassmann variables to
describe the atomic system and c-number variables to describe the cavity
mode can be used to treat the Jaynes-Cummings model and to obtain the same
results for treating phenomena such as pure Rabi oscillations and collapse,
revival effects as those from standard quantum optics methods. The
Liouville-von Neumann equation for the density operator was converted into a
Fokker-Planck equation for the canonical positive $P$ distribution function
using the correspondence rules associated with this choice of distribution
function. The distribution function is a Grassmann function involving
Grassmann phase space variables $g_{1}$, $g_{1}^{+}$ and $g_{2}$, $g_{2}^{+}$
for the two fermionic modes associated with the two atomic states, with six
c-number functions of the bosonic phase space variables $\alpha $, $\alpha
^{+}$ associated with the cavity mode being involved as coefficients in
specifying the distribution function. In the context of a general mixed
state where there may be zero, one or two atoms present, expressions for the
probabilities of finding one atom in one of the two atomic states, one atom
in both atomic states and no atom in either atomic state were obtained as
bosonic phase space integrals involving the six bosonic coefficients.
Coupled equations for the six bosonic coefficients for the canonical
distribution function were obtained from the Fokker-Planck equation \ These
equations were solved for the one atom Jaynes-Cummings model using an ansatz
similar to that applied by Stenholm \cite{Stenholm81a} in an earlier
Bargmann state treatment of the Jaynes-Cummings model, and the results shown
to be equivalent to the standard quantum optics treatment based on state
vectors and coupled amplitude equations.

Positive $P$ distribution functions have the feature of being non-unique,
with different correspondence rules applying in the derivation of the
specific Fokker-Planck equation. In this application we also found that
applying the standard correspondence rules (rather than those for the
canonical positive $P$ case) leads to a Fokker-Planck equation where the
solution for the coupled equations for the bosonic coefficients via a
similar ansatz was quite unsuitable. Not only did the solutions diverge for
large time $t$, but the distribution function diverged for large phase space
variables $\alpha $, $\alpha ^{+}$, thereby throwing into question the
derivation of the Fokker-Planck equation. The standard treatment requires
the distribution function to vanish on the phase space boundary. As the
correspondence rules for the canonical positive $P$ distribution do not
require this feature, it is suggested that Fokker-Planck equations based on
the canonical positive $P$ distribution may be more reliable. Furthermore,
in terms of matching a general solution of the Fokker-Planck equation to the
initial conditions, the use of the canonical form of the distribution
function is easiest since initial conditions are usually specified via the
initial density operator from which the canonical distribution function is
directly determined. However, more general Fokker-Planck equations involving
derivatives higher than second order may occur using the canonical
distribution function \cite{Schack91a}, so that no replacement by Langevin
stochastic equations is then possible. In fact even if the Fokker-Planck
equation is only second order, the diffusion matrix may not be positive
definite.

The successful treatment of this classic quantum optics system based on
phase space methods using Grassmann variables represents an important step
in applying such methods to treat more complex problems involving fermionic
systems.\medskip

\section{Acknowledgements}

This work was supported by the Australian Research Council Centre of
Excellence for Quantum Atom Optics. The authors thank J. Corney, P.
Drummond, M. Olsen and L. Plimak for helpful discussions. \pagebreak

\pagebreak

\section{Appendix 1 - Grassmann Numbers and Calculus}

\label{Appendix Grassmann}

Since Grassmann variables and their calculus may be unfamiliar to many
physicists a short summary may be desirable. More extensive accounts of the
properties of Grassmann variables and their calculus are given in Refs. \cite%
{Cahill99a, Berezin66a}.

\subsection{1.1 Grassmann Algebra}

Grassmann variables satisfy the following \emph{anti-commutation rules} with
each other and with fermion annihilation, creation operators%
\begin{eqnarray}
g_{i}g_{j} &=&-g_{j}g_{i}\qquad g_{i}\widehat{c}_{j}=-\widehat{c}%
_{j}g_{i}\qquad g_{i}\widehat{c}_{j}^{\dag }=-\widehat{c}_{j}^{\dag }g_{i} 
\nonumber \\
\{g_{i},g_{j}\} &=&\{g_{i},\widehat{c}_{j}\}=\{g_{i},\widehat{c}_{j}^{\dag
}\}=0  \label{Eq.AntiCommGrassmanVariables}
\end{eqnarray}%
and Grassmann variables and fermion operators commute with c-numbers and
boson operators. A key feature of Grassmann variables that immediately
follows is that their square and hence all higher powers are zero. 
\begin{equation}
g_{i}^{2}=g_{i}^{3}=...=0  \label{Eq.PowersGrassmanVariables}
\end{equation}%
Grassmann variables also have no inverses, and hence division is undefined.

\emph{Grassmann functions} involving linear combinations of products of
several Grassmann variables with c-numbers coefficients can be defined in an
obvious way and are of the form%
\begin{eqnarray}
f(h_{1},h_{2},..,h_{n})
&=&f_{0}+\tsum\limits_{i}f_{i}\,h_{i}+\tsum\limits_{i<j}f_{ij}\,h_{i}h_{j}+%
\tsum\limits_{i<j<k}f_{ijk}\,h_{i}h_{j}h_{k}+..  \nonumber \\
&&+f_{123..n}\,h_{1}h_{2}..h_{n}  \label{Eq.GrassmannFn}
\end{eqnarray}%
where $f_{0},f_{i},f_{ij},..,f_{12..n}$\ are c-numbers. Functions may be 
\emph{even} or \emph{odd} depending on whether all their terms contain an
even or odd number of Grassmann variables. A general function is the sum of
an even and an odd function. Two even or two odd Grassmann functions commute
with each other, whilst an even and an odd function anti-commute.

Grassmann functions have a \emph{linearity} feature in that each Grassmann
variable can at most appear in a linear form due to (\ref%
{Eq.PowersGrassmanVariables}). Thus 
\begin{eqnarray}
&&f(h_{1},h_{2},..,h_{n})  \nonumber \\
&=&a_{i}(h_{1},..,h_{i-1},h_{i+1},..h_{n})+b_{i}(h_{1},..,h_{i-1},h_{i+1},..h_{n})\,h_{i}
\label{Eq.LinearityGrassFns} \\
&=&a_{i}(h_{1},..,h_{i-1},h_{i+1},..h_{n})+h_{i}%
\,c_{i}(h_{1},..,h_{i-1},h_{i+1},..h_{n})\,  \nonumber
\end{eqnarray}%
where $a_{i}$, $b_{i}$ and $c_{i}$ are Grassmann functions that are
independent of $h_{i}$. In general $b_{i}$ and $c_{i}$ are not the same.
Linearity has the effect of truncating functions, thus the Grassmann
function $\exp g$ is just equal to $1+g$.

\emph{Complex conjugation} can also be defined with the property%
\begin{equation}
(g_{i}g_{j})^{\ast }=g_{j}^{\ast }g_{i}^{\ast }
\label{Eq.ConjugateGrassmannVariables}
\end{equation}%
Apart from the conjugation, the anti-commuting multiplication rules and the
lack of an inverse, Grassmann variables satisfy the normal algebraic rules
(associative laws of addition, multiplication etc).\medskip

\subsection{1.2 Grassmann Differentiation}

The basic rules for \emph{Grassmann differentiation} are 
\begin{eqnarray}
\frac{\overrightarrow{\partial }}{\partial g_{i}}1 &=&1\frac{\overleftarrow{%
\partial }}{\partial g_{i}}=0  \nonumber \\
\frac{\overrightarrow{\partial }}{\partial g_{i}}g_{j} &=&g_{j}\frac{%
\overleftarrow{\partial }}{\partial g_{i}}=\delta _{i,j}
\label{Eq.GrassmannDiffnBasic2}
\end{eqnarray}%
where both left and right differentiation are defined.

Differentiation proceeds via moving the variable to be differentiated to the
left or right of all the other Grassmann variables and then applying the
above rules. Thus for the Grassmann function in (\ref{Eq.LinearityGrassFns})
we have%
\begin{eqnarray}
\frac{\overrightarrow{\partial }}{\partial h_{i}}f(h_{1},h_{2},..,h_{n})
&=&c_{i}(h_{1},..,h_{i-1},h_{i+1},..h_{n})  \nonumber \\
f(h_{1},h_{2},..,h_{n})\frac{\overleftarrow{\partial }}{\partial h_{i}}
&=&b_{i}(h_{1},..,h_{i-1},h_{i+1},..h_{n})  \label{Eq.GrassmannDiffnFn}
\end{eqnarray}%
Note that in general, left and right differentiation do not give the same
result.

Multiple differentiation is carried out in the order of the derivatives.
Thus for left differentiation%
\begin{equation}
\frac{\overrightarrow{\partial }}{\partial h_{i}}\frac{\overrightarrow{%
\partial }}{\partial h_{j}}f(h_{1},h_{2},..,h_{n})=\frac{\overrightarrow{%
\partial }}{\partial h_{i}}\left( \frac{\overrightarrow{\partial }}{\partial
h_{j}}f(h_{1},h_{2},..,h_{n})\right)  \label{Eq.MultipleGrassDiffn}
\end{equation}%
with equivalent results for right differentiation. Mixed left and right
differentiation also occurs and it turns out the different orders for
carrying this out give the same result. Also left and right differentiation
are related for even and odd functions.

\begin{eqnarray}
\frac{\overrightarrow{\partial }}{\partial g_{i}}f_{E}(g) &=&(-1)f_{E}(g)%
\frac{\overleftarrow{\partial }}{\partial g_{i}}
\label{Eq.GrassmannDifferentn9} \\
\frac{\overrightarrow{\partial }}{\partial g_{i}}f_{O}(g) &=&(+1)f_{O}(g)%
\frac{\overleftarrow{\partial }}{\partial g_{i}}
\label{Eq.GrassmannDifferentn10}
\end{eqnarray}

Product rules for differentiation can be derived. These depend on whether
the factors are even or odd Grassmann functions\textbf{\ }%
\begin{eqnarray}
\frac{\overrightarrow{\partial }}{\partial g_{i}}(f_{1}^{E}f_{2}) &=&(\frac{%
\overrightarrow{\partial }}{\partial g_{i}}f_{1}^{E})f_{2}+f_{1}^{E}(\frac{%
\overrightarrow{\partial }}{\partial g_{i}}f_{2})  \nonumber \\
\frac{\overrightarrow{\partial }}{\partial g_{i}}(f_{1}^{O}f_{2}) &=&(\frac{%
\overrightarrow{\partial }}{\partial g_{i}}f_{1}^{O})f_{2}-f_{1}^{O}(\frac{%
\overrightarrow{\partial }}{\partial g_{i}}f_{2})  \nonumber \\
(f_{2}f_{1}^{E})\frac{\overleftarrow{\partial }}{\partial g_{i}}
&=&f_{2}(f_{1}^{E}\frac{\overleftarrow{\partial }}{\partial g_{i}})+(f_{2}%
\frac{\overleftarrow{\partial }}{\partial g_{i}})f_{1}^{E}  \nonumber \\
(f_{2}f_{1}^{O})\frac{\overleftarrow{\partial }}{\partial g_{i}}
&=&f_{2}(f_{1}^{O}\frac{\overleftarrow{\partial }}{\partial g_{i}})-(f_{2}%
\frac{\overleftarrow{\partial }}{\partial g_{i}})f_{1}^{O}
\label{Eq.ProdRuleGrassmannDiffn}
\end{eqnarray}%
Thus the product rule is different in general from that in ordinary
calculus. For Grassmann functions that are neither even nor odd the
derivative of a product can be obtained from (\ref{Eq.ProdRuleGrassmannDiffn}%
) after writing the function as the sum of its even and odd
components.\medskip

\subsection{1.3 Grassmann Integration}

The basic rules for \emph{Grassmann integration} are 
\begin{eqnarray}
\dint dg_{i}1 &=&\dint 1\,dg_{i}=0  \nonumber \\
\dint dg_{i}\,g_{j} &=&\delta _{ij}.\qquad \dint g_{j}\,dg_{i}\,=-\delta
_{ij}  \label{Eq.GrassmannIntegnBasic2}
\end{eqnarray}%
where both left and right integration are defined. The different results for
left and right integration are due to the differentials thems being
anti-commuting Grassmann variables. In the present paper only left
integration will be used.

The same outcome from $\frac{\overrightarrow{\partial }}{\partial g_{i}}%
g_{j}=\delta _{i,j},\frac{\overrightarrow{\partial }}{\partial g_{i}}1=0$
and $\dint dg_{i}\,g_{j}=\delta _{ij},\dint dg_{i}1=0$ provokes the comment
that differentiation and integration are the same, but this is not really
the case in view of the different results for right integration and
differentiation.

Integration proceeds via moving the variable to be integrated to the left or
right of all the other Grassmann variables and then applying the above
rules. Thus for the Grassmann function in (\ref{Eq.LinearityGrassFns}) we
have%
\begin{eqnarray}
\dint dh_{i}\,f(h_{1},h_{2},..,h_{n})
&=&c_{i}(h_{1},..,h_{i-1},h_{i+1},..h_{n})  \nonumber \\
\dint f(h_{1},h_{2},..,h_{n})\,dh_{i}\,f(h_{1},h_{2},..,h_{n})
&=&-b_{i}(h_{1},..,h_{i-1},h_{i+1},..h_{n})  \label{Eq.GrassmannIntegnFn}
\end{eqnarray}

Multiple integration is carried out in the order of the differentials, for
example in left integration%
\begin{equation}
\int \dint dh_{i}\,dh_{j}\,f(h_{1},h_{2},..,h_{n})=\int dh_{i}\,\left( \int
dh_{j}\,f(h_{1},h_{2},..,h_{n})\right)  \label{Eq.MultipleGrassIntegn}
\end{equation}%
with equivalent results for right integration.

An important results is for the complete Grassmann integral 
\begin{equation}
\int \dint ..\int dh_{1}\,dh_{2}..dh_{n}\,h_{n}...h_{2}\,h_{1}=1
\label{Eq.CompleteGrassInteg}
\end{equation}%
\medskip

\subsection{1.4 Grassmann States and Grassmann Operators}

For fermion systems, vectors can be defined in a generalised form of Hilbert
space, which involve linear combinations of basis vectors such as Eq.(\ref%
{Eq.BasisStates}) but now with Grassmann numbers as the coefficients. Such 
\emph{Grassmann vectors} are not taken to represent physical states (where
the coefficients must be c-numbers) even if a fixed number of fermions are
involved, but they have uses in the mathematical manipulations. The fermion
coherent states are Grassmann vectors.

Similarly, we may introduce generalised operators in this new Hilbert space
by taking linear combinations of the products of the fermion operators with
Grassmann numbers as the coefficients, such as in Eq.(\ref%
{Eq.FermionOmegaOprs}). Such \emph{Grassmann operators} do not represent
physical quantities (which must involve c-numbers as coefficients) or
symmetry operations, but again are useful mathematically.

Many of the results for Grassmann functions also apply for Grassmann
operators and states, in which some of the Grassmann variables are replaced
by fermion annihilation or creation operators. The anti-commuting feature of
the fermion operators with Grassmann variables enables the same proofs to be
made.

In addition to the rules (\ref{Eq.AntiCommGrassmanVariables}) for Grassmann
numbers, it is necessary to state the basic rules for multiplying the vacuum
state with Grassmann number, since the Fock states involve products of
creation operators acting on the vacuum state $\left\vert 0\right\rangle $,
. The rule is that the Grassmann numbers commute with $\left\vert
0\right\rangle $ or $\left\langle 0\right\vert $.%
\begin{equation}
g\left\vert 0\right\rangle =\left\vert 0\right\rangle g\qquad g\left\langle
0\right\vert =\left\langle 0\right\vert g  \label{Eq.CommRuleVacuumState}
\end{equation}%
Note that as a consequence%
\begin{eqnarray}
g\left\vert m_{1};m_{2};n\right\rangle &=&(-1)^{m_{1}+m_{2}}\left\vert
m_{1};m_{2};n\right\rangle g  \nonumber \\
\left\langle m_{1};m_{2};n\right\vert g &=&(-1)^{m_{1}+m_{2}}g\left\langle
m_{1};m_{2};n\right\vert  \label{Eq.CommRuleFermionFockStates}
\end{eqnarray}%
so clearly for $N$ fermion states, a Grassmann number anti-commutes with the
Fock states for fermion systems with an odd number of fermions, and commutes
if the number of fermions is even. The bra and ket vectors for fermion
states may be classified as \emph{even} or \emph{odd vectors} depending on
whether they only contain terms with even or odd numbers of fermions. Thus $%
(\left\vert 0;0;n\right\rangle +\left\vert 1;1;n\right\rangle )$ would be an
even vector, whilst $(\left\vert 0;1;n\right\rangle +\left\vert
1;0;n\right\rangle )$ would be odd. These concepts may be extended to
include \emph{Grassmann vectors} as well as ordinary state vectors. In this
case as well as the vacuum state $\left\vert 0\right\rangle $ or $%
\left\langle 0\right\vert $, all terms in even (odd) vectors contain an even
(odd) number of fermion creation operators, annihilation operators \emph{and}
Grassmann numbers, and consequently commute (anti-commute) with any
Grassmann number. Thus the Grassmann vectors $(\widehat{1}+\widehat{c}%
_{1}^{\dag }h_{1})\left\vert 0;0;n\right\rangle =\left\vert
0;0;n\right\rangle -h_{1}\left\vert 1;0;n\right\rangle $ or $\left\langle
0;0;n\right\vert (\widehat{1}+h_{1}^{+}\widehat{c}_{1})=\left\langle
0;0;n\right\vert -\left\langle 1;0;n\right\vert h_{1}^{+}$ are even vectors.

Some operators such as those that represent physical quantities contain an
even number of fermion creation and annihilation operators, usually the same
number of each (see the Hamiltonian $\widehat{H}$ in (\ref{Eq.Hamiltonian})
for example, which is the sum of terms each containing the same numbers of
annihilation and creation operators). These operators are called \emph{even} 
\emph{operators}. Consequently even operators will commute with a Grassmann
number. Other operators of interest such as the fermion cretaion and
annihilation operators themselves involve odd numbers of creation and
annihilation operators. These operators are called \emph{odd} \emph{operators%
}. Operators either commute or anti-commute with a Grassmann number
depending on whether they are even or odd. Arbitary operators can always be
expressed as the sum of an even operator and an odd operator and a Grassmann
number will commute with the even component and anti-commute with the odd
component. These concepts may be extended to include \emph{Grassmann
operators} as well as ordinary quantum operators. In this case all terms in
even (odd) operators contain an even (odd) number of fermion creation
operators, annihilation operators \emph{and} Grassmann numbers, and
consequently commute (anti-commute) with a Grassmann number. Thus the
operators $\hat{\Omega}_{f}^{+}(h^{+})$, $\hat{\Omega}_{f}^{-}(h)$ in (\ref%
{Eq.FermionOmegaOprs}) are even operators. Operators for which there are no
Grassmann variables involved are just a special case where evenness or
oddness only depends on the number of fermion creation, annihilation
operators.\pagebreak

\section{Appendix 2 - Bargmann Coherent States}

\label{Appendix Bargmann Coherent States}

For both bosonic and fermionic systems we can define the \emph{coherent
states}, which are defined by the effect of unitary displacement operators
on the vacuum state. These states are parameterised via c-number variables $%
\alpha _{i},\alpha _{i}^{\ast }$ for the bosonic modes and via Grassmann
variables $g_{i},g_{i}^{\ast }$ for the fermionic modes. The coherent states
are normalised to unity and are eigenstates of the boson or fermion
annihilation operators with eigenvalues $\alpha _{i}$ or $g_{i}$
respectively. A full description of the coherent states is given in \cite%
{Cahill69a}, \cite{Cahill99a} for the bosonic and fermionic cases. As in
these papers, we will treat the general case of multi-mode systems. For our
purposes it will be convenient to use a related set of un-normalised states
called the \emph{Bargmann states}, which have the property of only depending
on $\alpha _{i}$ or $g_{i}$, and not on the complex conjugates.\medskip

\subsection{2.1 Bosons}

For bosons the Bargmann states are defined as 
\begin{equation}
\left\vert \alpha \right\rangle _{B}=\exp \left( \sum\limits_{i=1}^{n}\hat{a}%
_{i}^{\dag }\alpha _{i}\right) \left\vert 0\right\rangle
=\prod\limits_{i}\exp \left( \hat{a}_{i}^{\dag }\alpha _{i}\right)
\left\vert 0\right\rangle  \label{Eq.BargmannBosonState2}
\end{equation}%
which only depend on the c-number variables $\alpha \equiv \{\alpha
_{1},\alpha _{2},..,\alpha _{i},..,\alpha _{n}\}$, and not on the complex
conjugates $\alpha ^{\ast }\equiv \{\alpha _{1}^{\ast },\alpha _{2}^{\ast
},..,\alpha _{i}^{\ast },..,\alpha _{n}^{\ast }\}$. These are related to the
coherent states $\left\vert \alpha ,\alpha ^{\ast }\right\rangle $ via 
\begin{eqnarray}
\left\vert \alpha ,\alpha ^{\ast }\right\rangle &=&\exp (-\frac{1}{2}\mathbf{%
\alpha }^{\ast }\cdot \mathbf{\alpha })\left\vert \alpha \right\rangle _{B}
\label{Eq.BosonCoherState} \\
&=&\exp (-\frac{1}{2}\mathbf{\alpha }^{\ast }\cdot \mathbf{\alpha }%
)\prod\limits_{i}\sum\limits_{\nu _{i}=0}^{\infty }\frac{\left( \alpha
_{i}\right) ^{\nu _{i}}}{\sqrt{\nu _{i}!}}\left\vert \nu _{i}\right\rangle ,
\label{Eq.BosonCoherState2}
\end{eqnarray}%
where $\mathbf{\alpha }^{\ast }\cdot \mathbf{\beta =}\sum\limits_{i=1}^{n}%
\alpha _{i}^{\ast }\beta _{i}$. The second expression is the well-known
expansion of the coherent state in terms of Fock states. This c-number
expansion represents a Poisson distribution of number states with a mean
boson number $\left\langle \widehat{n}_{i}\right\rangle =|\alpha _{i}|^{2}$
and a variance $\left\langle \Delta \widehat{n}_{i}^{2}\right\rangle
=\left\langle \widehat{n}_{i}\right\rangle $. The coherent state is not a
physical state except in the case of photons.

As in the case of coherent states, the Bargmann states are eigenstates of
the annihilation operator 
\begin{eqnarray}
\hat{a}_{i}\left\vert \alpha \right\rangle _{B} &=&\alpha _{i}\left\vert
\alpha \right\rangle _{B}  \label{Eq.BoseBargEigenv1} \\
_{B}\left\langle \alpha \right\vert \hat{a}_{i}^{\dag } &=&_{B}\left\langle
\alpha \right\vert \alpha _{i}^{\ast }.  \label{Eq.BoseBargEigenv2}
\end{eqnarray}%
They satisfy normalisation and orthogonality conditions 
\begin{equation}
_{B}\left\langle \alpha |\beta \right\rangle _{B}=\exp \{\mathbf{\alpha }%
^{\ast }\cdot \mathbf{\beta \}}  \label{Eq.OrthogBoseBarg2}
\end{equation}%
Hence Bargmann states are unnormalised versions of the coherent states.

The operation of a creation operator on a Bargmann ket vector or an
annihilation operator on a Bargmann bra vector can be written in terms of
derivatives of these vectors. 
\begin{eqnarray}
\widehat{a}_{i}^{\dag }\,\left\vert \alpha \right\rangle _{B} &=&\left( 
\frac{\partial }{\partial \alpha _{i}}\right) \left\vert \alpha
\right\rangle _{B}  \label{Eq.BoseBargDeriv1} \\
\left\langle \alpha \right\vert _{B}\,\widehat{a}_{i} &=&\left( \frac{%
\partial }{\partial \alpha _{i}^{\ast }}\right) \left\langle \alpha
\right\vert _{B}\,  \label{Eq.BoseBargDeriv2}
\end{eqnarray}%
there being no distinction between left and right differentiation.\medskip

\subsection{2.2 Fermions}

For fermions we define the Bargmann states via 
\begin{equation}
\left\vert g\right\rangle _{B}=\exp \left( \sum\limits_{i=1}^{n}\hat{c}%
_{i}^{\dag }g_{i}\right) \left\vert 0\right\rangle =\prod\limits_{i}(1+\hat{c%
}_{i}^{\dag }g_{i})\left\vert 0\right\rangle =\prod\limits_{i}(\left\vert
0_{i}\right\rangle -g_{i}\left\vert 1_{i}\right\rangle )
\label{Eq.BargmannFermionState2}
\end{equation}%
which only depend on $n$ Grassmann numbers $g\equiv
\{g_{1},g_{2},..,g_{i},..,g_{n}\}$, and not on the complex conjugates $%
g^{\ast }\equiv \{g_{1}^{\ast },g_{2}^{\ast },..,g_{i}^{\ast
},..,g_{n}^{\ast }\}$. These are related to the fermion coherent states $%
\left\vert g,g^{\ast }\right\rangle $ in a similar way as in the boson case. 
\begin{eqnarray}
\left\vert g,g^{\ast }\right\rangle &=&\exp (-\frac{1}{2}\mathbf{g}^{\ast
}\cdot \mathbf{g})\left\vert g\right\rangle _{B}
\label{Eq.FermionCoherState1} \\
&=&\exp (-\frac{1}{2}\mathbf{g}^{\ast }\cdot \mathbf{g})\prod\limits_{i}(%
\left\vert 0_{i}\right\rangle -g_{i}\left\vert 1_{i}\right\rangle ).
\label{Eq.FermionCoherState2}
\end{eqnarray}
where $\mathbf{g}^{\ast }\cdot \mathbf{h=}\sum\limits_{i=1}^{n}g_{i}^{\ast
}h_{i}$. This differs from the corresponding bosonic expansion because it
only involves a superposition of a zero fermion state with a one fermion
state. This is to be expected from the Pauli exclusion principle since any
mode can only be occupied by at most one fermion. The fermion coherent state
is of course unphysical as it involves Grassmann numbers as expansion
coefficients. The Bargmann states are employed by Plimak et al. \cite%
{Plimak01a} rather than the fermion coherent states as in Cahill and Glauber 
\cite{Cahill99a}. Bargmann states are even Grassmann vectors.

As for the coherent states, the fermion Bargmann states are eigenstates of
the annihilation operator 
\begin{eqnarray}
\hat{c}_{i}\left\vert g\right\rangle _{B} &=&g_{i}\left\vert g\right\rangle
_{B}  \label{Eq.FermiBargEigenv1} \\
_{B}\left\langle g\right\vert \hat{c}_{i}^{\dag } &=&_{B}\left\langle
g\right\vert g_{i}^{\ast }.  \label{Eq.FermiBargEigenv2}
\end{eqnarray}%
In the fermion case we can also find eigenstates of the creation operator,
see \cite{Cahill99a}. The Bargmann states satisfy normalisation and
orthogonality conditions 
\begin{equation}
_{B}\left\langle g|h\right\rangle _{B}=\exp \{\mathbf{g}^{\ast }\cdot 
\mathbf{h}\}.  \label{Eq.OrthogFermionBarg}
\end{equation}%
Note the similarity of these results to those for the boson states.

The operation of a creation operator on a Bargmann ket vector or an
annihilation operator on a Bargmann bra vector can be written in terms of
derivatives of these vectors. In the fermion case 
\begin{eqnarray}
\widehat{c}_{i}^{\dag }\,\left\vert g\right\rangle _{B} &=&\left( -\frac{%
\overrightarrow{\partial }}{\partial g_{i}}\right) \left\vert g\right\rangle
_{B}=\left\vert g\right\rangle _{B}\left( +\frac{\overleftarrow{\partial }}{%
\partial g_{i}}\right)  \label{Eq.FermiBargDeriv1} \\
\left\langle g\right\vert _{B}\,\widehat{c}_{i} &=&\left\langle g\right\vert
_{B}\,\left( -\frac{\overleftarrow{\partial }}{\partial g_{i}^{\ast }}%
\right) =\left( +\frac{\overrightarrow{\partial }}{\partial g_{i}^{\ast }}%
\right) \left\langle g\right\vert _{B}\,  \label{Eq.FermiBargDeriv2}
\end{eqnarray}%
where we note that both left and right derivatives forms apply for each of $%
\widehat{c}_{i}^{\dag }\,\left\vert g\right\rangle _{B}$ and $\left\langle
g\right\vert _{B}\,\widehat{c}_{i}$. This is analogous to the feature that
in (\ref{Eq.FermiBargEigenv1}) and (\ref{Eq.FermiBargEigenv2}) the
eigenvalues can be placed on either side of the bra or ket Bargmann vector.

The Bargmann states can be used for a representation of quantum operators.
For the fermion operator $\hat{\Omega}_{f}$ we introduce completeness
relationships for two sets of fermion Fock states $\left\vert \nu _{1},\nu
_{2},..,\nu _{n}\right\rangle $, $\left\vert \xi _{1},\xi _{2},..,\xi
_{n}\right\rangle ,$ where the occupation numbers for the various single
particle states $\nu _{i}$, $\xi _{i}$ are $0,1$ only. We have 
\[
\left\langle g\right\vert _{B}\hat{\Omega}_{f}\left\vert h\right\rangle
_{B}=\sum\limits_{\nu _{1},\nu _{2},..,\nu _{n}}\sum\limits_{\xi _{1},\xi
_{2},..,\xi _{n}}\left\langle g|\nu _{1},\nu _{2},..,\nu _{n}\right\rangle
_{B}\left\langle \nu _{1},\nu _{2},..,\nu _{n}\right\vert \hat{\Omega}%
_{f}\left\vert \xi _{1},\xi _{2},..,\xi _{n}\right\rangle \left\langle \xi
_{1},\xi _{2},..,\xi _{n}|h\right\rangle _{B} 
\]%
Then from the expression for the Bargmann states 
\[
\left\langle \xi _{1},\xi _{2},..,\xi _{n}|h\right\rangle
_{B}=\dprod\limits_{j}\left\langle \xi _{j}\right\vert
\prod\limits_{i}(\left\vert 0_{i}\right\rangle +\left\vert
1_{i}\right\rangle h_{i})=\prod\limits_{i}(\delta _{\xi _{j}0}+\delta _{\xi
_{j}1}h_{i})=\prod\limits_{i}(h_{i})^{\xi _{i}} 
\]%
giving the result%
\begin{equation}
\left\langle g\right\vert _{B}\hat{\Omega}_{f}\left\vert h\right\rangle
_{B}=\sum\limits_{\nu _{1},\nu _{2},..,\nu _{n}}\sum\limits_{\xi _{1},\xi
_{2},..,\xi _{n}}\Omega _{f}(\{\nu \};\{\xi \})(g_{m_{1}}^{\ast })^{\nu
_{1}}(g_{m_{2}}^{\ast })^{\nu _{2}}..(g_{m_{n}}^{\ast })^{\nu
_{n}}(h_{m_{n}})^{\xi _{n}}..(h_{m_{2}})^{\xi _{2}}(h_{m_{1}})^{\xi _{1}}
\label{Eq.BargStateRepnOpr}
\end{equation}%
where $\Omega _{f}(\{\nu \};\{\xi \})=\left\langle \nu _{1},\nu _{2},..,\nu
_{n}\right\vert \hat{\Omega}_{f}\left\vert \xi _{1},\xi _{2},..,\xi
_{n}\right\rangle $.\medskip

\subsection{2.3 Projectors}

The Bargmann states can be used to define normalised projectors as 
\begin{eqnarray}
\widehat{\Lambda }_{b}(\alpha ,\beta ^{\ast }) &=&\frac{\left\vert \alpha
\right\rangle _{B}\left\langle \beta \right\vert _{B}}{Tr_{b}(\left\vert
\alpha \right\rangle _{B}\left\langle \beta \right\vert _{B})}
\label{Eq.NormBoseProjector3} \\
\widehat{\Lambda }_{f}(g,h^{\ast }) &=&\frac{\left\vert g\right\rangle
_{B}\left\langle h\right\vert _{B}}{Tr_{f}(\left\vert g\right\rangle
_{B}\left\langle h\right\vert _{B})}.  \label{Eq.NormFermiProjector3}
\end{eqnarray}%
where the trace of the projectors $\left\vert \alpha \right\rangle
_{B}\left\langle \beta \right\vert _{B}$ and $\left\vert g\right\rangle
_{B}\left\langle h\right\vert _{B}$ have a simple form 
\begin{eqnarray}
Tr_{f}(\left\vert g\right\rangle _{B}\left\langle h\right\vert _{B}) &=&\exp
\{\mathbf{g}\cdot \mathbf{h}^{\ast }\}=_{B}\left\langle -h|g\right\rangle
_{B}=_{B}\left\langle h|-g\right\rangle _{B}  \label{Eq.TraceFermiProjector2}
\\
Tr_{b}(\left\vert \alpha \right\rangle _{B}\left\langle \beta \right\vert
_{B}) &=&\exp \{\mathbf{\alpha }\cdot \mathbf{\beta }^{\ast
}\}=_{B}\left\langle \beta |\alpha \right\rangle _{B}
\label{Eq.TraceBoseProjector2}
\end{eqnarray}%
The normalised projectors have the property that their trace is unity. 
\begin{equation}
Tr_{b}\widehat{\Lambda }_{b}(\alpha ,\beta ^{\ast })=1\qquad Tr_{f}\widehat{%
\Lambda }_{f}(g,h^{\ast })=1.  \label{Eq. TraceNormProjectors}
\end{equation}%
\medskip

\subsection{2.4 Completeness}

One of the important features of both the boson and fermion coherent states
is that they satisfy completeness relationships. In terms of Bargmann states
the completeness relationships are%
\begin{eqnarray}
\int d^{2}\mathbf{g\exp (-\mathbf{g}^{\ast }\cdot \mathbf{g})\,}\left\vert
g\right\rangle _{B}\left\langle g\right\vert _{B} &=&\hat{1}  \nonumber \\
\int d^{2}\mathbf{\alpha \exp (-\mathbf{\alpha }^{\ast }\cdot \mathbf{\alpha 
})\,}\left\vert \alpha \right\rangle _{B}\left\langle \alpha \right\vert
_{B} &=&\pi \hat{1}  \label{Eq.CompletenessBargmannStates}
\end{eqnarray}%
for fermions and bosons respectively. Here $d^{2}\mathbf{g\equiv }%
\tprod\limits_{i}dg_{i}^{\ast }dg_{i}$ and $d^{2}\mathbf{\alpha \equiv }%
\tprod\limits_{i}d\alpha _{ix}d\alpha _{iy}\medskip $

\subsection{2.5 Trace Properties}

Cyclic properties of Grassmann operators can be established. For a pair of
even Grassmann operators

\begin{equation}
Tr(\hat{\Omega}_{f}^{E}(h)\hat{\Delta}_{f}^{E}(k))=Tr(\hat{\Delta}_{f}^{E}(k)%
\hat{\Omega}_{f}^{E}(-h))=Tr(\hat{\Delta}_{f}^{E}(-k)\hat{\Omega}_{f}^{E}(h))
\label{Eq.CyclicTraceEvenOprs}
\end{equation}%
For the product of two odd Grassmann operators. 
\begin{equation}
Tr(\hat{\Omega}_{f}^{O}(h)\hat{\Delta}_{f}^{O}(k))=Tr(\hat{\Delta}_{f}^{O}(k)%
\hat{\Omega}_{f}^{O}(-h))=Tr(\hat{\Delta}_{f}^{O}(-k)\hat{\Omega}_{f}^{O}(h))
\label{Eq.CyclicTraceOddOprs}
\end{equation}

\subsection{2.6 Operator Identities}

The effect of bosonic annihilation and creation operators on the Bargmann
state projectors involved in the canonical form for the density operator are

\begin{eqnarray}
\widehat{a}_{i}\,\widehat{\Lambda }_{b}(\alpha ,\alpha ^{+}) &=&\alpha _{i}%
\widehat{\Lambda }_{b}(\alpha ,\alpha ^{+})  \nonumber \\
\widehat{\Lambda }_{b}(\alpha ,\alpha ^{+})\,\widehat{a}_{i}^{\dag }
&=&\alpha _{i}^{+}\,\widehat{\Lambda }_{b}(\alpha ,\alpha ^{+})\,  \nonumber
\\
\widehat{a}_{i}^{\dag }\,\widehat{\Lambda }_{b}(\alpha ,\alpha ^{+})
&=&\left( \frac{\partial }{\partial \alpha _{i}}+\alpha _{i}^{+}\right) 
\widehat{\Lambda }_{b}(\alpha ,\alpha ^{+})  \nonumber \\
\widehat{\Lambda }_{b}(\alpha ,\alpha ^{+})\,\widehat{a}_{i} &=&\left( \frac{%
\partial }{\partial \alpha _{i}^{+}}+\alpha _{i}\right) \widehat{\Lambda }%
_{b}(\alpha ,\alpha ^{+})\,  \label{Eq.BosonBargProjectorResults}
\end{eqnarray}%
The effect of fermionic annihilation and creation operators on the Bargmann
state projectors involved in the canonical form for the density operator are%
\begin{eqnarray}
\widehat{c}_{i}\,\widehat{\Lambda }_{f}(g,g^{+}) &=&g_{i}\,\widehat{\Lambda }%
_{f}(g,g^{+})=\widehat{\Lambda }_{f}(g,g^{+})\,g_{i}  \nonumber \\
\widehat{\Lambda }_{f}(g,g^{+})\,\widehat{c}_{i}^{\dag } &=&\widehat{\Lambda 
}_{f}(g,g^{+})\,g_{i}^{+}=g_{i}^{+}\,\widehat{\Lambda }_{f}(g,g^{+})\, 
\nonumber \\
\widehat{c}_{i}^{\dag }\,\widehat{\Lambda }_{f}(g,g^{+}) &=&\left( -\frac{%
\overrightarrow{\partial }}{\partial g_{i}}-g_{i}^{+}\right) \widehat{%
\Lambda }_{f}(g,g^{+})=\widehat{\Lambda }_{f}(g,g^{+})\left( +\frac{%
\overleftarrow{\partial }}{\partial g_{i}}-g_{i}^{+}\right)  \nonumber \\
\widehat{\Lambda }_{f}(g,g^{+})\,\widehat{c}_{i} &=&\widehat{\Lambda }%
_{f}(g,g^{+})\,\left( -\frac{\overleftarrow{\partial }}{\partial g_{i}^{+}}%
-g_{i}\right) =\left( +\frac{\overrightarrow{\partial }}{\partial g_{i}^{+}}%
-g_{i}\right) \widehat{\Lambda }_{f}(g,g^{+})\,  \nonumber \\
&&  \label{Eq.FermionBargProjectorResults}
\end{eqnarray}%
the second form of the result following from $\widehat{\Lambda }%
_{f}(g,g^{+}) $ being an even Grassmann operator.\pagebreak

\section{Appendix 5 - Results for Standard Distribution Function}

\label{Appendix - Results for Standard Distn Fn}

\subsection{5.1 Fokker-Planck Equation}

The \emph{Fokker-Planck equation }for the general distribution function $%
P(\alpha ,\alpha ^{+},\alpha ^{\ast },\alpha ^{+\ast },g,g^{+})$ is%
\begin{eqnarray}
&&\frac{\partial P(\alpha ,\alpha ^{+},g,g^{+})}{\partial t}  \nonumber \\
&=&-i\frac{E_{A}}{\hbar }\left( \frac{\overrightarrow{\partial }}{\partial
g_{2}}\{g_{2}\,P(\alpha ,\alpha ^{+},g,g^{+})\}+\frac{\overrightarrow{%
\partial }}{\partial g_{1}}\{g_{1}P(\alpha ,\alpha ^{+},g,g^{+})\}\right) 
\nonumber \\
&&+i\frac{E_{A}}{\hbar }\left( \{P(\alpha ,\alpha ^{+},g,g^{+})\,g_{2}^{+}\}%
\frac{\overleftarrow{\partial }}{\partial g_{2}^{+}}+\{P(\alpha ,\alpha
^{+},g,g^{+})\,g_{1}^{+}\}\frac{\overleftarrow{\partial }}{\partial g_{1}^{+}%
}\right)  \nonumber \\
&&-\frac{1}{2}i\,\omega _{0}\,\left( \frac{\overrightarrow{\partial }}{%
\partial g_{2}}\{g_{2}\,P(\alpha ,\alpha ^{+},g,g^{+})\}-\frac{%
\overrightarrow{\partial }}{\partial g_{1}}\{g_{1}P(\alpha ,\alpha
^{+},g,g^{+})\}\right)  \nonumber \\
&&+\frac{1}{2}i\,\omega _{0}\,\left( \{P(\alpha ,\alpha
^{+},g,g^{+})\,g_{2}^{+}\}\frac{\overleftarrow{\partial }}{\partial g_{2}^{+}%
}-\{P(\alpha ,\alpha ^{+},g,g^{+})\,g_{1}^{+}\}\frac{\overleftarrow{\partial 
}}{\partial g_{1}^{+}}\right)  \nonumber \\
&&-i\,\omega \,\left( -\frac{\partial }{\partial \alpha }\{\alpha \,P(\alpha
,\alpha ^{+},g,g^{+})\}\,+\frac{\partial }{\partial \alpha ^{+}}\{\alpha
^{+}\,P(\alpha ,\alpha ^{+},g,g^{+})\}\right)  \nonumber \\
&&-\frac{1}{2}i\,\Omega \,\left( \frac{\overrightarrow{\partial }}{\partial
g_{1}}\{g_{2}\alpha ^{+}\,P(\alpha ,\alpha ^{+},g,g^{+})\}+\frac{%
\overrightarrow{\partial }}{\partial g_{2}}\{g_{1}\alpha \,P(\alpha ,\alpha
^{+},g,g^{+})\}\right)  \nonumber \\
&&+\frac{1}{2}i\,\Omega \left( \{P(\alpha ,\alpha
^{+},g,g^{+})\,g_{2}^{+}\,\alpha \}\frac{\overleftarrow{\partial }}{\partial
g_{1}^{+}}+\{P(\alpha ,\alpha ^{+},g,g^{+})\,g_{1}^{+}\,\alpha ^{+}\}\frac{%
\overleftarrow{\partial }}{\partial g_{2}^{+}}\right)  \nonumber \\
&&+\frac{1}{2}i\,\Omega \,\left( \frac{\overrightarrow{\partial }}{\partial
g_{1}}\{g_{2}\,\left( \frac{\partial }{\partial \alpha }P(\alpha ,\alpha
^{+},g,g^{+})\right) \}-\{\left( \frac{\partial }{\partial \alpha ^{+}}%
P(\alpha ,\alpha ^{+},g,g^{+})\,\right) g_{2}^{+}\,\}\frac{\overleftarrow{%
\partial }}{\partial g_{1}^{+}}\right)  \nonumber \\
&&-\frac{1}{2}i\,\Omega \,\left( \,\left( g_{1}^{+}g_{2}\right) \left( \frac{%
\partial }{\partial \alpha }P(\alpha ,\alpha ^{+},g,g^{+})\right) -\left( 
\frac{\partial }{\partial \alpha ^{+}}P(\alpha ,\alpha ^{+},g,g^{+})\right)
\,(g_{2}^{+}g_{1})\right)  \nonumber \\
&&  \label{Eq.Fokker-Planck}
\end{eqnarray}%
where for simplicity we write $P(\alpha ,\alpha ^{+},g,g^{+})$ instead of
the full expression $P(\alpha ,\alpha ^{+},\alpha ^{\ast },\alpha ^{+\ast
},g,g^{+})$ using the notation $\alpha ,\alpha ^{+}$ for $\alpha ,\alpha
^{+},\alpha ^{\ast },\alpha ^{+\ast }$. \medskip

\subsection{5.2 Coupled Expansion Coefficients}

The coupled equations for the coefficients specifying the general
distribution function are obtained from the Fokker-Planck equation (\ref%
{Eq.Fokker-Planck}) and as follows.

For the \emph{zeroth order} terms%
\begin{equation}
\frac{\partial }{\partial t}P_{0}(\alpha ,\alpha ^{+})=-i\,\omega \,\left( -%
\frac{\partial }{\partial \alpha }\alpha \,+\frac{\partial }{\partial \alpha
^{+}}\alpha ^{+}\right) \,P_{0}(\alpha ,\alpha ^{+})
\label{Eq.ZerothOrderCoeft}
\end{equation}

For the \emph{second order} terms 
\begin{eqnarray}
\frac{\partial }{\partial t}P_{2}^{1;1}(\alpha ,\alpha ^{+}) &=&-i\,\omega
\,\left( -\frac{\partial }{\partial \alpha }\alpha \,+\frac{\partial }{%
\partial \alpha ^{+}}\alpha ^{+}\right) \,P_{2}^{1;1}(\alpha ,\alpha ^{+}) 
\nonumber \\
&&+\frac{1}{2}i\,\Omega \,\left( \alpha \,P_{2}^{2;1}(\alpha ,\alpha
^{+})-\alpha ^{+}P_{2}^{1;2}(\alpha ,\alpha ^{+})\right)
\label{Eq.SecondOrderCoeft11}
\end{eqnarray}

\begin{eqnarray}
\frac{\partial }{\partial t}P_{2}^{1;2}(\alpha ,\alpha ^{+}) &=&-i\,\omega
\,\left( -\frac{\partial }{\partial \alpha }\alpha \,+\frac{\partial }{%
\partial \alpha ^{+}}\alpha ^{+}\right) \,P_{2}^{1;2}(\alpha ,\alpha ^{+}) 
\nonumber \\
&&-i\,\omega _{0}\,P_{2}^{1;2}(\alpha ,\alpha ^{+})\,  \nonumber \\
&&+\frac{1}{2}i\,\Omega \,\left( \alpha \,P_{2}^{2;2}(\alpha ,\alpha
^{+})-\alpha P_{2}^{1;1}(\alpha ,\alpha ^{+})\right)  \nonumber \\
&&+\,\frac{1}{2}i\,\Omega \left( \frac{\partial }{\partial \alpha ^{+}}%
\left\{ P_{2}^{1;1}(\alpha ,\alpha ^{+})-P_{0}(\alpha ,\alpha ^{+})\right\}
\right)  \label{Eq.SecondOrderCoeft12}
\end{eqnarray}

\begin{eqnarray}
\frac{\partial }{\partial t}P_{2}^{2;1}(\alpha ,\alpha ^{+}) &=&-i\,\omega
\,\left( -\frac{\partial }{\partial \alpha }\alpha \,+\frac{\partial }{%
\partial \alpha ^{+}}\alpha ^{+}\right) \,P_{2}^{2;1}(\alpha ,\alpha ^{+}) 
\nonumber \\
&&+i\,\omega _{0}\,P_{2}^{2;1}(\alpha ,\alpha ^{+})\,  \nonumber \\
&&+\frac{1}{2}i\,\Omega \,\left( \alpha ^{+}P_{2}^{1;1}(\alpha ,\alpha
^{+})-\alpha ^{+}P_{2}^{2;2}(\alpha ,\alpha ^{+})\right)  \nonumber \\
&&-\frac{1}{2}i\,\Omega \,\left( \frac{\partial }{\partial \alpha }\left\{
P_{2}^{1;1}(\alpha ,\alpha ^{+})-P_{0}(\alpha ,\alpha ^{+})\right\} \right)
\label{Eq.SecondOrderCoeft21}
\end{eqnarray}

\begin{eqnarray}
\frac{\partial }{\partial t}P_{2}^{2;2}(\alpha ,\alpha ^{+}) &=&-i\,\omega
\,\left( -\frac{\partial }{\partial \alpha }\alpha \,+\frac{\partial }{%
\partial \alpha ^{+}}\alpha ^{+}\right) \,P_{2}^{2;2}(\alpha ,\alpha ^{+}) 
\nonumber \\
&&+\frac{1}{2}i\,\Omega \,\left( \alpha ^{+}P_{2}^{1;2}(\alpha ,\alpha
^{+})\,-\alpha P_{2}^{2;1}(\alpha ,\alpha ^{+})\right)  \nonumber \\
&&+\frac{1}{2}i\,\Omega \left( -\frac{\partial }{\partial \alpha }%
P_{2}^{1;2}(\alpha ,\alpha ^{+})+\frac{\partial }{\partial \alpha ^{+}}%
P_{2}^{2;1}(\alpha ,\alpha ^{+})\right)  \label{Eq.SecondOrderCoeft22}
\end{eqnarray}

For the \emph{fourth order} term%
\begin{eqnarray}
\frac{\partial }{\partial t}P_{4}^{12;21}(\alpha ,\alpha ^{+})\,
&=&-i\,\omega \,\left( -\frac{\partial }{\partial \alpha }\alpha \,+\frac{%
\partial }{\partial \alpha ^{+}}\alpha ^{+}\right) P_{4}^{12;21}(\alpha
,\alpha ^{+})  \nonumber \\
&&-\frac{1}{2}i\,\Omega \left( \,\frac{\partial }{\partial \alpha }%
P_{2}^{1;2}(\alpha ,\alpha ^{+})\,-\,\frac{\partial }{\partial \alpha ^{+}}%
P_{2}^{2;1}(\alpha ,\alpha ^{+})\,\right)  \label{Eq.FourthOrderCoeft}
\end{eqnarray}%
Thus we see that for the general distribution $P(\alpha ,\alpha ^{+},\alpha
^{\ast },\alpha ^{+\ast },g,g^{+})$ the coefficients satisfy similar sets of
coupled equations as in the canonical distribution case.\pagebreak

\subsection{3.3 Rotating Phase Variables and Coefficients}

The transformation to rotating phase variables is as in Eq. (\ref%
{Eq.RotatingPhaseVariables}). The new expansion coefficients will be
designated $R(\underrightarrow{\beta })$ where $\underrightarrow{\beta }%
=\{\beta ,\beta ^{+},\beta ^{\ast },\beta ^{+\ast }\}$ and are related to
the original $P(\underrightarrow{\alpha })$ via 
\begin{eqnarray}
P_{2}^{1;1}(\underrightarrow{\mathbf{\alpha }}) &=&R_{2}^{1;1}(%
\underrightarrow{\mathbf{\beta }})\qquad P_{2}^{2;2}(\underrightarrow{%
\mathbf{\alpha }})=R_{2}^{2;2}(\underrightarrow{\mathbf{\beta }})  \nonumber
\\
P_{2}^{1;2}(\underrightarrow{\mathbf{\alpha }}) &=&R_{2}^{1;2}(%
\underrightarrow{\mathbf{\beta }})\exp (-i\omega t)\qquad P_{2}^{2;1}(%
\underrightarrow{\mathbf{\alpha }})=R_{2}^{2;1}(\underrightarrow{\mathbf{%
\beta }})\exp (+i\omega t)  \nonumber \\
P_{0}(\underrightarrow{\mathbf{\alpha }}) &=&R_{0}(\underrightarrow{\mathbf{%
\beta }})\qquad P_{4}^{12;21}(\underrightarrow{\mathbf{\alpha }}%
)=R_{4}^{12;21}(\underrightarrow{\mathbf{\beta }})
\label{Eq.FinalQuantitiesB}
\end{eqnarray}%
Coupled equations for $R_{0}(\underrightarrow{\beta }),R_{2}^{i;j}(%
\underrightarrow{\beta })$ and $R_{4}^{12;21}(\underrightarrow{\beta })$ are
obtained which only involve $\Omega $ and $\Delta $. These are as follows.

The \emph{second order} equations are 
\begin{equation}
\frac{\partial }{\partial t}\left( R_{2}^{1;1}(\underrightarrow{\mathbf{%
\beta }})-R_{0}(\underrightarrow{\mathbf{\beta }})\right) =+\frac{1}{2}%
i\,\Omega \,\left( \beta \,R_{2}^{2;1}(\underrightarrow{\mathbf{\beta }}%
)-\beta ^{+}\,R_{2}^{1;2}(\underrightarrow{\mathbf{\beta }})\right)
\label{Eq.FinalSecondOrderCoeftR11}
\end{equation}%
\begin{eqnarray}
\frac{\partial }{\partial t}R_{2}^{1;2}(\underrightarrow{\mathbf{\beta }})
&=&-i\,\Delta \,R_{2}^{1;2}(\underrightarrow{\mathbf{\beta }})\,  \nonumber
\\
&&-\frac{1}{2}i\,\Omega \,\beta \,\left( \left\{ R_{2}^{1;1}(%
\underrightarrow{\mathbf{\beta }})-R_{0}(\underrightarrow{\mathbf{\beta }}%
)\right\} -\left\{ R_{2}^{2;2}(\underrightarrow{\mathbf{\beta }})-R_{0}(%
\underrightarrow{\mathbf{\beta }})\right\} \right)  \nonumber \\
&&+\,\frac{1}{2}i\,\Omega \left( \frac{\partial }{\partial \beta ^{+}}%
\left\{ R_{2}^{1;1}(\underrightarrow{\mathbf{\beta }})-R_{0}(%
\underrightarrow{\mathbf{\beta }})\right\} \right)
\label{Eq.FinalSecondOrderR12}
\end{eqnarray}%
\begin{eqnarray}
\frac{\partial }{\partial t}R_{2}^{2;1}(\underrightarrow{\mathbf{\beta }})
&=&+i\,\Delta \,R_{2}^{2;1}(\underrightarrow{\mathbf{\beta }})\,  \nonumber
\\
&&+\frac{1}{2}i\,\Omega \,\beta ^{+}\left( \left\{ R_{2}^{1;1}(%
\underrightarrow{\mathbf{\beta }})-R_{0}(\underrightarrow{\mathbf{\beta }}%
)\right\} -\left\{ R_{2}^{2;2}(\underrightarrow{\mathbf{\beta }})-R_{0}(%
\underrightarrow{\mathbf{\beta }})\right\} \right)  \nonumber \\
&&-\frac{1}{2}i\,\Omega \,\left( \frac{\partial }{\partial \beta }\left\{
R_{2}^{1;1}(\underrightarrow{\mathbf{\beta }})-R_{0}(\underrightarrow{%
\mathbf{\beta }})\right\} \right)  \label{Eq.FinalSecondOrderR21}
\end{eqnarray}%
\begin{eqnarray}
\frac{\partial }{\partial t}\left( R_{2}^{2;2}(\underrightarrow{\mathbf{%
\beta }})-R_{0}(\underrightarrow{\mathbf{\beta }})\right) &=&+\frac{1}{2}%
i\,\Omega \,\left( \beta ^{+}\,R_{2}^{1;2}(\underrightarrow{\mathbf{\beta }}%
)\,-\beta \,R_{2}^{2;1}(\underrightarrow{\mathbf{\beta }})\right)  \nonumber
\\
&&+\frac{1}{2}i\,\Omega \left( -\frac{\partial }{\partial \beta }R_{2}^{1;2}(%
\underrightarrow{\mathbf{\beta }})+\frac{\partial }{\partial \beta ^{+}}%
R_{2}^{2;1}(\underrightarrow{\mathbf{\beta }})\right)
\label{Eq.FinalSecondOrderR22}
\end{eqnarray}

The \emph{zeroth} and \emph{fourth order} equations are%
\begin{equation}
\frac{\partial }{\partial t}R_{0}(\underrightarrow{\mathbf{\beta }})=0
\label{Eq.FinalZeroOrderCoefftR0}
\end{equation}%
\begin{equation}
\frac{\partial }{\partial t}R_{4}^{12;21}(\underrightarrow{\mathbf{\beta }}%
)\,=-\frac{1}{2}i\,\Omega \left( \,\frac{\partial }{\partial \beta }%
R_{2}^{1;2}(\underrightarrow{\mathbf{\beta }})\,-\,\frac{\partial }{\partial
\beta ^{+}}R_{2}^{2;1}(\underrightarrow{\mathbf{\beta }})\,\right)
\label{Eq.FinalFourthOrderCoeftR1221}
\end{equation}%
\medskip

\subsection{3.4 Solution for Distribution Function}

The solution to the coupled equation for the rotating distribution function
coefficients is found using the ansatz%
\begin{equation}
R_{2}^{i;j}(\underrightarrow{\mathbf{\beta }})=\exp (\beta \beta ^{+})\;\Phi
_{i}^{\ast }(\beta ^{+})\,\Phi _{j}(\beta )  \label{Eq.AnsatzGenDistnFn}
\end{equation}%
where the $\Phi _{i}^{\ast }(\beta ^{+})$ are functions of the $\beta ^{+}$
and the $\Phi _{i}(\beta )$ are functions of the $\beta $, and where $\Phi
_{i}(\beta )$ satisfy the coupled equations%
\begin{eqnarray}
\frac{\partial }{\partial t}\Phi _{1}(\beta ) &=&\frac{1}{2}i\Delta \,\Phi
_{1}(\beta )-\frac{1}{2}i\Omega \,\left( \frac{\partial }{\partial \beta }%
\right) \Phi _{2}(\beta )  \nonumber \\
\frac{\partial }{\partial t}\Phi _{2}(\beta ) &=&-\frac{1}{2}i\Delta \,\Phi
_{2}(\beta )-\frac{1}{2}i\Omega \,\beta \,\Phi _{1}(\beta )
\label{Eq.CoupledAnsatz}
\end{eqnarray}%
This ansatz is consistent with the original equations (\ref%
{Eq.FinalSecondOrderCoeftR11}) -(\ref{Eq.FinalSecondOrderR22}) for the $%
R_{2}^{i;j}(\underrightarrow{\mathbf{\beta }})$. As indicated previously we
set $R_{0}(\underrightarrow{\mathbf{\beta }}))=0$ for the one atom case.

Following a similar proceedure as for the canonical distribution function we
find a solution for the $\Phi _{i}(\beta )$ that is single valued and not
divergent at the origin which is of the form%
\begin{equation}
\Phi _{1}(\beta )=\beta ^{m}\left( A_{m}\cos \frac{1}{2}\nu _{m}t+B_{m}\sin 
\frac{1}{2}\nu _{m}t\right)
\end{equation}%
where 
\begin{equation}
\nu _{m}=\sqrt{-m\,\Omega ^{2}+\Delta ^{2}}\qquad m=0,1,2,..
\end{equation}%
However these are \emph{not} the frequenies associated with population and
coherence oscillations in the one atom Jaynes-Cummings model. In fact for
any detuning and Rabi frequency the quantity $\nu _{m}$ becomes purely
imaginary when $m$ is large, leading to solutions for $\Phi _{1}(\beta )$
and $\Phi _{2}(\beta )$ involving \emph{hyperbolic functions}. For example
with zero detuning we have 
\begin{equation}
\Phi _{1}(\beta )=\beta ^{m}\left( A_{m}\cosh \frac{1}{2}\sqrt{m}\Omega
t-iB_{m}\sinh \frac{1}{2}\sqrt{m}\Omega t\right)
\end{equation}%
These solutions \emph{diverge} as $t$ becomes large. This indicates that the
standard distribution function obtained from the Fokker-Planck equation via
the standard correspondence rules becomes infinite for large time. The
corresponding solution for $\Phi _{2}(\beta )$ is obtained from Eq.(\ref%
{Eq.CoupledAnsatz}) and thus%
\begin{equation}
\Phi _{2}(\beta )=i\;\beta ^{(m-1)}\left( \cos \frac{1}{2}\nu _{m}t\left\{ 
\frac{\nu _{m}B_{m}+i\Delta A_{m}}{\Omega }\right\} +\sin \frac{1}{2}\nu
_{m}t\left\{ \frac{-\nu _{m}A_{m}+i\Delta B_{m}}{\Omega }\right\} \right)
\end{equation}%
However, a solution with $m=0$ leads to a singular $\beta ^{-1}$ behaviour,
so it follows that $m$ is restricted to the positive integers. Also, as the
ansatz equations are linear the general solution is a sum of terms with
differing $m$ so that we finally have the solution in the form 
\begin{eqnarray}
\Phi _{1}(\beta ) &=&\dsum\limits_{m=1}\beta ^{m}\left( A_{m}\cos \frac{1}{2}%
\nu _{m}t+B_{m}\sin \frac{1}{2}\nu _{m}t\right)  \nonumber \\
\Phi _{2}(\beta ) &=&i\;\dsum\limits_{m=1}\beta ^{(m-1)}\left( \cos \frac{1}{%
2}\nu _{m}t\left\{ \frac{\nu _{m}B_{m}+i\Delta A_{m}}{\Omega }\right\} +\sin 
\frac{1}{2}\nu _{m}t\left\{ \frac{-\nu _{m}A_{m}+i\Delta B_{m}}{\Omega }%
\right\} \right)  \nonumber \\
&&  \label{Eq.GenSolnGeneralP}
\end{eqnarray}

Overall, this solution in (\ref{Eq.AnsatzGenDistnFn}) and (\ref%
{Eq.GenSolnGeneralP}) for the distribution function is unsatisfactory. Not
only does it diverge for large $t$ but also the distribution function found
diverges at all times as $\beta $, $\beta ^{+}$ become large due to the $%
\exp (\beta \beta ^{+})\,$\ factor. This contradicts the requirement in
deriving the Fokker-Planck equation that the distribution function goes to
zero for large $\beta $, $\beta ^{+}$. The conclusion is that the
Fokker-Planck equation (\ref{Eq.Fokker-Planck}) obtained via the standard
correspondence rules must be invalid, since if it was valid then the general
solution to it that we have found would have produced a distribution
function that was not divergent either in phase space or in time. There is
also the question of how to choose the quantities $A_{n}$ and $B_{n}$ so
that a valid overall distribution function at time $t=0$ is obtained. The
form of the initial distribution function is not known, since the canonical
distribution function is not being used in this case where the Fokker-Planck
equation is based on the standard correspondence rules. Of course the only
requirement is that the choice of distribution function gives the correct
characteristic function - the latter being uniquely determined from the
density operator. Since the characteristic function is obtained via the
phase space integral of Eq. (\ref{Eq.BoseFermiDistnFn}) it is clear that
there is a problem - the bosonic integral of the distribution function will
not converge due to the $\exp (\beta \beta ^{+})\,$\ factor. No matter how
the quantities $A_{n}$ and $B_{n}$ are chosen no valid positive $P$
distribution function can therefore be found via the solution given by (\ref%
{Eq.GenSolnGeneralP}) and (\ref{Eq.AnsatzGenDistnFn}).\medskip \pagebreak


\begin{thebibliography}{99}
\bibitem{JaynesCummings63a} E. T. Jaynes and F. W. Cummings, \textit{Proc.
I. E. E. E}. \textbf{51}, 89 (1963).

\bibitem{Mollow69a} B. R. Mollow, \textit{Phys. Rev. }\textbf{188}, 1969
(1969).

\bibitem{Cohen76a} C. Cohen-Tannoudji and S. Reynaud, \textit{J. Phys. B:
Atom. Molec. Opt. Phys. }\textbf{10}, 345 (1976).

\bibitem{Schuda74a} F. Scuda, C. R. Stroud Jr. and M. Hercher, \textit{J.
Phys. B: Atom. Molec. Opt. Phys. }\textbf{7}, L198 (1974).

\bibitem{Wu75a} F. Y. Wu, R. E. Grove and S. Ezekiel, \textit{Phys. Rev. Lett%
}. \textbf{35}, 1426 (1975).

\bibitem{Hartig76a} W. Hartig, W. Rasmussen, R. Schieder and H. Walther, 
\textit{Z. Phys. A, }\textbf{278}, 205 (1976).

\bibitem{Eberly80a} J. H. Eberly, N. B. Narozhny and J. J.
Sanchez-Mondragon, \textit{Phys. Rev. Lett}. \textbf{44}, 1323 (1980).

\bibitem{Barnett86a} S. M. Barnett, P. Filipowicz, J. Javanainen, P. L.
Knight and P. Meystre, \textit{The Jaynes-Cummings Model and Beyond, }in 
\textit{Frontiers in Quantum Optics, }editors E. R. Pike and S. Sarkar,
(Adam Hilger, Bristol, UK, 1986) p 485.

\bibitem{Gerry05a} C. C. Gerry and P. L. Knight, \textit{Introductory
Quantum Optics, }(Cambridge University Press, Cambridge, UK, 2005).

\bibitem{Rempe87a} G. Rempe, H. Walther and N. Klein, \textit{Phys. Rev. Lett%
}. \textbf{58}, 353 (1987).

\bibitem{Haroche96a} M. Brune, F. Schmidt-Kaler, A. Maaili, J. Dreyer, E.
Hagley, J. M. Raimond and S. Haroche, \textit{Phys. Rev. Lett}. \textbf{76},
1800 (1996).

\bibitem{Shore93a} B. W. Shore and P. L. Knight, \textit{J. Mod. Opt. }%
\textbf{40}, 1195 (1993).

\bibitem{Barnett97a} S. M. Barnett and P. M. Radmore, \textit{Methods of
Theoretical Quantum Optics, }(Clarendon, Oxford, UK, 1997).

\bibitem{Stenholm81a} S. Stenholm, \textit{Opt. Comm. }\textbf{36}, 75
(1981).

\bibitem{Gardiner91a} C. W. Gardiner, \textit{Quantum Noise}
(Springer-Verlag, Berlin, 1991).

\bibitem{Walls94a} D. F. Walls and G. J. Milburn, \textit{Quantum Optics, }%
(Springer-Verlag, Berlin, Germany, 1994).

\bibitem{Scully97a} M. O. Scully and M. S. Zubairy, \textit{Quantum Optics, }%
(Cambridge, Cambridge, UK, 1997).

\bibitem{Drummond80a} P. D. Drummond and C. W. Gardiner, \textit{J. Phys. A:
Math. Gen. Phys. }\textbf{13}, 2353 (1980).

\bibitem{Berezin66a} F. A. Berezin, \textit{The Method of Second Quantization%
}, (Academic Press, New York, USA, 1966).

\bibitem{Cahill99a} K.E. Cahill and R. J. Glauber, \textit{Phys. Rev. A }%
\textbf{59}, 1538 (1999).

\bibitem{Plimak01a} L. Plimak, M.J. Collett and M.K. Olsen, \textit{Phys.
Rev. A} \textbf{64}, 063409 (2001).

\bibitem{Anastopoulos00a} C. Anastopoulos \& B.L. Hu, \textit{Phys. Rev. A }%
\textbf{62}, 033821 (2000).

\bibitem{Shresta05a} S. Shresta, C. Anastopoulos, A. Dragulescu and B. L.
Hu, \textit{Phys. Rev. A} \textbf{71}, 022109 (2005).

\bibitem{Plimak09a} L. Plimak and S. Stenholm, \textit{Ann. Phys. }\textbf{%
324}, 600 (2009).

\bibitem{Zinn-Justin02a} J. Zinn-Justin, \textit{Quantum Field Theory and
Critical Phenomena}, (Clarendon Press, Oxford, UK, 2002).

\bibitem{Rivers87a} R. J. Rivers, \textit{Path Integral Methods in Quantum
Field Theory, (}Cambridge University Press, Cambridge, UK,1987).

\bibitem{Blaizot86a} J-P. Blaizot and G. Ripka, \textit{Quantum Theory of
Finite Systems, }(MIT Press, Cambridge, Massachusetts, USA, 1986).

\bibitem{Schack91a} R. Schack and A. Schenzle, \textit{Phys. Rev. A }\textbf{%
44}, 682 (1991).

\bibitem{Eiselt91a} J. Eiselt and H. Risken, \textit{Phys. Rev. A} \textbf{43%
}, 346 (1991).

\bibitem{Drummond87a} P. D. Drummond and S. J. Carter, \textit{J. Opt. Soc.
Amer. B}, \textbf{4}, 1565 (1987).

\bibitem{Kennedy88a} T. A. B. Kennedy and E. M. Wright, \textit{Phys. Rev. A}
\textbf{38}, 212 (1988).

\bibitem{Gatti97a} A. Gatti, H. Wiedemann, L.Lugiato, I. Marzoli, G-L. Oppo
and S. M. Barnett, \textit{Phys. Rev. A} \textbf{56}, 877 (1997).

\bibitem{Steel98a} M. J. Steel, M. K. Olsen, L. I. Plimak, P. D. Drummond,
S. M. Tan, M. J. Collett, D. F. Walls and R. Graham, \textit{Phys. Rev. A} 
\textbf{58}, 4824 (1998).

\bibitem{Graham70a} R. Graham and H. Haken, \textit{Z. Phys. }\textbf{235},
166 (1970).

\bibitem{Corney06a} J. F. Corney and P. D. Drummond, \textit{J. Phys. A:
Math. Gen. Phys. }\textbf{39}, 269 (2006).

\bibitem{Corney03a} J. F. Corney and P. D. Drummond, \textit{Phys. Rev. A} 
\textbf{68}, 063822 (2003).

\bibitem{Stenholm73a} S. Stenholm. \textit{Phys. Rep. }\textbf{6}, 1 (1973).

\bibitem{Kaku93a} M. Kaku, \textit{Quantum Field Theory}, (Oxford University
Press, Oxford, UK, 1993) p52.

\bibitem{Glauber65a} R. J. Glauber, in \textit{Quantum Optics and
Electronics, }edited by C. De Witt \textit{et al }(Gordon and Breach, New
Yrk, 1965), pp65-185.

\bibitem{Gradshteyn65a} I. S. Gradshteyn and I. M. Ryzhik, \textit{Tables of
Integrals, Series and Products}, (Academic Press, New York, USA, 1965), p3,
4.

\bibitem{Gilchrist97a} A. Gilchrist, C. W. Gardiner and P. D. Drummond, 
\textit{Phys. Rev. A }\textbf{55}, 3014 (1997).

\bibitem{Cahill69a} K.E. Cahill and R. J. Glauber, \textit{Phys. Rev. }%
\textbf{177}, 1857, 1882 (1969).
\end{thebibliography}
\end{document}